\documentclass[12pt, twocolappendix, appendixfloats, numberedappendix]{emulateapj}

\def\eg{e.g., }

\newcommand{\mpc}{{\rm\,Mpc}}
\newcommand{\kpc}{{\rm\,kpc}}
\newcommand{\pc}{{\rm\,pc}}

\newcommand{\beq}{\begin{equation}}
\newcommand{\eeq}{\end{equation}}

\def\hyi{\ion{H}{1}}
\def\hyii{\ion{H}{2}}
\def\hei{\ion{He}{1}}

\def\ni{[\ion{N}{1}]}
\def\nii{[\ion{N}{2}]}
\def\niilam{[\ion{N}{2}]\,$\lambda6584$}
\def\niidoublet{[\ion{N}{2}]\,$\lambda\lambda6548,6583$}
\def\oii{[\ion{O}{2}]}

\def\oiidoublet{[\ion{O}{2}]\,$\lambda\lambda3727,3730$}
\def\oiiiperm{\ion{O}{3}}
\def\oiii{[\ion{O}{3}]}
\def\oiiilam{[\ion{O}{3}]\,$\lambda5008$}

\def\mnii{\ion{Mn}{2}}
\def\mniitriplet{\ion{Mn}{2}\,$\lambda\lambda\lambda2577,2594,2606$}
\def\mgi{\ion{Mg}{1}}
\def\mgii{\ion{Mg}{2}}

\def\mgiidoublet{\ion{Mg}{2}\,$\lambda\lambda2796,2804$}
\def\fei{\ion{Fe}{1}}
\def\feii{\ion{Fe}{2}}
\def\feiiast{\ion{Fe}{2}*}
\def\tiii{\ion{Ti}{2}}

\def\cii{\ion{C}{2}}
\def\ciisemi{\ion{C}{2}]}
\def\ciiforb{[\ion{C}{2}]}
\def\ciii{\ion{C}{3}}

\def\neiii{[\ion{Ne}{3}]}
\def\neiv{[\ion{Ne}{4}]}
\def\suii{[\ion{S}{2}]}
\def\suiidoublet{[\ion{S}{2}]\,$\lambda\lambda6718,6733$}

\def\oi{[\ion{O}{1}]}
\def\nai{\ion{Na}{1}}

\def\caii{\ion{Ca}{2}}

\def\ariii{[\ion{Ar}{3}]}

\def\ha{H$\alpha$}
\def\hb{H$\beta$}

\newcommand{\kms}{\ensuremath{{\rm km\,s}^{-1}}}

\newcommand{\rewmgiione}{\ensuremath{W_0^{\lambda2796}}}

\newcommand{\MSun}{\ensuremath{\rm M_\odot}}

\newcommand{\ud}{\ensuremath{{\rm d}}}

\usepackage{natbib}
\usepackage{rotating}
\begin{document}

\shorttitle{NUV spectroscopy of star-forming galaxies}
\shortauthors{Zhu et al.}
\title {Near-ultraviolet Spectroscopy of Star-forming Galaxies from \lowercase{e}BOSS: \\
Signatures of Ubiquitous Galactic-scale Outflows}

\author{
Guangtun Ben Zhu\altaffilmark{1,2}, Johan Comparat\altaffilmark{3,4}, Jean-Paul Kneib\altaffilmark{5,6}, Timoth{\'e}e Delubac\altaffilmark{5}, Anand Raichoor\altaffilmark{7}, Kyle S. Dawson\altaffilmark{8}, Jeffrey Newman\altaffilmark{9}, Christophe Y{\`e}che\altaffilmark{7}, Xu Zhou\altaffilmark{10}, 
and Donald P. Schneider\altaffilmark{11, 12}
} 
\altaffiltext{1}{Department of Physics \& Astronomy, Johns Hopkins University, 3400 N. Charles Street, Baltimore, MD 21218, USA, guangtun@jhu.edu}
\altaffiltext{2}{Hubble Fellow}
\altaffiltext{3}{Instituto de F\'{\i}sica Te\'orica, UAM/CSIC, Universidad Aut\'onoma de Madrid, Cantoblanco, E-28049 Madrid, Spain}
\altaffiltext{4}{Departamento de F\'{\i}sica Te\'orica, Universidad Aut\'onoma de Madrid, Cantoblanco, E-28049 Madrid, Spain}
\altaffiltext{5}{Laboratoire d\'astrophysique, Ecole Polytechnique F\'ed\'erale de Lausanne, Observatoire de Sauverny, 1290 Versoix, Switzerland}
\altaffiltext{6}{Aix Marseille Universit\'e, CNRS, LAM (Laboratoire d'Astrophysique de Marseille), UMR 7326, 13388, Marseille, France}
\altaffiltext{7}{CEA, Centre de Saclay, Irfu/SPP,  F-91191 Gif-sur-Yvette, France}
\altaffiltext{8}{Department of Physics and Astronomy, University of Utah, Salt Lake City, UT 84112, USA}
\altaffiltext{9}{PITT PACC, Department of Physics and Astronomy, University of Pittsburgh, Pittsburgh, PA 15260, USA}
\altaffiltext{10}{Key Laboratory of Optical Astronomy, National Astronomical Observatories, Chinese Academy of Sciences, Beijing, 100012, China}
\altaffiltext{11}{Department of Astronomy and Astrophysics, The Pennsylvania State University, University Park, PA 16802}
\altaffiltext{12}{Institute for Gravitation and the Cosmos, The Pennsylvania State University, University Park, PA 16802}

\begin{abstract}
We present the rest-frame near-ultraviolet (NUV) spectroscopy of star-forming galaxies (SFGs) at $0.6<z<1.2$ from the Extended Baryon Oscillation Spectroscopic Survey (eBOSS) in SDSS-IV. 
One of the eBOSS programs is to obtain $2\arcsec$ (about $15\,\kpc$) fiber spectra of about $200,000$ emission-line galaxies (ELGs) at redshift $z\gtrsim0.6$. We use the data from the pilot observations of this program, including $8620$ spectra of SFGs at $0.6<z<1.2$. 
The median composite spectra of these SFGs at $2200\,{\rm \AA}<\lambda<4000\,{\rm \AA}$ feature asymmetric, preferentially blueshifted non-resonant emission, \feiiast, and blueshifted resonant absorption, e.g., \feii\ and \mgii, indicating ubiquitous outflows driven by star formation at these redshifts. For the absorption lines, we find a variety of velocity profiles with different degrees of blueshift.
Comparing our new observations with the literature, we do not observe the non-resonant emission in the small-aperture ($<40\,{\rm pc}$) spectra of local star-forming regions with the \textit{Hubble Space Telescope} (\textit{HST}), and find the observed line ratios in the SFG spectra to be different from those in the spectra of local star-forming regions, as well as those of quasar absorption-line systems in the same redshift range. 
We introduce an outflow model that can simultaneously explain the multiple observed properties and suggest that the variety of absorption velocity profiles and the line ratio differences are caused by scattered fluorescent emission filling in on top of the absorption in the large-aperture eBOSS spectra. We develop an observation-driven, model-independent method to correct the emission-infill to reveal the true absorption profiles. 
Finally, we show that the strengths of both the non-resonant emission and the emission-corrected resonant absorption increase with \oiidoublet\ rest equivalent width and luminosity, with a slightly larger dependence on the former.
Our results show that eBOSS and future dark-energy surveys (e.g., DESI and PFS) will provide rich datasets of NUV spectroscopy for astrophysical applications.
\end{abstract}


\section {Introduction}

Spectroscopy is one of the most important tools in modern astronomy. Historically, ground-based observations have mostly been focused on the optical window of the spectrum because of the extinction effects of the Earth's atmosphere and the low brightness of most cosmological sources. In the near-ultraviolet (NUV) range at $2000\,{\rm \AA}\lesssim\lambda\lesssim3000\,{\rm \AA}$, the most-studied sources are quasars \citep[\eg][]{schmidt63a, francis91a, vandenberk01a} and quasar absorption-line systems \citep[\eg][]{burbidge66a, bergeron86a, steidel92a, churchill00a} at moderate redshifts ($0.5\lesssim z \lesssim 2.5$), thanks to the high energy-conversion efficiency of supermassive black holes. After space-based telescopes became available, such as the \textit{International Ultraviolet Explorer} (\textit{IUE}) and the \textit{Hubble Space Telescope} (HST), astronomers started to explore the UV part of the spectral energy distributions (SEDs) of stars and galaxies. Most of the space-based observations are dedicated to the far-UV (FUV) window at $\lambda\lesssim2000\,{\rm \AA}$ because of the large number of transitions available at those energies, leaving the NUV window of the SEDs of stars and galaxies largely-unexplored territory. Most of our knowledge about galaxy SEDs in the NUV have originated from most recent observations with the $8$-meter or larger telescopes such as the Keck and VLT telescopes \citep[\eg][]{martin09a, weiner09a, rubin11a, talia12a, erb12a}.

The most useful spectral features in the NUV have been the resonant\footnote{We refer to a transition between two terms as a resonant transition if and only if the lower energy level is the lowest (ground) level of the ground term. See Appendix~\ref{app:atomic} for clarification.} absorption lines induced by the low-ionization species \mgii\ and \feii\ in the interstellar medium (ISM) and/or circumgalactic medium(CGM). In the spectra of star-forming galaxies (SFGs), these absorption lines are observed to be almost always blueshifted from the systemic velocity of the galaxy, which serves as direct evidence for ubiquitous outflows driven by star formation \citep[][]{tremonti07a, martin09a, weiner09a, rubin10a, rubin14a, talia12a, erb12a, martin12a, kornei12a}. 

More recently, observations have also revealed the non-resonant emission features \feii$^*$ in the NUV spectra of SFGs \citep[][]{rubin11a, talia12a, erb12a, kornei12a, kornei13a}. These non-resonant emission lines are likely caused by scattered fluorescent photons after the occurrence of resonant absorption. As the non-resonant emission lines in the FUV \citep[\eg][]{shapley03a, jones12a}, they have provided a new means to study the radiative transfer physics in the complex baryon processes in galaxy evolution \citep[\eg][]{rubin11a, prochaska11a, scarlata15a}. 

The non-resonant emission, e.g., \feii$^*$, was sometimes referred to as fine-structure emission in the literature, because the lower energy level of the transition is an excited level due to fine-structure splitting, as opposed to the ground level for the corresponding resonant transition. The term \textit{fine-structure emission}, however, is used more often for the emission between two fine-structure levels (with different $J$-values) in the same term (with the same $L, S$), which usually occurs in the infrared (IR), such as \ciiforb$\,\lambda157.7\,\micron$. To avoid confusion, we here simply use the term \textit{non-resonant emission} and reserve \textit{fine-structure emission} for those infrared emission lines. We also label the (permitted) non-resonant emission lines with a right superscript asterisk ``$*$'', although the origin of this convention appears to have been lost in the literature\footnote{In \textit{SYMBOLS, UNITS, NOMENCLATURE AND FUNDAMENTAL CONSTANTS IN PHYSICS}, or the ``Red Book'', compiled by International Union of Pure and Applied Physics (IUPAP), the authors recommended this convention for excited atomic levels \citep[see Section~{2.1},][]{cohen87a}.}.

The fine-structure splitting causes multiple transitions, i.e., multiplets, with energies only slightly different from each other. In her pioneering multiplet tables \citep[][]{moore45a, moore50a},  Charlotte E. Moore grouped the multiplets in the ascending order of the excited terms and this numbering convention has been widely used since. We follow the same convention for the multiplets in the UV \citep[][]{moore50a, moore52a}, though we skip the numbering of multiplets in the optical as in the recent literature. In Appendix~\ref{app:atomic}, we summarize our choices of conventions and also present the atomic data needed in this analysis for references.

In the last decade, the detection of the baryon acoustic oscillation (BAO) in the large-scale distribution of galaxies \citep[][]{eisenstein05a, cole05a} has shifted the focus of observational cosmology to massive redshift surveys over the entire observable sky, and we are now in the course of measuring the scale of BAO as a function of cosmic time and chronicling the expansion history of the Universe. Among them are the Extended Baryon Oscillation Spectroscopic Survey \citep[eBOSS,][]{dawson15a} within the fourth phase of the Sloan Digital Sky Survey (SDSS-IV, Blanton et al. in prep.), the Dark Energy Spectroscopic Instrument survey \citep[DESI,][]{schlegel11a, levi13a}, and the Prime Focus Spectrograph survey \citep[PFS,][]{takada14a}. These programs will obtain optical spectra in the observer frame for tens of millions of galaxies at redshift $0.6\lesssim z \lesssim 2.0$, providing unprecedentedly rich spectroscopic datasets with the coverage in the rest-frame NUV.

Among the new BAO projects, the eBOSS survey started taking data in Fall 2014. One of the large-scale structure tracers in eBOSS is emission-line galaxies (ELGs). We have obtained about $12,000$ ELG spectra across the redshift range $0<z\lesssim1.5$ in the pilot observations, conducted to optimize targeting strategies. We discuss the whole sample briefly and focus on the NUV spectroscopy of $8620$ ELGs at redshift $0.6<z<1.2$. We describe the observations and the dataset in detail in Section~\ref{sec:data}.

The rest of the paper is organized as follows. In Section~\ref{sec:results}, we present the composite spectrum of ELGs in the NUV and compare it with those of quasar absorption-line systems and local star-forming (SF) regions. We interpret the observations with a spherical gas outflow model in Section~\ref{sec:interpretation}. In Section~\ref{sec:oii}, we investigate the dependence of the spectral features on the \oiidoublet\ properties. We summarize our results in Section~\ref{sec:summary}. When necessary, we assume the $\Lambda$CDM cosmogony, with $\Omega_\Lambda=0.7$, $\Omega_{\mathrm m}=0.3$, and ${\mathrm H}_0=70\,\kms\,\mpc^{-1}$. Throughout the paper, we use vacuum wavelength and some of the lines are therefore labeled as $1\,$\AA\ longer than if labeled with the air wavelength, such as \oiii$\,\lambda5008$.

\section{Data}\label{sec:data}

\subsection{The eBOSS Survey}\label{sec:ebosssurvey}

The eBOSS survey is the cosmology survey within SDSS-IV, an extension of the BOSS survey \citep[][]{dawson13a} in SDSS-III \citep[][]{eisenstein11a}. Over a six-year period beginning in Fall 2014, eBOSS will observe four independent tracers of the underlying density field: luminous red galaxies (LRGs), quasars, Ly$\alpha$ forest, and ELGs, and measure the cosmological distances as a function of redshift with BAO as a standard ruler and record the expansion history of the Universe at $0.6\lesssim z \lesssim 2.3$.

The eBOSS survey uses the same two identical, multi-object spectrographs as in BOSS \citep[][]{smee13a} on the $2.5$-meter SDSS Telescope \citep[][]{gunn06a} at the Apache Point Observatory in New Mexico. Each spectrograph is equipped with two cameras, one blue and one red, with a dichroic splitting the light at roughly $6000\,$\AA\ and a combined coverage spanning between $3600\,$\AA\ and $10400\,$\AA. The spectral resolution $\mathcal{R}$ is $1560-2270$ in the blue and $1850-2650$ in the red channel, with a mean approximately $\bar{\mathcal{R}}\approx2000$. The spectrographs are fed by $1000$ optical fibers ($500$ for each), covering a field-of-view (FoV) of about $7.5$ square degrees. The aperture diameter of the fibers is $2\,$\arcsec, smaller than $3\,$\arcsec\ in SDSS-I/II. The typical total exposure time for each pointing is about $75$ minutes.

eBOSS selects targets primarily based on the SDSS imaging, obtained through a set of $ugriz$ filters \citep[][]{fukugita96a} with a wide-field camera \citep[][]{gunn98a} in a drift-scan mode. As the desired targets are fainter than those in SDSS I-III, in order to achieve high targeting efficiency, eBOSS includes supplementary imaging data with other instruments, including the infrared photometry \citep[][]{lang14a} by the \textit{Wide-field Infrared Survey Explorer} \citep[\textit{WISE},][]{wright10a}, the U-band imaging from the South Galactic Cap $U$-band Sky Survey (SCUSS)\footnote{\texttt{http://batc.bao.ac.cn/Uband/}} conducted at the 2.3-meter Bok telescope at the Kitt Peak National Observatory, and the deep $grz$ imaging with the Dark Energy Camera \citep[DECam,][]{flaugher12a} on the 4-meter Blanco telescope at the Cerro Tololo Inter-American Observatory (CTIO) in Chile. 

The spectral-reduction and redshift-fitting pipeline in eBOSS is a continuation of the BOSS pipeline \citep[][]{bolton12a}, improved to yield required performance on spectra with lower signal-to-noise ratio (S/N) than in BOSS. As BOSS, eBOSS classifies objects and derives redshifts based upon principal-component-analysis (PCA) templates for quasars and galaxies, and archetypal templates for stars, though the team is currently exploring an alternative approach fully based on archetypal templates. For the analysis presented here, we use the results derived from the pipeline version \texttt{v5\_7\_8}, which still uses the PCA templates for galaxies and quasars. The pipeline flags a redshift with a warning (\texttt{ZWARNING}) based on various quantitative criteria, such as a small $\chi^2$ difference between the best and second-best fits. Tests have shown that the \texttt{ZWARNING} is conservatively defined and the redshift success rate for objects with \texttt{ZWARNING==0} is better than $99\%$. We only use objects classified as a \texttt{GALAXY} with \texttt{ZWARNING==0} in this analysis.

For more details regarding the eBOSS survey, we refer the reader to \citet[][]{dawson15a}.

\subsection{Emission-line Galaxies}\label{sec:elgs}

Emission-line galaxies are one of the four tracers targeted in eBOSS. The primary tracer of the large-scale structures in BAO observations (at $z\lesssim1$) has been LRGs \citep[][]{eisenstein01a}, because of their broadband brightness and well-understood SEDs. At higher redshift, however, most of the distinctive features of LRGs, such as the $4000\,$\AA\ break, G4300-band and MgH/Mg b absorption bands, are redshifted into the Meinel hydroxyl forest. Blueward of the $4000\,$\AA\ break, the cool giant stars dominating the continuum emit little flux, which makes spectral classification and redshift determination difficult. Moreover, the number density of red galaxies is lower at higher redshift due to galaxy evolution \citep[\eg][]{bell04a, faber07a, moustakas13a}, rendering them less useful for cosmological purposes. On the other hand, the cosmic star-formation rate (SFR) density increases precipitously with redshift, and at $z\sim1$, is about an order-of-magnitude higher than at present \citep[\eg][]{madau96a, hopkins04a, madau14a}. Star-forming galaxies (SFGs) exhibit the strong \oiidoublet\ emission feature, which is a doublet distinguishable at a medium spectral resolution $\mathcal{R}\sim4000$. The number density of \oiidoublet\ emitters increases steadily with redshift up to $z\sim2$ \citep[\eg][]{zhu09a, comparat15a, sobral15a}. The next-generation BAO surveys, e.g., DESI \citep[][]{schlegel11a, levi13a} and PFS \citep[][]{takada14a}, will primarily target these \oii\ emitters at redshift $z>1$.

At redshift $z\lesssim1$, the strong emission lines of SFGs enhance their brightness in the optical (in the observer frame), facilitating their detection even in the relatively shallow SDSS imaging. \citet{comparat13a} demonstrated that it is feasible to conduct an ELG survey with the BOSS spectrographs that can achieve the number density and the volume coverage required for BAO measurements at $0.6\lesssim z \lesssim 1.0$, with targets selected from the SDSS imaging. As the continuum of SFGs is dominated by the emission of hot O/B stars in the rest-frame NUV and that of dust and polycyclic aromatic hydrocarbons (PAHs) in the IR, $u$/$U$-band and/or IR data can help improve the targeting efficiency of these objects. We have been working on optimizing the ELG selection strategies with additional data, including the $U$-band photometry from SCUSS, the IR photometry from \textit{WISE}, and the deeper $grz$ imaging with DECam. 

Favorable weather during SDSS-III led to an early completion of the BOSS survey. A fraction of the remaining time was allocated to an eBOSS pilot program known as the Sloan Extended Quasar, ELG, and LRG Survey \citep[SEQUELS,][]{alam15a}. In Fall 2014, within eBOSS, we also conducted a series of pilot observations to test possible techniques for the ELG target selection. With data from these pilot observations, the team is currently investigating different selection algorithms to maximize the targeting efficiency \citep[][]{raichoor15a, comparat15b, delubac15a, jouvel15a}. The ELG cosmology survey will begin in Fall 2016, the third year of eBOSS, with the optimal selection strategy to be defined from these pilot data and investigations. 
The survey aims at obtaining secure redshift for about $200,000$ ELGs at $0.6\lesssim z \lesssim 1.0$ and measuring the BAO scale with an accuracy of about $2\%$ at an effective redshift $\left<z\right>\sim0.8$. 

For more details regarding the ELG target selections and the cosmological applications, we refer the reader to references above and \citet[][]{dawson15a} and \citet[][]{zhao15a}.

\subsection{The ELGs from the eBOSS Pilot Observations}\label{sec:ebosselg}

\begin{figure}
\epsscale{1.28}
\plotone{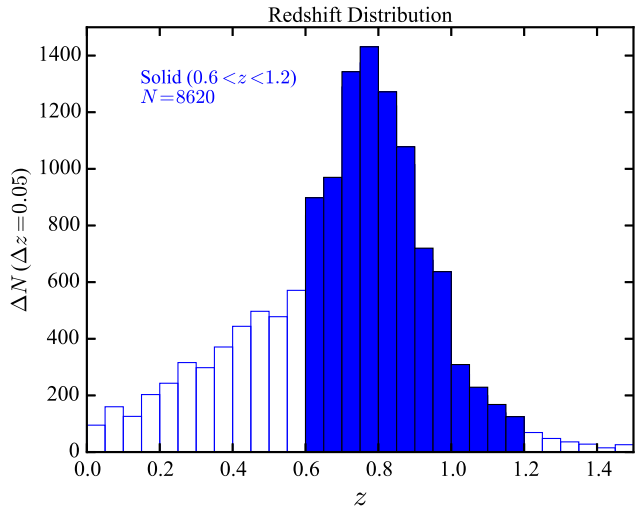}
\caption{The redshift distribution of all emission-line galaxies (ELGs) from the eBOSS pilot observations. The solid area shows $8620$ galaxies at $0.6<z<1.2$.}
\vspace{0.2cm}
\label{fig:redshift}
\end{figure}

In total, the BOSS/SEQUELS ancillary program and the early eBOSS pilot observations provided about $12,000$ ELGs spanning $0<z\lesssim1.5$, peaked at $z\sim0.8$. We show the redshift distribution in Figure~\ref{fig:redshift}.  In Figure~\ref{fig:fullcomposite}, we present the median composite spectrum of all the ELGs, in the wavelength range $2000<\lambda<7500\,$\AA. As we included \textit{all} the ELGs in this composite spectrum, the objects contributing at each wavelength are different, with high-redshift objects dominating in the NUV and low-redshift ones in the optical. 
The composite spectrum appears as a typical active SFG spectrum \citep[\eg][]{kennicutt92a}. We have labeled the prominent features in the figure. In the optical, it features strong hydrogen Balmer recombination emission on top of relatively weak stellar Balmer absorption, strong nebular forbidden lines due to collisionally-excited metal atoms, e.g., N, O, $\mathrm{N}^+$, $\mathrm{O}^+$, $\mathrm{S}^+$, $\mathrm{O}^{2+}$, $\mathrm{Ne}^{2+}$, and $\mathrm{Ar}^{2+}$. As the BOSS spectrographs cover the \oiii$\,\lambda5008$ up to redshift $z\sim1.0$, we expect that eBOSS will provide an important opportunity for studies of the ISM properties, such as the gas-phase metallicity, in strong SFGs at moderate redshift. The full composite spectrum also displays some relatively weak stellar metal absorption lines, e.g., the Fraunhofer \caii\ H \& K, G4300-band, \mgi\ b, and \nai\ D lines. We summarize the identified lines in the list given in Appendix~\ref{app:atomic}.

Comparat et al. (2015b) describes the full sample in detail and we refer the reader to that paper for more information. The mean precision of the redshift at $z\sim0.8$ is about $20\,\kms$ thanks to the strength of the emission lines. From the SED fitting of the composite spectra with different line strengths, the average stellar mass $\left<M_*\right>$ at $z\sim0.8$ is about $10^{10}\,\MSun$. The investigations of the physical properties of individual galaxies, such as stellar mass ($M_*$), SFR, and metallicity, are ongoing and will be presented in future papers. 

The redshift coverage of the ELG observations at $z>0.6$ allows us to probe the NUV part of their SEDs, as shown in Figure~\ref{fig:fullcomposite}. The composite spectrum between $2900\,{\rm \AA}$ and the Balmer Break at $3646\,{\rm \AA}$ is essentially featureless except for the weak \hei$\,\lambda3189$ emission line. Blueward of $2900\,{\rm\AA}$, we see strong absorption lines similar to those in intervening quasar absorption-line systems, \mgi, \mgii\ and \feii, and non-resonant \feii$^*$ emission as well as the \ciisemi\ emission. The underlying stellar continuum rises towards higher energy, as typical of hot O/B stellar spectra. We fit a power law, $f(\lambda)\propto\lambda^\beta$, to the continuum between $2000\,{\rm \AA}$ and $2200\,{\rm \AA}$ and obtain a slope $\beta\sim-2.1$, as expected for an SED dominated by O/B stellar emission in the UV. In the rest of the paper, we will focus on the absorption and emission features in the NUV.

\section{Near-ultraviolet Spectroscopy}\label{sec:results}

\subsection{The NUV ELG Sample}\label{sec:nuvsample}

\begin{sidewaysfigure*}
\hspace{0.0in}
\includegraphics[width=1.00\textheight]{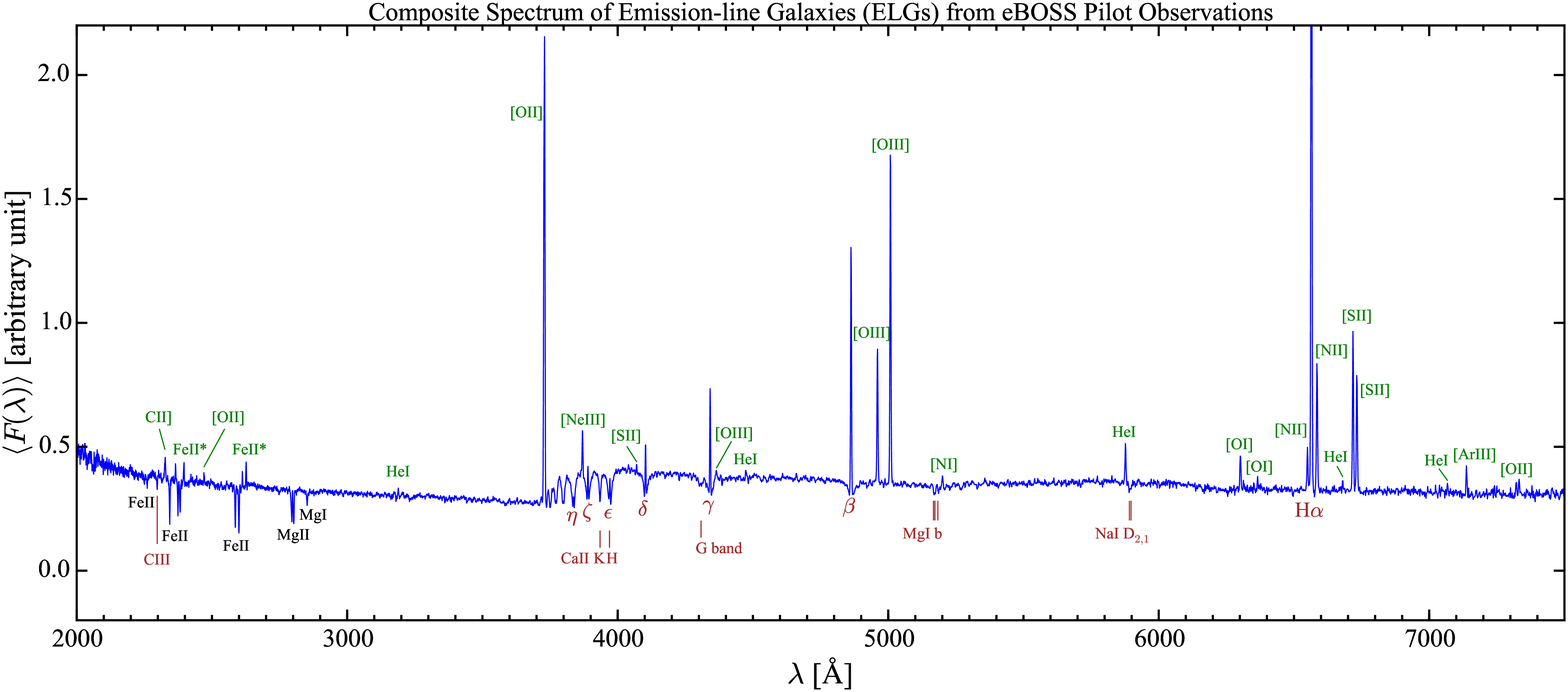}
\caption{The median composite spectrum of all ELGs (at $0<z\lesssim1.5$) at $2000\,\mathrm{\AA}<\lambda<7500\,\mathrm{\AA}$ from the eBOSS pilot observations. We label emission features in green, stellar absorption features in brown, and ISM/CGM absorption lines in black. The Greek symbols indicate hydrogen Balmer lines, which appear as nebular recombination emission on top of stellar absorption.} 
\vspace{-3.5in}
\label{fig:fullcomposite}
\end{sidewaysfigure*}

The eBOSS pilot observations have obtained spectra for a large sample of ELGs at redshift $z>0.6$ with rest-frame NUV coverage, providing us with a good opportunity to investigate the gas processes associated with these objects. We here focus on the wavelength range between $2200\,{\rm \AA}$ and $4000\,{\rm \AA}$. We choose the shorter wavelength limit so as to cover \feii$\,\lambda2250$ and \feii$\,\lambda2261$, and the longer limit to cover \oiidoublet. To ensure the same wavelength coverage for all the objects and thus the same contributing objects at all the wavelengths, we select ELGs between redshift $0.6$ and $1.2$, which include $8620$ objects.

As discussed in Section~\ref{sec:ebosssurvey}, we use the reduction outputs based on the spectroscopic pipeline version \texttt{v5\_7\_8}, and consider only objects classified as a \texttt{GALAXY} with no redshift warning, i.e., \texttt{ZWARNING==0}. The classification selection automatically rejects broad-line active galactic nuclei (AGNs). 
We do not make further cuts in our sample selection. At the redshifts we are interested in, some of the lines required in the narrow-line AGN classification schemes \citep[\eg][]{baldwin81a}, such as \ha\ and \niilam,  are not covered by eBOSS. 
For those with \oiiilam\ and \hb\ measurements (at $z\lesssim1$), based on the blue optical color and line ratios \citep[\eg][]{yan11a, trouille11a}, we expect the fraction of narrow-line AGNs to be at most a few percent.
We also do not expect many low-ionization nuclear emission-line regions \citep[LINERs,][]{heckman80a} in our sample. The ELGs are selected to be blue galaxies, while the majority of LINERs are found in red galaxies \citep[\eg][]{ho97a}. We therefore expect the line emission in the integrated spectra of the eBOSS ELGs to be dominated by contributions from SF activities and will use the terms ELGs and SFGs interchangeably.

\subsection{The Method}\label{sec:method}

The average S/N per pixel in the continuum region of the individual ELG spectra is low ($\lesssim1$) and does not allow precise measurements of the absorption features for single objects. To study the gas associated with the ELGs, we construct high-S/N composite continuum-normalized spectra. We use a median estimator, which is less prone to extreme outliers. However, we also tested our analysis with the arithmetic mean estimator and found consistent results, with differences in relative dependences smaller than $1\sigma$. 

For each observed spectrum, $F(\lambda)$, we first blueshift it back to its rest frame on a common wavelength grid. We choose the common wavelength grid to have the same logarithmic (or equivalently, velocity) spacing as in the observer frame, i.e., with $\ud \log_{10} \lambda=10^{-4}$ or $\ud v=69\,\kms$. In the blueshifting process, we interpolate the spectrum with the cubic-B spline method\footnote{As the spacing is identical before and after interpolation, linear interpolation yields almost the same results.}, as in the standard SDSS pipeline. We then mask out absorption and emission features and fit a cubic polynomial function through the rest of the spectrum. Using the best-fit polynomial function as an estimate of the underlying continuum, $\hat F_{\rm cont}(\lambda)$, we normalize the observed spectrum to obtain the continuum-normalized spectrum suited for absorption studies:
\beq
R(\lambda) \equiv \frac{F(\lambda)}{\hat F_{\rm cont}(\lambda)} \, \mathrm{.}
\label{eq:residual}
\eeq
For a given sample, we construct a median composite spectrum of all the continuum-normalized spectra (in the rest frame). Finally, we fit a quadratic polynomial function to the composite spectrum, again with absorption and emission features masked out, to remove any large-scale residuals, though skipping this final step has a negligible effect on the results. We designate the final composite as $\left<R(\lambda)\right>$:
\beq
\left<R(\lambda)\right> \equiv \left<\frac{F(\lambda)}{\hat F_{\rm cont}(\lambda)}\right> \, \mathrm{.}
\label{eq:composite}
\eeq
Since we mostly work with composite spectra, we will drop the ensemble symbol $\left<\,\right>$ in the text for simplicity.

We quantify the absorption and emission strength in the continuum-normalized spectra with the rest equivalent width $W_0$. We define the rest equivalent widths for absorption and emission in such a way that they are both positive. For absorption, the rest equivalent width is given by 
\beq
W^{\rm absorption}_0 \equiv \int_{\lambda_{\rm min}}^{\lambda_{\rm max}}\left[1-R(\lambda)\right]\,\ud \lambda \,\mathrm{,}
\label{eq:rewabs}
\eeq
and for emission, it is defined as 
\beq
W^{\rm emission}_0 \equiv \int_{\lambda_{\rm min}}^{\lambda_{\rm max}}\left[R(\lambda)-1\right]\,\ud \lambda \,\mathrm{,}
\label{eq:rewemi}
\eeq
where the integration range ($\lambda_{\rm min}<\lambda<\lambda_{\rm max}$) encloses the absorption/emission profile.

Throughout the paper, unless otherwise specified, we estimate the measurement uncertainties for a given sample by bootstrapping (i.e., with replacement) $100$ times.

\begin{figure*}[t]
\epsscale{1.2}
\plotone{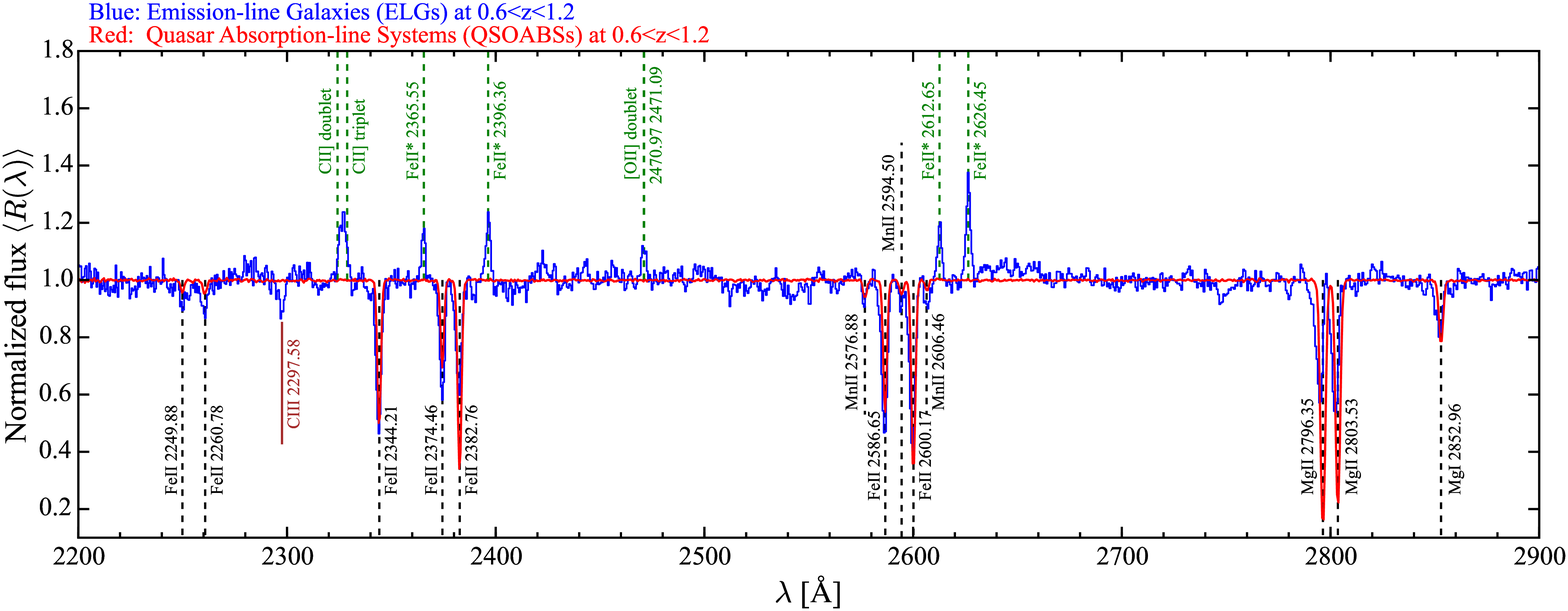}
\caption{The composite continuum-normalized spectrum of $8620$ ELGs at $0.6<z<1.2$ (\textit{blue}). We show the rest-frame positions of emission features with vertical green dashed lines and those of ISM/CGM absorption lines with black dashed lines. The brown solid line marks the stellar photospheric absorption feature \ciii$\,\lambda2298$. The red line shows the composite spectrum of $2310$ quasar intervening absorption-line systems with $\rewmgiione>2\,\mathrm{\AA}$ in the same redshift range.
}
\label{fig:nuvcomposite}
\vspace{0.05in}
\end{figure*}

\subsection{The Composite Continuum-normalized Spectrum}\label{sec:nuvcomposite}

\begin{figure*}
\epsscale{1.04}
\plotone{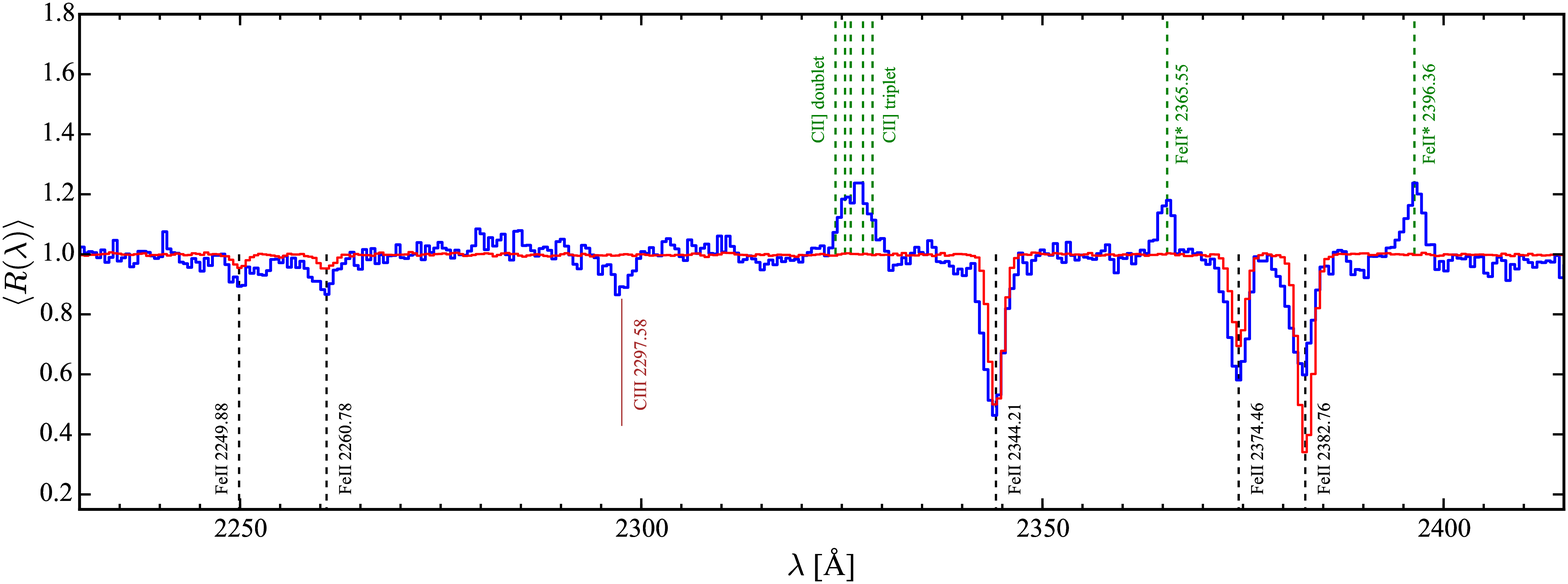}
\vspace{-0.05in}
\epsscale{1.04}
\plotone{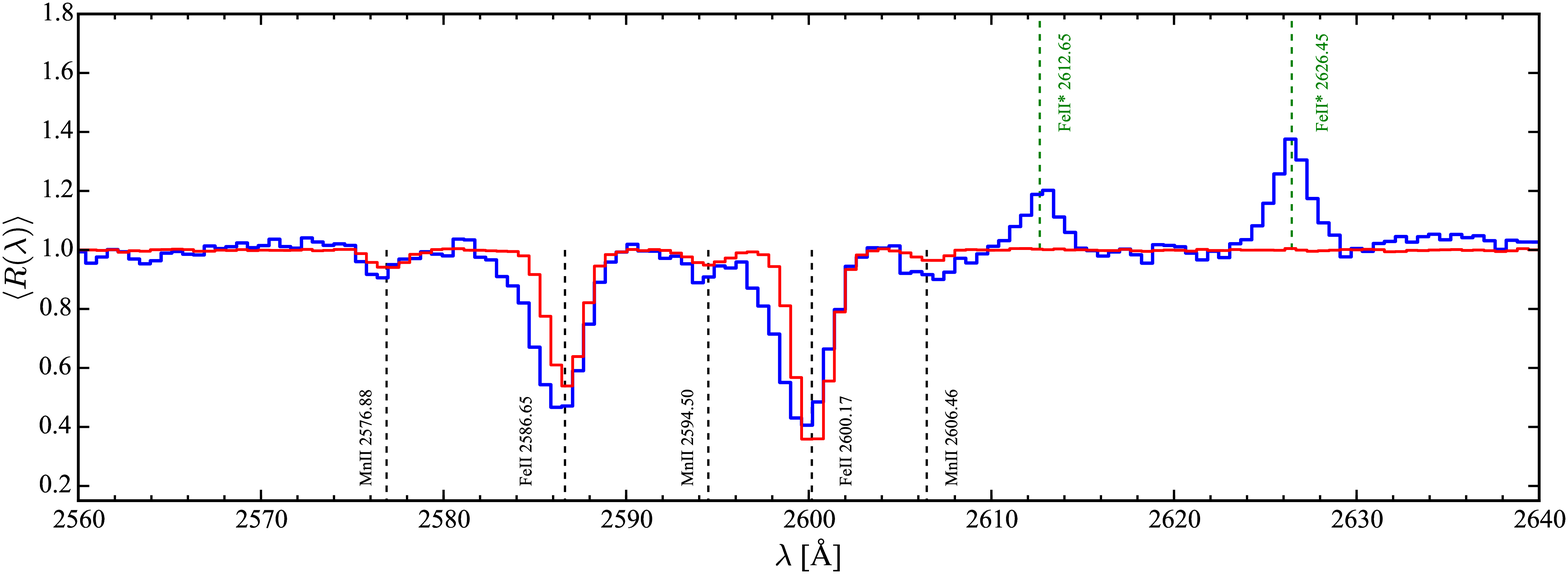}
\vspace{-0.05in}
\epsscale{1.04}
\plotone{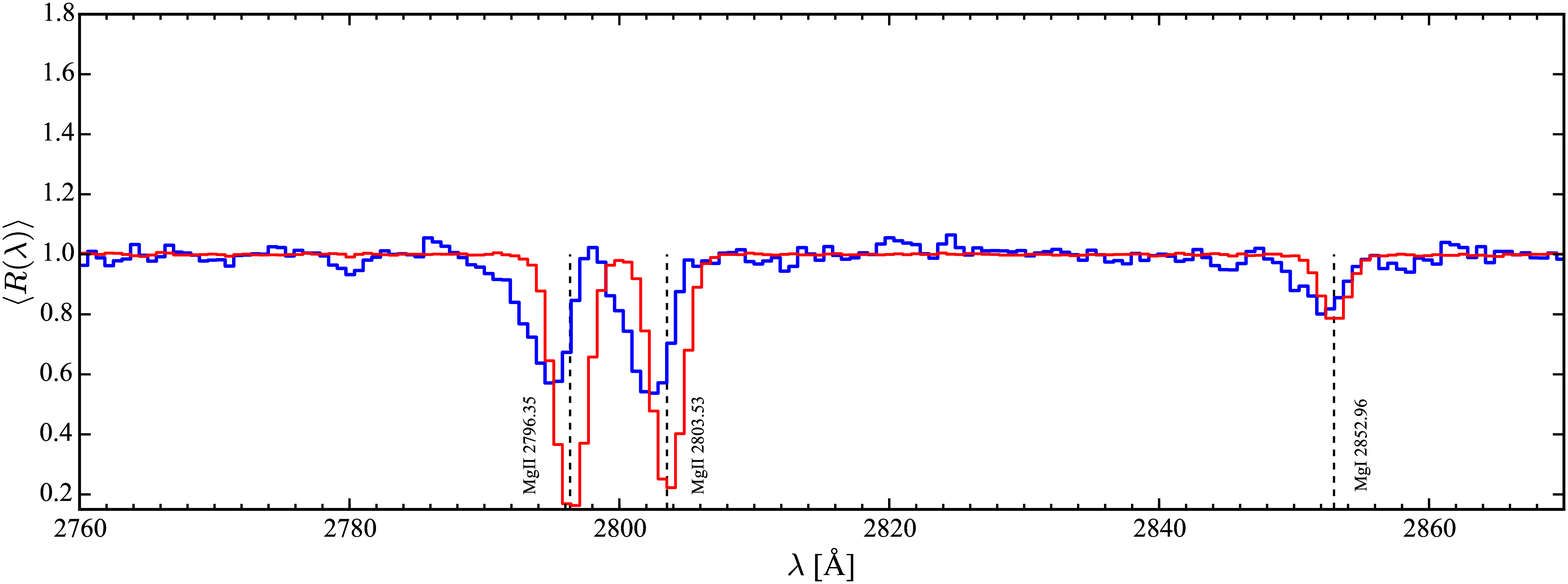}
\vspace{-0.05in}
\epsscale{1.04}
\plotone{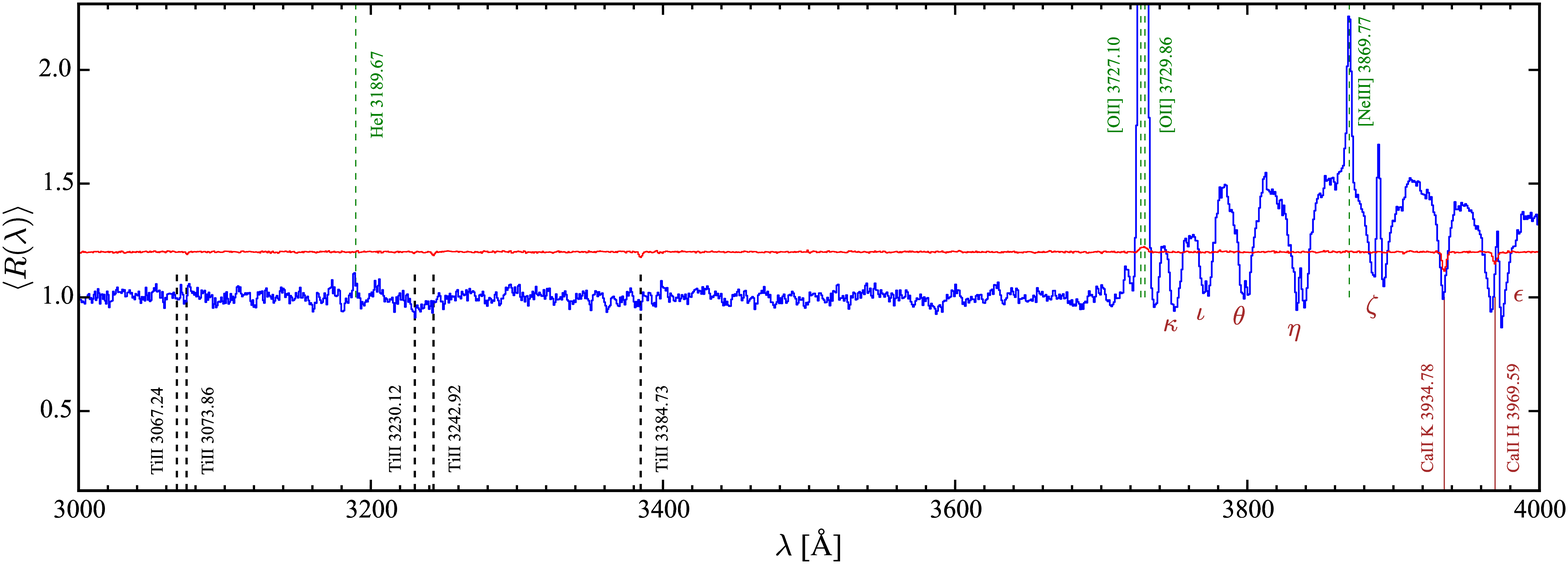}
\caption{The composite continuum-normalized spectrum of ELGs (\textit{blue}), zoomed in on prominent features and compared with the composite spectrum of quasar absorbers (\textit{red}) as in Figure~\ref{fig:nuvcomposite}. In addition, the bottom panel shows the spectra at $3000\,\mathrm{\AA}<\lambda<4000\,\mathrm{\AA}$, where we have shifted the quasar absorber composite upward by $0.2$ for clarity.
}
\label{fig:nuvfine}
\end{figure*}

Figure~\ref{fig:nuvcomposite} presents the median composite spectrum of the $8620$ ELGs at $0.6<z<1.2$ in the wavelength range $2200\,{\rm \AA}<\lambda<2900\,{\rm \AA}$. In Figure~\ref{fig:nuvfine}, we zoom in on the most prominent absorption and emission features for a more detailed illustration, and also include the part in the wavelength range $3000\,{\rm \AA}<\lambda<4000\,{\rm \AA}$. We have omitted the part between $2900\,{\rm \AA}$ and $3000\,{\rm \AA}$ since it is featureless (see Figure~\ref{fig:fullcomposite}). To guide the eye, we mark the rest-frame positions of the identified features.

The line features in the NUV can be categorized into three primary groups\footnote{In the FUV, stellar wind features comprise another major class of spectral features.} based upon their origins: stellar photospheric absorption lines, nebular emission lines, and absorption/emission lines due to gas in the ISM and CGM. The last category includes combined effects caused by both the ISM and CGM and is the focus of this work. We briefly discuss the observed features in these groups before investigating the ISM/CGM features in detail.

\vspace{0.15in}

\noindent $\bullet$ \textbf{Stellar photospheric absorption features}

\vspace{0.05in}

At $2200\,{\rm \AA}<\lambda<2900\,{\rm \AA}$, we are able to identify only one photospheric absorption line, \ciii\ at $2297.58\,$\AA. However, in subsampling exercises with bootstrapping and jackknife, we notice that there are some weak but persistent features that are not due to noise (also see Section \ref{sec:localsf} below). We do not identify these weak stellar features because the NUV part of the O/B star SEDs has not been sufficiently explored in either theory or observation. The most recent work was by \citet[][the UVBLUE library]{rodriguez05a}\footnote{\texttt{http://www.inaoep.mx/$\sim$modelos/uvblue/uvblue.html}}, who built a suite of stellar spectral templates in the NUV based on the atmospheric model code ATLAS9 \citep[][]{kurucz92a, castelli97a} and, for O/B stars, compared the model spectra with a few low-resolution spectra taken by \textit{IUE} in 1980s. Although the shape of the underlying stellar continuum is well-understood ($\beta\sim-2.0$, \eg \citealt{kinney93a}, and Section~\ref{sec:ebosselg}), the absorption features in theoretical calculations and observations do not match each other. For example, we do not detect the \oiiiperm\ line at $2496\,$\AA\ predicted by the models in our composite spectrum, nor did in the \textit{IUE} observations of O/B stars \citep[][]{fanelli92a, rodriguez05a}.

At $3000<\lambda<4000\,$\AA, the prominent stellar absorption features are the well-known Balmer series at $\lambda>3646\,$\AA, and \caii\ H ($3969.59\,$\AA) \& K ($3934.77\,$\AA) lines. The \caii\ doublet, however, likely has a large contribution from the gaseous ${\rm Ca}^+$ in the ISM/CGM \citep[see, \eg][]{zhu13a, murga15a}.

\vspace{0.15in}

\noindent $\bullet$ \textbf{Nebular emission features}

\vspace{0.05in}

Between $2200$ and $2900\,$\AA, we identify two nebular emission features with high confidence: semi-forbidden \ciisemi\ at about $2326\,{\rm \AA}$ and forbidden \oii\ at about $2471\,{\rm \AA}$. The \ciisemi\ feature is a blend including five transitions from the doublet and the triplet between the ground state and first excited state: \ciisemi$\,\lambda\lambda2324,2325$ and \ciisemi$\,\lambda\lambda\lambda2326, 2327, 2329$. The \oii\ feature is a doublet at $2470.90\,{\rm \AA}$ and $2471.10\,{\rm \AA}$, due to the transitions between the ground state and the second excited state of $\mathrm{O^+}$. For comparison, the \oiidoublet\ doublet is due to the transitions between the ground state and the first excited state of $\mathrm{O^+}$. Both the \ciisemi\ and \oii\ emission lines were observed in \hyii\ regions by \textit{IUE} \citep[][]{dufour87a}. 
As the low-ionization nebular emission lines in the optical spectra of SFGs, e.g., \oiidoublet, \niidoublet\ and \suiidoublet, they must be dominated by the emission from \hyii\ regions, where the UV photons from O/B stars ionized the elements in the surrounding gas to low-ionization states \citep[\eg][]{stromgren39a}, though diffuse ionized gas in the ISM, e.g., the warm ionized medium \citep[WIM,][]{mckee77a}, also contributes to the integrated emission \citep[\eg][]{reynolds84a}. $\mathrm{C^+}$ can also be abundant in a cooler medium (e.g., photodissociation regions, or PDRs, \citealt{tielens85a}) because the ionization potential of neutral carbon ($11.26\,{\rm eV}$) is lower than that of hydrogen or oxygen (both $\sim13.6\,{\rm eV}$, see Appendix~\ref{app:atomic}). It is interesting to note that, though \ciisemi\ in the NUV has been little studied, the fine-structure emission of the $\mathrm{C^+}$ ground term at $157.7\,\micron$ in the IR is known to be a major coolant in the ISM \citep[][]{dalgarno72a}. \ciiforb$\,\lambda157.7\,\micron$ can be observed for objects at $z\sim1$ in the submillimeter \citep[\eg][]{stacey10a}, thus for the ELG targets in eBOSS and future BAO surveys. 

We tentatively identify \neiv\ (not labeled) at about $2424\,{\rm \AA}$, a doublet at $2422.56\,{\rm \AA}$ and $2425.14\,{\rm \AA}$, due to the transitions between the ground state and the first excited state of the triply-ionized $\mathrm{Ne^{3+}}$. In SFGs, this doublet is observed mostly in planetary nebulae \citep[\eg][]{koeppen87a}, supernova remnants \citep[\eg][]{blair87a}, and other environments hotter than \hyii\ regions as the ionization potential of $\mathrm{Ne^{2+}}$ ($63.4\,{\rm eV}$) is larger than $\mathrm{He^{+}}$ ($54.4\,{\rm eV}$).

At $3000<\lambda<4000\,$\AA, besides Balmer recombination lines and \oiidoublet, we also observe \neiii\ at $\lambda=3869.86\,{\rm \AA}$. Highly sensitive to the ionization parameter, the \neiii\ emission can be combined with \oii\ and used to probe the metallicity and other properties of the ISM \citep[][]{nagao06a, levesque14a}. 
There is also a \hei\ line at $3889.74\,$\AA\ (not labeled), which is blended with Balmer H$\zeta$.
Between $2900\,{\rm \AA}$ and the Balmer break at $3646\,{\rm \AA}$, we only detect the weak \hei\ emission at $3188.67\,$\AA, which likely sits on top of \hei\ absorption commonly associated with O/B stars \citep[\eg][]{morrison75a}. 

\begin{figure}
\vspace{0.2in}
\epsscale{1.15}
\plotone{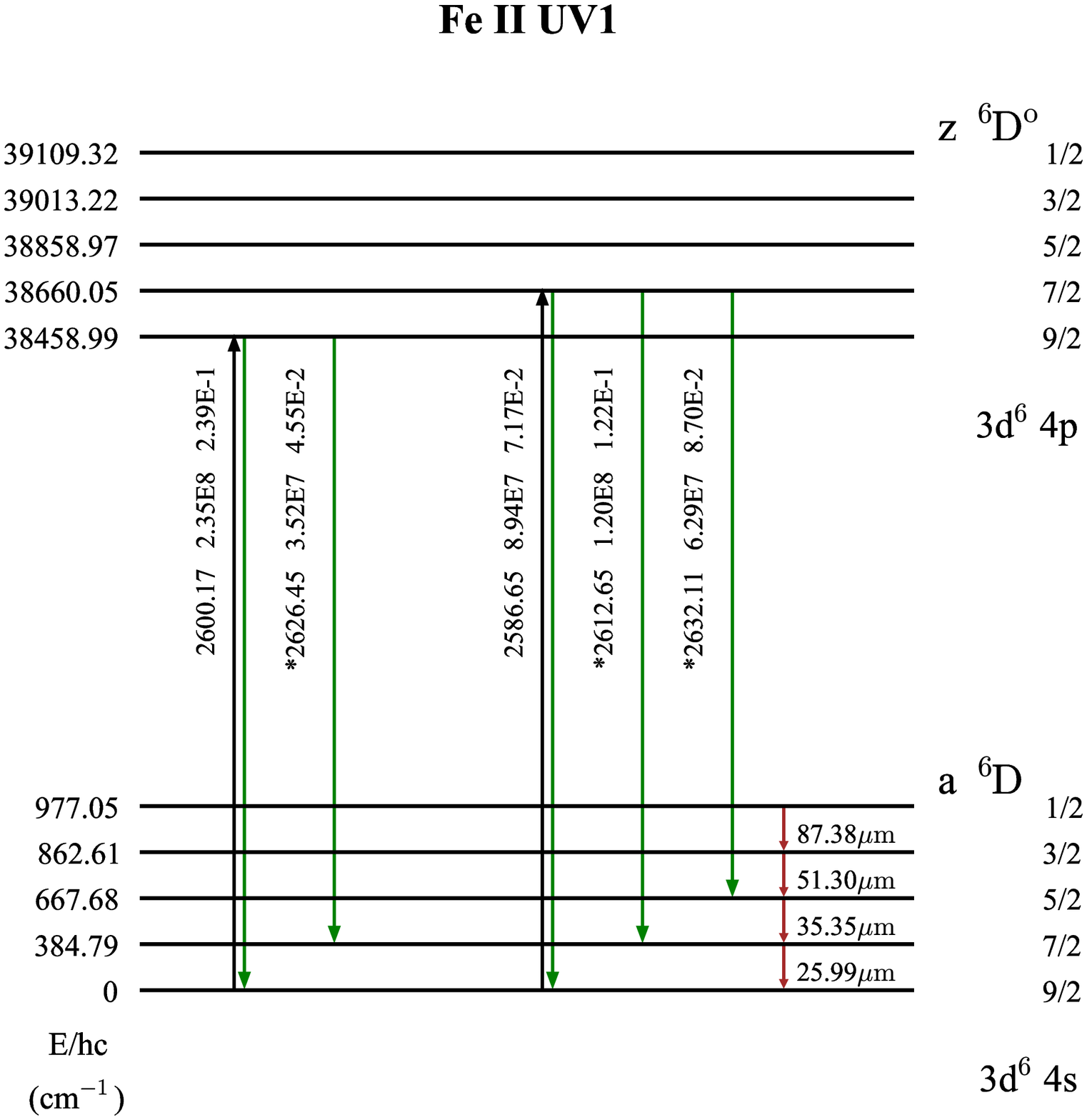}
\caption{The energy diagram for the transitions in the first UV multiplet group (UV1) of \feii, between the ground state and the first excited state. See Appendix A for a full description of the symbols and terms.
}
\vspace{0.03in}
\label{fig:feiiuv1}
\end{figure}

\vspace{0.15in}

\noindent $\bullet$ \textbf{ISM/CGM absorption and emission features} 

\vspace{0.05in}

Most of the absorption lines between $2200$ and $2900\,$\AA\ are induced by gas in the ISM and/or the CGM, providing us with a unique tool to probe the diffuse gas otherwise elusive. From low to high energy, we identify the following lines due to different species:
\vspace{-0.05in}
\begin{itemize}
\item \mgi$\,\lambda2853$ (UV1); 
\item the \mgiidoublet\ doublet (UV1)\footnote{Compared to the absorption induced by gas in the ISM/CGM, we expect the intrinsic \mgi$\,\lambda2853$ and \mgiidoublet\ absorption in the spectra of O/B stars to be very weak, even for metal-rich stars \citep[\eg][]{rodriguez05a}.}; 
\item the \mniitriplet\ triplet (UV1); 
\item \feii$\,\lambda2600$ and $\lambda2587$ (UV1), \feii$\,\lambda2383$ and $\lambda2374$ (UV2), \feii\,$\lambda2344$ (UV3), \feii\,$\lambda2261$ (UV4), and \feii\,$\lambda2250$ (UV5). 
\end{itemize}
\vspace{-0.05in}
Between $3000$ and $4000\,$\AA, the \caii\ H \& K lines also trace a significant fraction of gas in the ISM/CGM. 
All these absorption lines are also commonly seen in intervening quasar absorption-line systems (see next subsection). 

We identify four non-resonant \feii$^*$ emission lines with high confidence: 
\vspace{-0.05in}
\begin{itemize}
\item \feii$^*\,\lambda2626$ and $\lambda2613$ (UV1), \feii$^*\,\lambda2396$ (UV2), and \feii$^*\,\lambda2366$ (UV3). 
\end{itemize}
\vspace{-0.05in}
To illustrate the relationships between the resonant absorption and these non-resonant emission lines, we use the \feii\ UV1 multiplet group as an example.

Figure~\ref{fig:feiiuv1} presents the energy-level diagram of \feii\ UV1, showing the transitions between the ground state and the first excited state of \feii. We refer the reader to Appendix~\ref{app:atomic} for a detailed description of the symbols and terms. \feii$\,\lambda2600$ is the transition between the lowest (ground) energy level (with $J=9/2$) of the ground state and the lowest level (with $J=9/2$) of the excited state. Because the second lowest energy level of the ground state has the total angular momentum number $J=7/2$, the excited electron after the resonant absorption has a high probability, i.e., a high Einstein $A$ coefficient, to spontaneously decay to this level, releasing a fluorescent photon at a slightly longer wavelength $\lambda=2626.45\,$\AA. The pair \feii$\,\lambda2586$ and \feii$^*\,\lambda2613$ also belong to UV1, though with the second energy level of the excited state as the higher anchor level. 

The presence of these non-resonant emission lines imply that many of the resonant absorption lines above must be blended with emission filling in. For example, the Einstein $A$ coefficient for \feii$\,\lambda2600$ ($2.35\times10^8\,\mathrm{s}^{-1}$) is over six times higher than that for \feii$^*\,\lambda2626$ ($3.53\times10^7\,\mathrm{s}^{-1}$), so an excited electron has a higher probability to decay to the lowest level and release a \feii$\,\lambda2600$ photon, though if the \feii\ optical depth is high, the emitted photon will be absorbed again. The exact amount of emission infill depends on the optical depth of the relevant transitions and the balance between absorption and emission in the multiple-scattering process. 

We find the velocity profiles of both absorption and emission lines to be asymmetric, skewed towards negative values, i.e., in the blueshift direction, indicating a larger fraction of the gas sources that cause these features are flowing outwards than inwards. The profiles of the emission lines appear to be similar, while the degree of asymmetry of the absorption profiles varies from line to line. For example, compared to \feii$\,\lambda2586$ or \feii$\,\lambda2600$ as shown in the second panel in Figure~\ref{fig:nuvfine}, the \mgiidoublet\ lines are blueshifted from their rest-frame positions by a larger amount. Considering these species have similar ionization potentials ($7.90\,{\rm eV}$ for neutral Fe and $7.64\,{\rm eV}$ for Mg) and likely co-exist spatially, the difference must be due to a larger amount of emission infill on top of the \mgii\ absorption.

For a better understanding of the absorption and emission in the composite spectrum of ELGs, we present a comparison with a composite spectrum of intervening quasar absorption-line systems in Section~\ref{sec:qsoabsorber}, and with a composite spectrum of local SF regions in Section~\ref{sec:localsf}. We will then investigate the physical processes with a gas flow model in Section~\ref{sec:interpretation}.

\subsection{Comparison with Intervening Absorbers}\label{sec:qsoabsorber}

\begin{figure*}
\epsscale{1.20}
\plotone{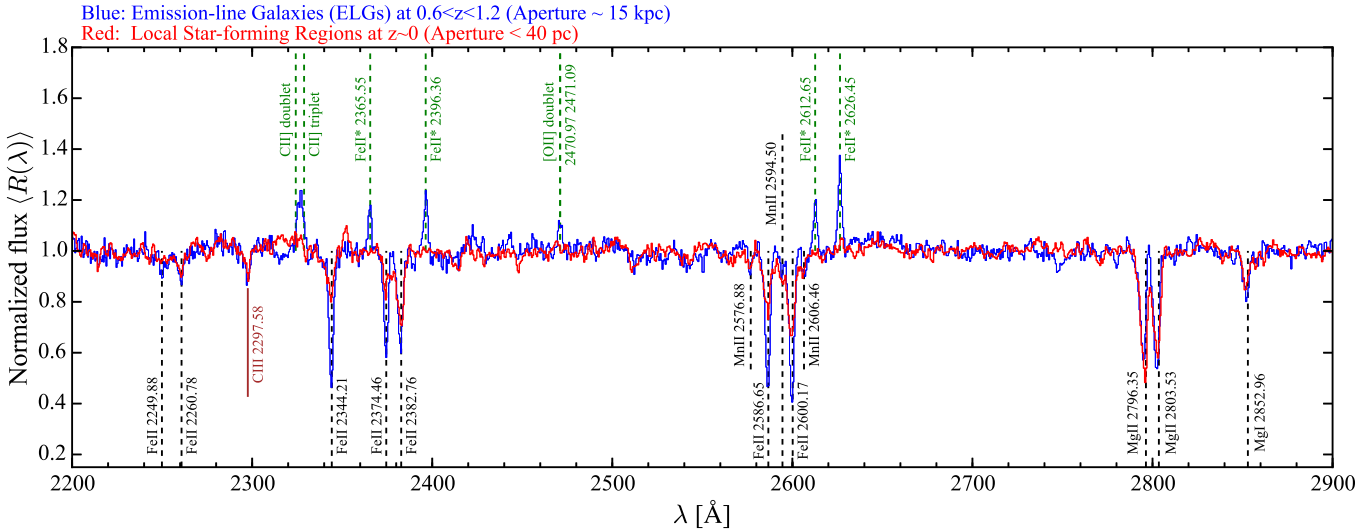}
\caption{The composite continuum-normalized spectrum of ELGs (\textit{blue}), the same as in Figure~\ref{fig:nuvcomposite} but compared with the composite spectrum of local star-forming regions (\textit{red}) from \citet{leitherer11a}.
}
\label{fig:localsf}
\vspace{0.05in}
\end{figure*}

The composite spectrum of ELGs exhibits the absorption lines commonly seen in intervening quasar absorption-line systems. It is interesting to compare the so-called ``down-the-barrel'' spectra of ELGs with the intervening quasar absorber spectra. We select the absorbers from the JHU-SDSS metal absorber catalog \citep[][]{zhu13b}, updated to the 12th Data Release \citep[DR12,][]{alam15a}\footnote{\texttt{http://www.pha.jhu.edu/$\sim$gz323/jhusdss}}. We choose strong absorbers with \mgii$\,\lambda2796$ rest equivalent width $\rewmgiione>2\,$\AA, because there has been evidence showing that a large fraction of these strong absorbers are physically associated with the CGM of strong SF galaxies \citep[\eg][]{bergeron91a, norman96a, bouche07a, nestor11a, lan14a}. We select the $2310$ such strong absorbers at $0.6<z<1.2$, the same redshift range as the ELGs and construct a median composite continuum-normalized spectrum. For more details regarding the quasar absorbers and how we estimate the continua of background quasars, we refer the reader to \citet{zhu13b}.

We overplot the composite spectrum of quasar absorbers in red in Figure~\ref{fig:nuvcomposite} and \ref{fig:nuvfine}. Note the absorber spectra are based on luminous quasar spectra from SDSS I-III, so the S/N of their composite is orders-of-magnitude higher than the ELG spectra. When constructing the composite spectrum, we shifted the absorber spectra to the rest frame of the absorbers, so the absorption profiles are centered on their rest-frame positions. We find two major differences between the ELG composite and the quasar absorber composite: (1) There is no detectable non-resonant emission in the quasar absorber composite, even though its S/N is orders-of-magnitude higher. (2) The ratios of the absorption lines are different.
For example, the strength of \feii$\,\lambda2344$ is similar in both composites, while \feii$\,\lambda2383$ and \mgiidoublet\ are much stronger in the quasar absorber one. We note that the line ratios of quasar absorbers depend both on the strength ($\rewmgiione$) and redshift (e.g., Figure~{3} in \citealt{dey15a}), but at $\rewmgiione>2\,$\AA\ and a given redshift, the strength dependence is weak and has no effect on any of our conclusions. In addition to the two main differences, we do not detect \tiii\ absorption lines in the ELG composite, which we suspect is due to the limited sample size and the low S/N. 

In Section~\ref{sec:interpretation}, we show that the line ratio difference is caused by the emission infill present in the ELG composite but absent in the quasar absorber composite. 

\subsection{Comparison with Local Star-forming Regions}\label{sec:localsf}


Spectroscopic observations of SFGs, or astronomical sources in general, have been scarce in the NUV. \citet{leitherer11a} compiled a UV spectroscopic atlas\footnote{\texttt{http://www.stsci.edu/science/starburst/templ.html}} of \textit{local} SF galaxies observed with the Faint Object Spectrograph (FOS) and the Goddard High Resolution Spectrograph (GHRS) on \textit{HST}. The atlas includes small-(physical) aperture spectra of 15 regions of nine SFGs with coverage between $2200\,{\rm \AA}$ and $3200\,{\rm \AA}$, providing a rare opportunity for a direct comparison of the integrated NUV SEDs at different physical scales. From this compilation, we select nine spectra of six galaxies with relatively high S/N and strong absorption: ${\rm NGC}\,1569$, ${\rm NGC}\,2403$ (all $3$ spectra), ${\rm NGC}\,4569$, ${\rm NGC}\,5055$, ${\rm NGC}\,5253$ ($1$ and $3$), and ${\rm NGC}\,5457$ (${\rm NGC}\,5455$). Note the different spectra of one galaxy are independent from each other, originating from different SF regions in that galaxy. For details regarding their atlas, we refer the reader to \citet{leitherer11a}. With the selected nine spectra, we then construct the composite continuum-normalized spectrum following the same procedure as for the eBOSS ELGs. Before a careful comparison, we here emphasize several characteristics of the observations:
\vspace{-0.05in}
\begin{itemize}
\item The aperture sizes of \textit{HST} FOS/GHRS are one to a few arcseconds, and for all the nine spectra, correspond to physical sizes smaller than $40\,{\rm pc}$ ($2-37\,{\rm pc}$, see Table~$6$ in \citealt{leitherer11a}).
They are nearly three orders-of-magnitude smaller compared to the aperture size for the eBOSS ELGs at $0.6<z<1.2$, which is about $15\,\kpc$ (for $2\,\arcsec$)\footnote{We note that the average seeing at the SDSS telescope is about $1.5\,\arcsec$.}.
\item The wavelength calibration of the FOS/GHRS spectra is largely based on the absorption lines induced by the gas in the ISM in the Milky Way, which are often blended with the lines of the low-redshift extragalactic sources. We find that, based on the position of \ciii$\,\lambda2298$, the composite spectrum is offset to longer wavelength by about $0.7\,$\AA\ (about $91\,\kms$), we therefore correct the wavelength by $-0.7\,$\AA. 
\item The spectral resolutions of FOS/GHRS are one to a few hundred $\kms$, a factor of $2-4$ lower than that of the BOSS spectrographs.
\end{itemize}
\vspace{-0.05in}
The last two characteristics prevent us from making a quantitative comparison of the observed velocity profiles. 

In Figure~\ref{fig:localsf}, we compare the ELG composite (in blue) with the composite spectrum of the nine local SF regions (in red) at $2200\,{\rm \AA}<\lambda<2900\,{\rm \AA}$. We present a zoomed-in version in Appendix B. The two spectra share common absorption features across the wavelength range, including both the stellar photospheric absorption line \ciii$\,\lambda2298$ and the absorption lines due to the ISM/CGM. We also observe some common weak absorption lines that must be intrinsic to the underlying stellar continuum. As discussed in Section~\ref{sec:nuvcomposite}, we do not identify these weak lines yet since they are still poorly-understood. We find the following main differences: (1) There is no detectable line emission, either the nebular lines (\ciisemi, \oii) or the non-resonant lines (\feii$^*$), in the composite of local SF regions. (2) The absorption line ratios are different. For instance, the strength of \mgiidoublet\ is similar in both composites, while \feii$\,\lambda2586$ and \feii$\,\lambda2600$ are about $50\%$ weaker in the composite of local SF regions. We note that in the composite of local SF regions the ratios of the lines are similar to those in the quasar absorber composite, though the absolute absorption strength is about $50\%$ weaker. 

\begin{figure*}[t]
\epsscale{1.10}
\plotone{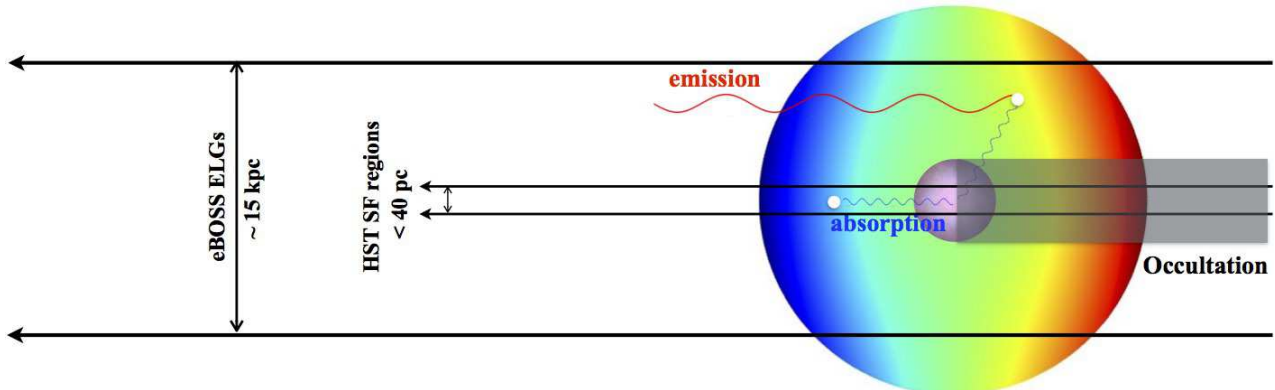}
\vspace{-0.10in}
\caption{The spherical outflow model. The color scale indicates the mean line-of-sight velocity of emission/absorption at a given position, assuming velocity $v(r) \propto r^{\alpha}$, where $\alpha=2$ is arbitrarily chosen. Resonant absorption takes place in front of the background light source (stars). Fluorescent photons, resonant or non-resonant, are scattered isotropically and only those scattered into the line of sight can be captured. The aperture size of the eBOSS fibers is $2\,\arcsec$, corresponding to about 15 \kpc\ at $0.6<z<1.2$. The aperture sizes of \textit{HST} FOS/GHRS are one to a few arcseconds, corresponding to less than $40\,$pc for the local star-forming regions observed. 
See \citet{scarlata15a} for a similar model.
}
\vspace{0.05in}
\label{fig:model}
\end{figure*}

The reasons for the non-detection of different emission lines are likely different. The \oii\ doublet at $2471\,{\rm \AA}$, mostly associated with \hyii\ regions, is very weak compared to its lower energy counterpart \oiidoublet\ and may be buried in the noise. The \ciisemi\ emission, which is strong in the ELG spectra, is more extended than \hyii\ regions (as observed through the fine-structure emission \ciiforb$\,\lambda157.7\,\micron$, \eg \citealt{pineda13a}) because of the lower ionization potential of neutral carbon, and the small (physical) apertures of FOS/GHRS did not capture enough \ciisemi\ photons. The non-resonant \feii$^*$ lines are not detected also because the emission is extended, though the emission mechanism is different from \ciisemi. 

As in the comparison with quasar absorbers, the difference in absorption line ratios must be due to emission infill in the ELG spectra. We present a more detailed discussion below in the context of a gas outflow model. 

\section{Interpretation}\label{sec:interpretation}

We have shown that the NUV composite spectrum of eBOSS ELGs display preferentially blueshifted absorption, induced by neutral and singly-ionized species, \mgi, \mgii, and \feii. In addition, we detected non-resonant \feii$^*$ emission lines, which also exhibit a preferentially blueshifted profile and are not detected in either quasar absorber spectra or small-aperture spectra of local SF regions. These observed properties indicate that the gas causing the absorption and emission is predominantly flowing outwards, and the outflows must be driven by the strong SF activities in the ELGs and extend to large galactic scales.

Galactic-scale outflows driven by star formation have been observed for over half a century, such as from the starburst galaxy M82 \citep[\eg][]{lynds63a, bland88a}. The physics of galactic winds has been extensively studied \citep[\eg][]{heckman90a, heckman00a}, though it is not yet conclusive due to the complexity of baryon processes involved. For our purposes, we circumvent some of the complex processes, such as the origins and properties of the wind and gas, instead we introduce a phenomenological gas outflow model and interpret our data with an observation-driven approach in the context of the model.

\subsection{A Spherical Gas Outflow Model}\label{sec:model}

We describe the model in three steps. First, we present the key geometrical and physical characteristics of the model. Second, we consider the properties of the observations of the integrated spectra along the line of sight. Finally, we discuss quantitatively the radiative transfer processes and the model predictions for the SFG observations.

\vspace{0.15in}

\noindent $\bullet$ \textbf{Basics of the gas outflow model}

\vspace{0.15in}

Because of the statistical fashion of our composite analysis, we construct our model for the average of an ensemble of galaxies, not for a single source. We illustrate the model in Figure~\ref{fig:model}\footnote{The image in the center is a composite image of the central region of M82 with different orientations. The original image is from \texttt{http://hubblesite.org/gallery/album/galaxy/pr2006014a/}, courtesy of NASA, ESA, and The Hubble Heritage Team (STScI/AURA).} and first emphasize the following two key characteristics.

\vspace{0.05in}
$\bullet$ Spherical symmetry -- A basic outflow model for an individual galaxy includes a density profile $n(\vec{r})$ (in number) and a velocity profile $v(\vec{r})$, both as a function of the vector position $\vec{r}$. Observations have shown that the large-scale outflows driven by star formation are often bipolar, in the form of an expanding envelope \citep[\eg][]{heckman90a}. In our composite analysis, we are averaging over all orientations randomly distributed and it is reasonable to assume spherically symmetric profiles $n(r)$ and $v(r)$. 

$\bullet$ Velocity distribution -- Also due to the statistical nature of our approach, at a given galactocentric distance, there is a distribution of velocities. If gas accretion takes place around the ELGs and if the infalling gas is enriched with the species we are interested in, it will also induce absorption and contributes to the statistical signatures. At small scales, the gas in the ISM also contributes to the absorption, and its motions, ordered or disordered, also affect the observed velocity distribution.
\vspace{0.05in}

A statistical gas flow model therefore includes an average density profile, $n(r)$, an average velocity profile, $v(r)$, and a velocity dispersion profile, $\sigma(r)$. We expect the direction of the average velocity $v(r)$ to be outwards as we expect there is more gas flowing outwards than inwards. The velocity dispersion then accounts for the contributions at different velocities from outflows, inflows, and motions of ISM at small scales. In addition, in observation, we need to consider the finite instrumental resolution and redshift precision, which can be effectively included in the velocity dispersion through convolution. For the eBOSS ELGs, the mean spectral resolution is about $\left<\mathcal{R}\right>\sim2000$ ($60$ - $70\,\kms$) and the mean redshift precision is about $20\,\kms$ at redshift $z\sim0.8$.

Our model is in principle similar to the one introduced by \citet[][see also \citealt{rubin11a, prochaska11a}]{scarlata15a}, which the authors used as the basis in their radiative transfer simulations to interpret space-based observations in the FUV, though here we emphasize the statistical aspect of our model. 
Considering the sample size and S/N of our current data, we cannot yet place strong constraints on the model details, e.g, the functional form of the profiles or the parameters. In Figure~\ref{fig:model}, for illustration purposes, we assume a power law for the average velocity profile $v(r) \approx r^{\alpha}$ with $\alpha=2$, which is arbitrarily chosen. We leave the full modeling to future work, and focus on the general properties of the model predictions below.

\vspace{0.15in}

\noindent $\bullet$ \textbf{General model predictions}

\vspace{0.15in}
Based on the model, we can predict the following general properties of the absorption and emission features in the integrated spectra along the line of sight.

\vspace{-0.05in}
\begin{itemize}
\item[i.] Origins and aperture dependence -- In the model, the observed absorption and emission have different origins. The absorption is induced by gas in front of the background light source, e.g., the stellar populations inside the galaxy. The re-emitted (fluorescent) photons, resonant or non-resonant, are scattered isotropically, so the observed emission originates from gas located everywhere within the aperture, except from behind the galaxy due to occultation (see below). The strength of the emission therefore increases with the aperture size until the aperture encloses all the absorbing gas, while the strength of absorption depends little on the aperture size unless the column density varies significantly across the galaxy. 

\item[ii.] Net effect -- The sum of absorption and emission in a given set of transitions, including all the resonant and non-resonant channels, would be zero if and only if the observer could collect all the re-emitted photons scattered into the line of sight with a very large aperture. However, because of the finite size of a galaxy, the photons behind the galaxy cannot penetrate the high-density regions of the galaxy to be captured. As \citet{scarlata15a}, we refer to this effect as occultation. The finite size of galaxies makes the outflow model more complicated than that for stellar winds in which the star can be considered as a point source. The net effect of absorption and emission is therefore \textit{always} absorption. At redshift $z\sim0.8$, the typical effective \textit{radius} of SFGs with $M_*\sim10^{10}\,\MSun$ is about $2-4\,\kpc$ \citep[\eg][]{williams10a, wuyts12a}.

\item[iii.] Velocity profiles -- If outflows are dominant (compared to inflows), the observed emission profile is asymmetric and preferentially blueshifted due to occultation of the redshifted emission behind the galaxy. The profile of the absorption (even without emission infill) is also blueshifted, since it is only induced by the gas in front of the galaxy's stellar populations (the light source). 
We expect the degree of blueshift is smaller for the emission profiles than absorption, because the re-emitted photons are scattered isotropically and only a negligible fraction of photons originating from the observed absorption are scattered into the line of sight, while other re-emitted photons within the aperture are less blueshifted.

\item[iv-1.] Emission infill -- Like emission via the non-resonant channels, the model also predicts re-emitted photons via the resonant channels, producing emission filling in on top of the absorption. The emission infill is not sufficient to compensate for all the absorption so an observer always sees absorption (see point ii). However, because the emission profile is less blueshifted than absorption (point iii), the \textit{observed} absorption profile is more blueshifted than the ``\textit{true}'' absorption profile before emission infill. If the emission and absorption profiles are significantly different, \eg due to large outflow velocities or large aperture (relative to the occultation), we also expect to see P-cygni-like profiles. The amount of emission infill depends on the aperture size, the galaxy (occultation) size, the permitted channels and their transition probabilities, and the optical depth, which determines whether the observed emission originates from a single or multiple scattering process and the relative strength of different channels. The degree of emission infill (relative to the absorption) is therefore different from line to line. Below we discuss this quantitatively for the lines in the NUV.
\end{itemize}
\vspace{-0.05in}

\vspace{0.15in}

\noindent $\bullet$ \textbf{General radiative transfer processes}

\vspace{0.15in}

In an expanding envelope, if the velocity gradient is large, the radiative transfer processes can be treated under the Sobolev approximation \citep[][]{sobolev60a, castor70a, rybicki78a}. \citet[][see also \citealt{prochaska11a}]{scarlata15a} included an excellent discussion of the general radiative transfer processes involved in a galactic outflow, and \citet[][]{prochaska11a} also pioneered theoretical considerations of the NUV transitions. We refer the reader to those papers for more details. We here present a summary of the formulae most relevant to our analysis.

\vspace{0.05in}
\noindent $\bullet$ Absorption: In the Sobolev approximation, at a given position, the optical depth is given by
\beq
\tau(r) = \frac{\pi e^2}{m_e c} f_{lu} \lambda_{lu} n_l(r)\left|\frac{\ud v(r)}{\ud r}\right|^{-1} \,\mathrm{,}
\eeq
which is proportional to the density $n_l(r)$ at the lower level, the rest-frame wavelength of the transition $\lambda_{lu}=\lambda_{ul}$, oscillator strength $f_{lu}$, and the inverse of the velocity gradient (i.e., the thickness of the thin shell with the same velocity). We have ignored stimulated emission and angular dependence. Along the line of sight, the optical depth above applies to the velocity $v(r)$, or equivalently, the wavelength $\lambda_r=\lambda_{lu}[1-v(r)/c]$, at which the absorption is given by $R_r=e^{-\tau(r)}$.

\vspace{0.05in}
\noindent $\bullet$ Emission:
In a single-scattering event, after the absorption of a photon, the probability that the excited electron can decay from the upper level ($u$) to a given lower level ($l$) is given by
\begin{equation}
p_{ul} = \frac{A_{ul}}{\sum\nolimits_{i} A_{ui}}\,\mathrm{,}
\label{eq:single}
\end{equation}
where $A_{ui}$ is the spontaneous emission coefficient from the upper level $u$ to the lower level $i$, and the summation is over all the possible channels in the lower state. The above equation ignores stimulated emission, collisional (de-)excitation, and also fine-structure emission within the same state. 

If the electron decays to the original level, in our case, the lowest level in the lower state, the re-emitted photon can be absorbed again, resulting in a multiple-scattering process. The escape probability of a resonant photon from a shell of optical depth $\tau(v)$ is given by \citep[][]{mathis72a}
\begin{equation}
\beta_{esc} = \frac{1-e^{-\tau}}{\tau}\,\mathrm{,}
\end{equation}
and the fraction of the absorbed photons that are eventually re-emitted via a non-resonant ($nr$)  channel to a lower level $l$ is
\begin{eqnarray}
f_{nr,l}(\tau) & = & p_{nr,l}\sum\limits_{n=0}^{\infty}\left[p_r(1-\beta_{esc})\right]^{n} \nonumber \\
 & = & \frac{p_{nr,l}}{1-p_r(1-\beta_{esc})}\,\mathrm{,}
\end{eqnarray}
where $p_{nr, l}$ and $p_{r}$ are the probabilities of decaying to the non-resonant lower level $l$ and the resonant lowest level, respectively (Eq.~\ref{eq:single}), and we have omitted the upper level symbol $u$ for simplicity.
The fraction of the absorbed photons that are eventually re-emitted via the resonant channel and \textit{escape} from the shell is
\begin{eqnarray}
f_{r}(\tau) & = & p_{r}\beta_{esc}\sum\limits_{n=0}^{\infty}\left[p_r(1-\beta_{esc})\right]^{n} \nonumber \\
 & = & \frac{p_{r}\beta_{esc}}{1-p_r(1-\beta_{esc})}\,\mathrm{.}
\label{eq:multiple}
\end{eqnarray}
When the optical depth of a given shell $\tau(v)$ is shallow and the escape probability $\beta_{esc}\approx1$, we reach the single-scattering approximation, i.e., Eq.~\ref{eq:single}:
\begin{eqnarray}
f_{nr,l} & \approx & p_{nr,l} \,\mathrm{, and }\nonumber \\
f_{r} & \approx & p_{r} \,\mathrm{.}
\label{eq:shallow}
\end{eqnarray}
When the optical depth $\tau(v)$ is deep and the escape probability $\beta_{esc}\approx0$, all the re-emitted photons will be through the non-resonant channels, if there are any, with:
\begin{eqnarray}
f_{nr,l} & \approx & \frac{p_{nr,l}}{1-p_{r}}  \,\mathrm{. and }\nonumber \\
f_{r} & \approx & 0 \,\mathrm{.}
\label{eq:deep}
\end{eqnarray}
Note in both cases, summing over all channels gives 
\begin{equation}
\sum\nolimits_{nr, i} f_{nr, i} + f_{r} = 1 \,\mathrm{.} 
\end{equation}

\vspace{0.15in}

\noindent $\bullet$ \textbf{Radiative transfer processes in the NUV}

\vspace{0.15in}
With the formalism above, we now investigate quantitatively the absorption and emission lines in the NUV and their correlations in the context of the outflow model. We focus on eight resonant absorption lines, among which four have non-resonant channels: 
\vspace{-0.05in}
\begin{itemize}
\item \feii$\,\lambda2600$ with \feii$^*\,\lambda2626$ (UV1), 
\item \feii$\,\lambda2587$ with \feii$^*\,\lambda2613$ and $\lambda2632$ (UV1), 
\item \feii$\,\lambda2374$ with \feii$^*\,\lambda2396$ (UV2), 
\item \feii$\,\lambda2344$ with \feii$^*\,\lambda2366$ and $\lambda2381$ (UV3), 
\end{itemize}
\vspace{-0.05in}
and the other four do not: 
\vspace{-0.05in}
\begin{itemize}
\item \mgi$\,\lambda2853$, \mgii$\,\lambda2804$, \mgii$\,\lambda2796$, and \feii$\,\lambda2383$ (UV2).
\end{itemize}
\vspace{-0.05in}
We do not consider \feii$\,\lambda2261$ (UV4), \feii$\,\lambda2250$ (UV5), and \mniitriplet\ here because of their lower S/N in the data.

We present the relevant atomic data in Appendix~\ref{app:atomic}. We note that we do not detect \feii$^*\,\lambda2632$ from UV1, whose Einstein $A$ coefficient is about half that of \feii$^*\lambda2613$, and \feii$^*\,\lambda2381$ from UV3 has Einstein $A$ coefficient about half that of \feii$^*\,\lambda2366$ and is blended with \feii$\,\lambda2383$ from UV2. 

Within the context of the outflow model, we can now estimate the degree of the emission infill effect, i.e., the ratio of emission to absorption in the observed spectra, for the eight absorption lines in the NUV.
\vspace{-0.05in}
\begin{itemize}
\item[iv-2.] Degree of emission infill -- The effect of emission infill depends on two main factors. For those with non-resonant channels, it depends on the fraction of resonant emission ($f_r$). In the single-scattering approximation, we have $f_r \approx p_r$ (Eqs.~\ref{eq:shallow} and \ref{eq:single}) and with the atomic data from Appendix~\ref{app:atomic}, we have
\begin{eqnarray}
 & p^{\lambda2374}_{\rm Fe\,II}<p^{\lambda2587}_{\rm Fe\,II}<p^{\lambda2344}_{\rm Fe\,II}<p^{\lambda2600}_{\rm Fe\,II}<p_{\rm no\,nr}=1 \,\mathrm{.} \nonumber \\
 & 
\label{eq:ordernonresonant}
\end{eqnarray}
When multiple-scattering events are considered (Eq.~\ref{eq:multiple}), this order does not change though the relative difference is smaller. We expect the effect of the emission infill to follow the same order. 

When there is no non-resonant channel, it mainly depends on the degree of saturation -- the more saturated the absorption line is, the larger an effect the emission infill has on the observation. Note the observed emission and absorption have different origins (point i). Based upon the elemental abundance \citep[][]{asplund09a} and oscillator strength (Appendix~\ref{app:atomic}), the degree of saturation should be in the order of absorption strength as
\beq
W^{\lambda2853}_{\rm Mg\,I}<W^{\lambda2383}_{\rm Fe\,II}<W^{\lambda2804}_{\rm Mg\,II}<W^{\lambda2796}_{\rm Mg\,II}\,\mathrm{.}
\label{eq:orderresonant}
\eeq
Finally, we expect the degree of emission infill, manifested by the degree of blueshift (due to the difference in the profiles of emission and absorption (point iii)) and the change in the observed absorption strength, to follow the same order given by Eqs.~\ref{eq:ordernonresonant} and \ref{eq:orderresonant}. 
\end{itemize}
\vspace{-0.05in}

For a direct comparison of observations with the model predictions, it is necessary to understand what the true absorption profiles (before the emission infill) are. In the next section (\ref{sec:trueabsorption}), we introduce an observation-driven method to reveal the true absorption profiles.

\subsection{Revealing the True Absorption Profiles}\label{sec:trueabsorption}

To reveal the true absorption profiles, we make two assumptions: 
\vspace{-0.03in}
\begin{enumerate}
\item All the emission lines share the same normalized velocity profile.
\item All the absorption lines share the same true normalized velocity profile, i.e., prior to the emission infill.
\end{enumerate}
\vspace{-0.03in}
In both cases, the lines are normalized to have the same amplitude. We make these assumptions according to our understanding of quasar absorption-line systems from high S/N high-resolution spectroscopic observations, which show that \feii\ and \mgii\ (as well as \mgi\ when detected) usually trace each other \citep[\eg][]{churchill00a}. We consider it reasonable to extrapolate these results to galaxy absorption lines. Since we work on the observed profile, i.e, $R(\lambda)=e^{-\tau(\lambda)}$, but not the optical depth $\tau(\lambda)$, saturation plays an important role and we will treat the \mgii\ lines separately in our method.

Under the assumptions, our observation-driven method consists of two steps.
\vspace{-0.03in}
\begin{enumerate}
\item We determine the common emission profile from the four non-resonant emission lines, which we call the unified profile and present in Section~\ref{sec:emission}.
\item With the unified emission profile, we determine the unified true absorption profile with an iterative approach. We describe this process in detail in Section~\ref{sec:absorption}.
\end{enumerate}
\vspace{-0.03in}

\subsubsection{The Unified Emission Line Profile}\label{sec:emission}

\begin{figure}
\vspace{0.12in}
\epsscale{1.25}
\plotone{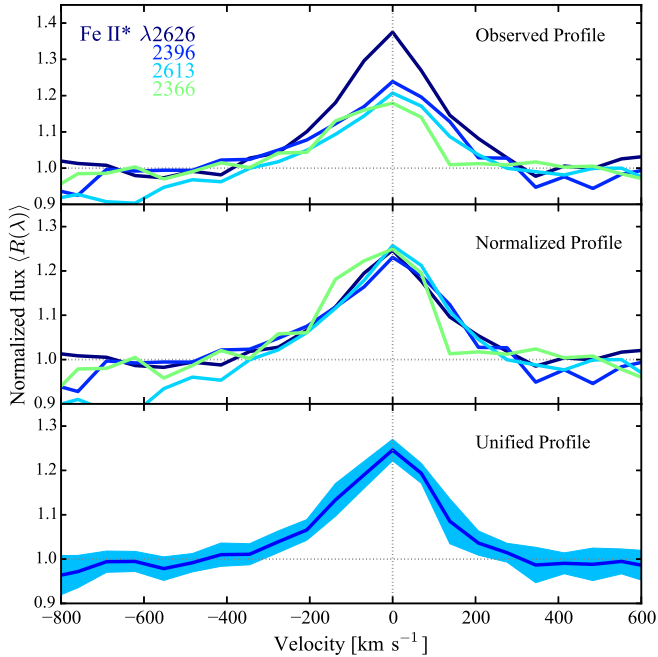}
\caption{The velocity profiles of the non-resonant emission lines, shown in the rest frame of the galaxies. \textit{Top panel}: The observed profiles. \textit{Middle panel}: The observed profiles normalized to the same amplitude. \textit{Bottom panel}: The mean unified emission profile. The shaded region indicates the $1\sigma$ uncertainties determined by bootstrapping.
}
\label{fig:emission}
\vspace{0.05in}
\end{figure}

\begin{figure}
\epsscale{1.25}
\plotone{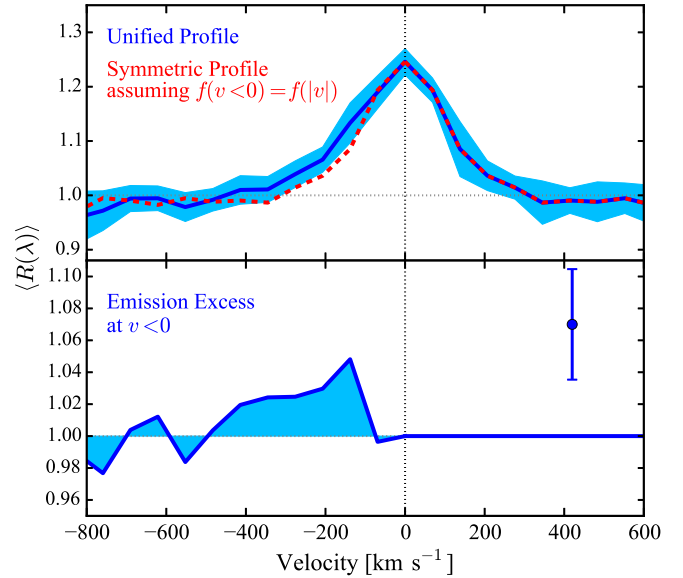}
\caption{The asymmetry of the unified emission profile. \textit{Upper panel}: The blue line and the shaded region show the mean unified emission profile and uncertainties as in the bottom panel of Figure~\ref{fig:emission}. The red dashed line shows a symmetric profile assuming the blueshifted side mirrors the observed redshifted side. \textit{Lower panel}: The emission excess on the blueshifted side. The errorbar at the top right indicates the mean uncertainty at a given pixel (velocity). 
}
\label{fig:emissionexcess}
\vspace{0.05in}
\end{figure}

We combine the four non-resonant emission lines, \feii$^*\,\lambda2626$, $\lambda2613$, $\lambda2396$, and $\lambda2366$, to determine the unified profile, as shown in Figure~\ref{fig:emission}. The top panel presents the observed velocity profiles, shown in the rest frame of the galaxies. In the middle panel, we normalize all the lines to have the same amplitude, and in the bottom panel, we present the mean normalized profile as the estimate of the unified emission profile. We also plot the $1\sigma$ uncertainties determined from bootstrapping. We note that, because the far blue side of \feii$^*\,\lambda2613$ at $v\lesssim-600\,\kms$ overlaps with the red side of \mnii$\,\lambda2606$, to avoid the contamination, we do not include the blue side of \feii$^*\,\lambda2613$ at $v<0\,\kms$ while calculating the mean profile.

The unified emission profile appears to be asymmetric. We investigate its asymmetry further in Figure~\ref{fig:emissionexcess}. In the upper panel, on top of the unified profile, we overlay a symmetric profile assuming the blue side mirrors the red side. We subtract the symmetric profile from the original one and show the result in the lower panel. The unified non-resonant \feii$^*$ emission exhibits an excess on the blue side, with a confidence level higher than $2\sigma$ when integrated over $-500\,\kms<v<0\,\kms$, indicating that a larger fraction of emission is blueshifted than redshifted. 

\subsubsection{The Unified Absorption Line Profile}\label{sec:absorption}

\begin{figure*}
\epsscale{1.20}
\plotone{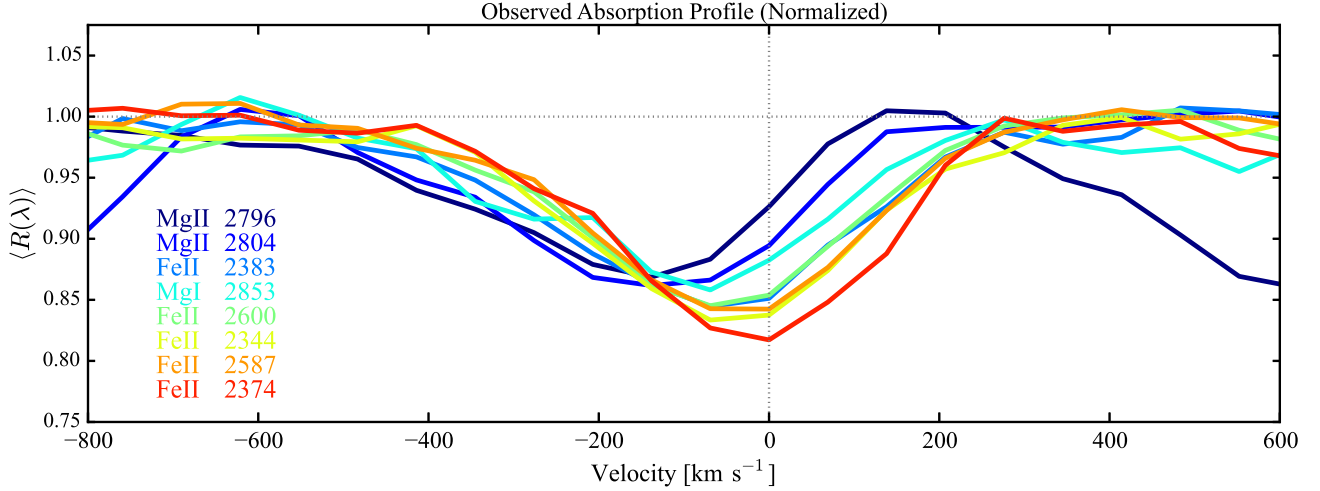}
\caption{The observed velocity profiles of the absorption lines, presented in the rest frame of the galaxies and normalized to the same amplitude. The color scales indicate the order given by Eqs.~\ref{eq:ordernonresonant} and \ref{eq:orderresonant}.
}
\label{fig:observedabsorption}
\end{figure*}

\begin{figure*}
\epsscale{0.565}
\plotone{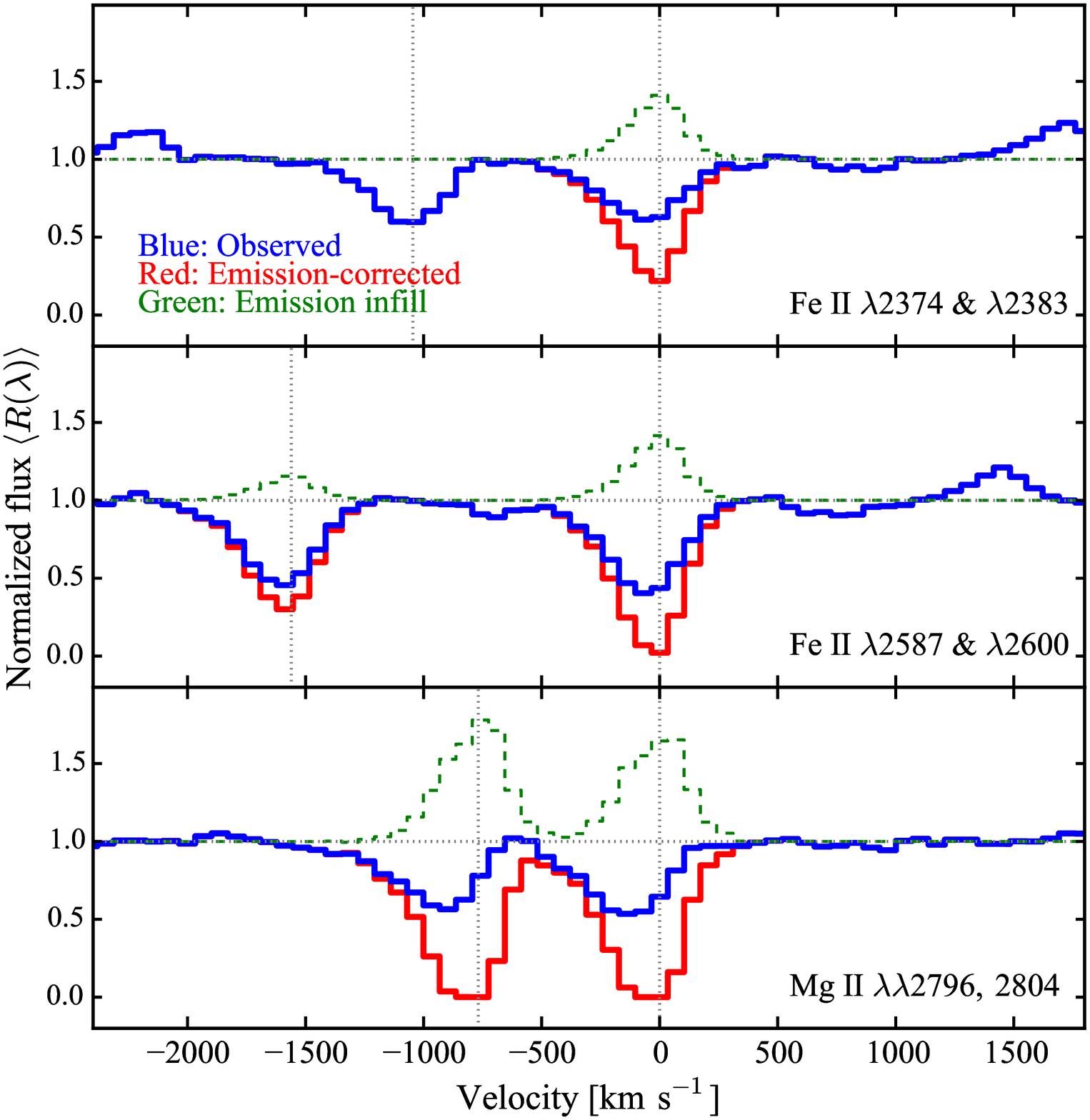}
\epsscale{0.565}
\plotone{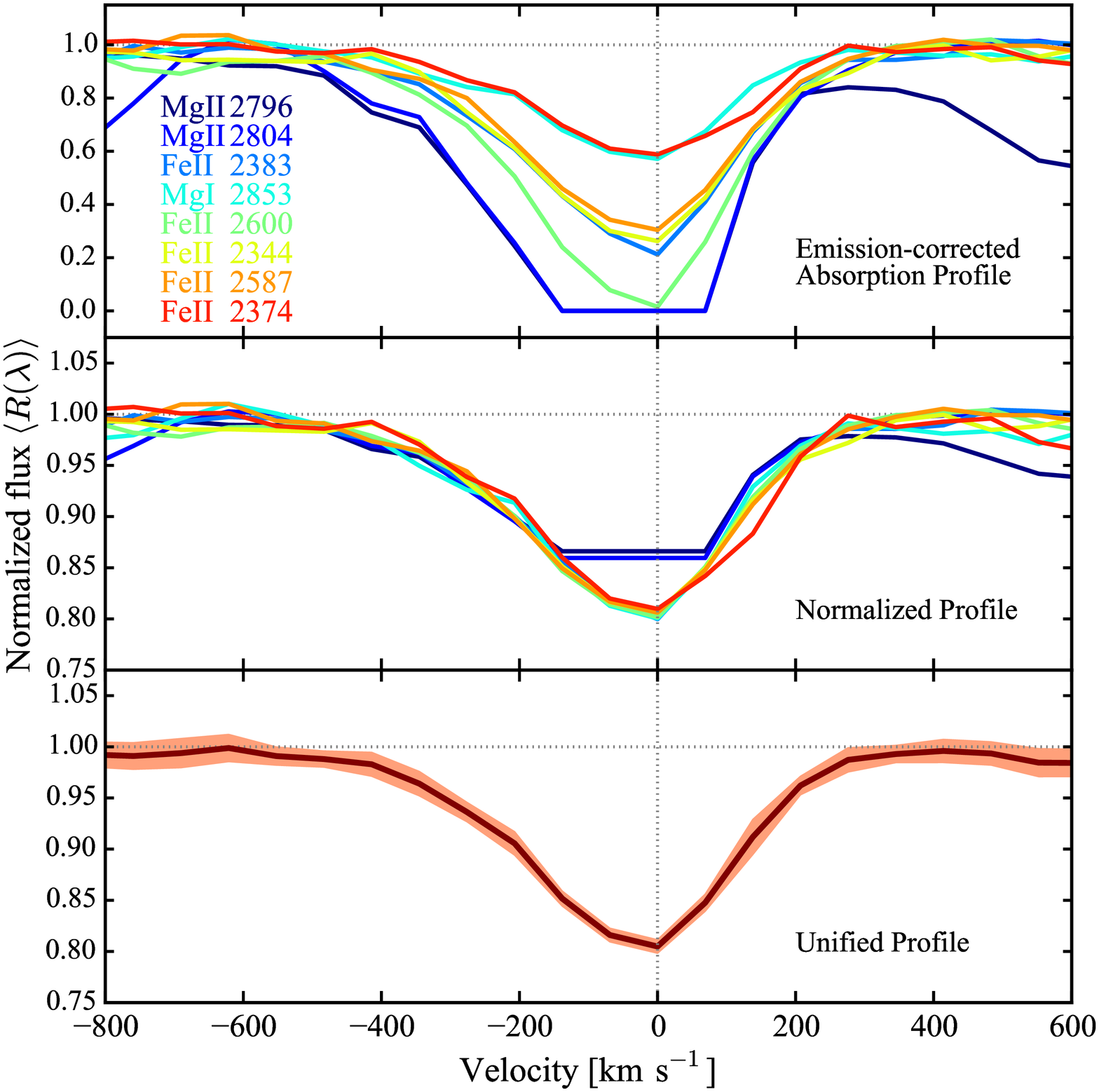}
\caption{\textit{Left panels}: Examples of emission-infill correction. The blue lines show the observed absorption velocity profiles, red the emission-corrected, and green the subtracted emission. The vertical dotted lines mark the rest-frame positions of the lines and the zero velocity corresponds to the wavelength of the one with lower energy in each panel.
\textit{Right panels:} \textit{Top} -- The emission-corrected absorption velocity profiles. \textit{Middle} -- The emission-corrected profiles normalized to the same amplitude. \textit{Bottom} -- The unified emission-corrected absorption profile. The shaded areas indicate the $1\sigma$ bootstrapping uncertainties. The color scales in the top two panels are the same as in Figure~\ref{fig:observedabsorption}.
}
\label{fig:trueabsorption}
\vspace{0.05in}
\end{figure*}

With the unified emission line profile, we determine the emission infill and the unified true absorption line profile simultaneously with an iterative approach. 

To proceed, we first investigate the observed velocity profiles in more detail.  In Figure~\ref{fig:observedabsorption}, we present the observed profiles of the eight absorption lines in the rest frame of the galaxies, normalized to have the same amplitude integrated over $-700\,\kms<v<300\,\kms$. We have ordered the lines according to Eqs.~\ref{eq:ordernonresonant} and \ref{eq:orderresonant}, with bluer color indicating a larger predicted effect from emission infill. Figure~\ref{fig:observedabsorption} shows that: (1) the observed absorption profiles are different; (2) the degree of blueshift follows the order predicted by the model, and (3) the most blueshifted line, \mgii$\,\lambda2796$, is about $200\,\kms$ more blueshifted than the least blueshifted \feii$\,\lambda2374$.

\begin{figure*}
\epsscale{0.57}
\plotone{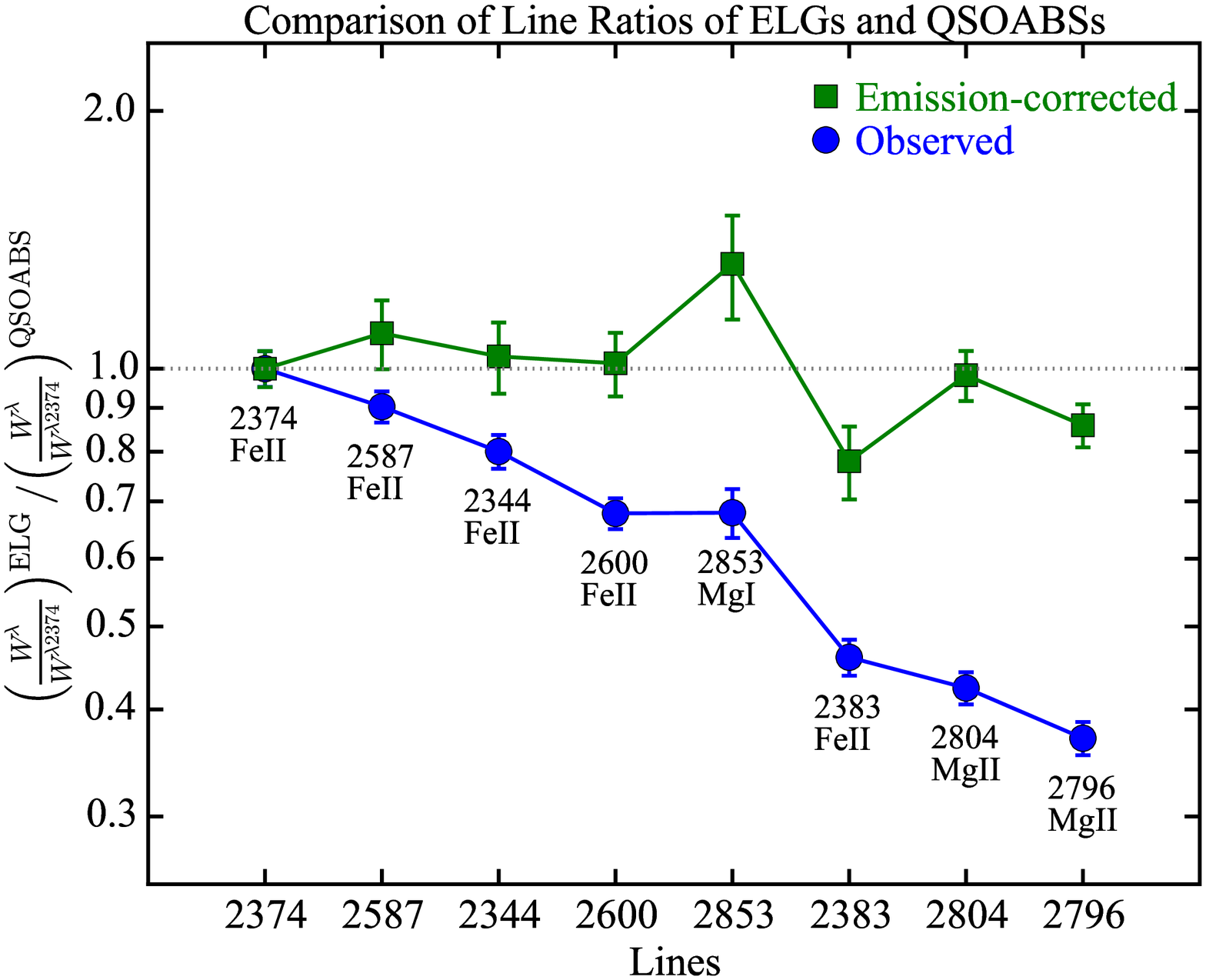}
\epsscale{0.57}
\plotone{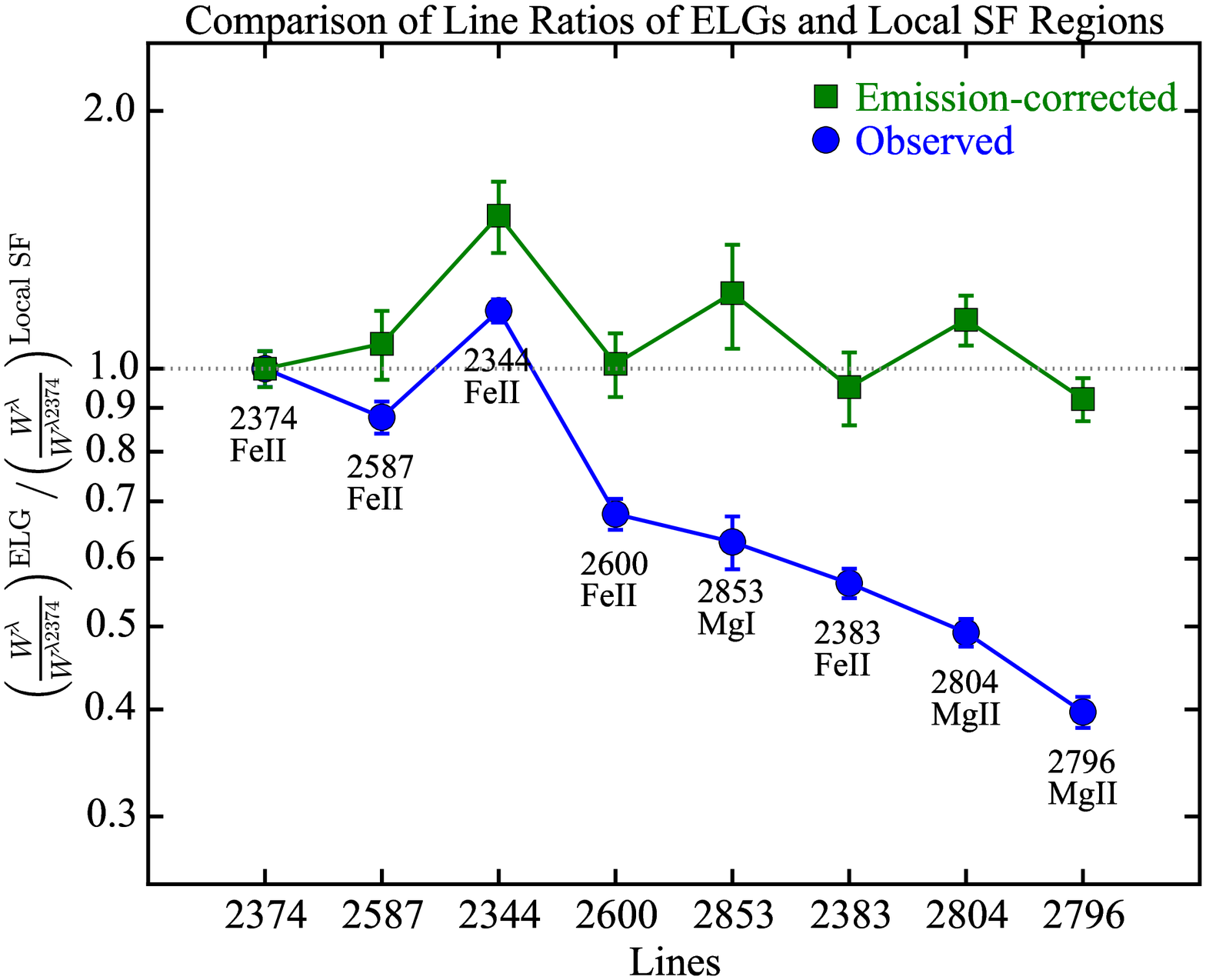}
\caption{The comparison of line ratios of ELGs with strong quasar absorbers (QSOABSs, \textit{left}) and local SF regions (\textit{right}). The blue circles show the observed line ratios, and the green squares the emission-corrected ones. We use \feii$\,\lambda2374$ as the anchor and order the lines according to Eq.~\ref{eq:ordernonresonant} and \ref{eq:orderresonant}. The errorbars indicate uncertainties in ELG line ratio measurements determined by bootstrapping, not including the uncertainties in the measurements of QSOABSs or local SF regions.
}
\vspace{0.1in}
\label{fig:absorptionlineratios}
\end{figure*}

Our iterative approach to determining the true absorption profile consists of the following steps.
\vspace{-0.03in}
\begin{enumerate}
\item We first use the normalized profile of \feii$\,\lambda2374$ as the initial guess of the unified true absorption profile, because the fluorescent emission after the \feii$\,\lambda2374$ absorption is dominated by the non-resonant channel \feii$^*\lambda 2396$ (Eq.~\ref{eq:ordernonresonant}).
\item With the unified absorption profile estimated from the previous step, we fit for the amount of emission that needs to be subtracted from the observed profile. More specifically, we express the observed profile $R^{\rm obs}_{\rm abs}(\lambda)$ by
\begin{eqnarray}
R^{\rm obs}_{\rm abs}(\lambda) & = & R^{\rm true}_{\rm abs}(\lambda)+\left[R_{\rm emi}(\lambda)-1\right] \, \nonumber \\
  & = & \left\{1-a\left[1-R^{\rm uni}_{\rm abs}(\lambda)\right]\right\}+b\left[R^{\rm uni}_{\rm emi}(\lambda)-1\right] \, \mathrm{,} \, \nonumber \\
  &   & 
\label{eq:decomp}
\end{eqnarray}
where $R^{\rm true}_{\rm abs}(\lambda)$ and $R_{\rm emi}(\lambda)$ are the unnormalized true absorption and emission profiles, respectively, and $R^{\rm uni}_{\rm abs}(\lambda)$ and $R^{\rm uni}_{\rm emi}(\lambda)$ are the unified normalized absorption profile from the previous step and the unified emission profile from Section~\ref{sec:emission}, respectively.
We perform a least-squares fit for the coefficients $a$ and $b$.
\item We normalize the new absorption profiles $R^{\rm true}_{\rm abs}(\lambda)$ from the fitting in Step $2$, calculate the mean as the new estimate of the unified absorption profile, and then repeat Step $2$.
\end{enumerate}
\vspace{-0.03in}
As discussed above, saturation requires special attention when it is severe. We set $R^{\rm true}_{\rm abs}$ at saturated pixels to be zero and do not include \mgii\ while estimating the unified profile in Step $3$. We iterate the steps until the unified absorption profile and the coefficients $a$ and $b$ converge. In practice, we find that three iterations are sufficient to reach convergence.

We show the results in Figure~\ref{fig:trueabsorption}. On the left, we show examples of the decomposition (Eq.~\ref{eq:decomp}), with the emission-infill indicated by the green dashed lines. The emission-corrected absorption profiles, shown with the red lines, are deeper but less blueshifted than the observed ones (blue). On the right, in the top panel, we show the emission-corrected profiles of all the eight absorption lines, with color scales the same as in Figure~\ref{fig:observedabsorption}. The middle panel shows the emission-corrected profiles normalized to the same amplitude, which we use to estimate the unified true absorption profile. We present the unified profile in the bottom panel, together with the uncertainties estimated via bootstrapping. Note that after emission-infill correction, the \mgii\ lines are heavily saturated at the line centers and are not used in the calculation of the mean profile.

The unified absorption profile is apparently asymmetric and preferentially blueshifted. With the emission-infill correction and the true absorption profiles estimated, we now investigate the details of the observed and emission-corrected profiles.

\subsubsection{Non-parametric characterization of the true absorption profiles}\label{sec:nonpar}

To characterize the absorption profiles, we choose to use non-parametric variables, line ratios based on rest equivalent width (Eq.~\ref{eq:rewabs}) and velocity offsets.

\vspace{0.05in}
\noindent $\bullet$ \textbf{Line ratios}
\vspace{0.05in}

Because of the different degrees of emission infill, the changes in the absorption strength vary from line to line. We compare line ratios before and after the emission-infill correction to study if the effect of emission infill follows the order of Eqs.~\ref{eq:ordernonresonant} and \ref{eq:orderresonant}.

We measure the rest equivalent width of the emission-corrected profiles by integrating over $-700\,\kms<v<300\,\kms$. For the observed profiles of lines other than the \mgii\ doublet, we integrate over the same velocity range. We select the integration velocity range $-700\,\kms<v<200\,\kms$ for the observed \mgii$\,\lambda2796$ and $-600\,\kms<v<300\,\kms$ for \mgii$\,\lambda2804$ to avoid the contaminations from each other (see Figure~\ref{fig:observedabsorption}). 

To calculate line ratios, we select \feii$\,\lambda2374$ as the anchor, i.e., the common denominator, which has negligible emission infill. In Figure~\ref{fig:absorptionlineratios}, we compare the line ratios before and after the emission-infill correction with the line ratios in the composite spectrum of strong quasar absorbers (QSOABSs, in the same redshift range) on the left and with those of local SF regions on the right. The variables plotted are the ratios of line ratios:
\begin{eqnarray}
\left(\frac{W^{\lambda_i}}{W^{\lambda2374}}\right)^{\rm ELG}{\bigg/}\left(\frac{W^{\lambda_i}}{W^{\lambda2374}}\right)^{\rm QSOABS} & \,\mathrm{and} \nonumber \\
\left(\frac{W^{\lambda_i}}{W^{\lambda2374}}\right)^{\rm ELG}{\bigg/}\left(\frac{W^{\lambda_i}}{W^{\lambda2374}}\right)^{\rm Local\,SF} \,\mathrm{,} & 
\end{eqnarray}
where $\lambda_i$ represents the line names. In the figure, we have ordered the lines according to the predictions of Eqs.~\ref{eq:ordernonresonant} and \ref{eq:orderresonant}.

Figure~\ref{fig:absorptionlineratios} shows that the observed line ratios in the composite spectrum of ELGs differ from those of strong quasar absorbers and also local SF regions, and the degree of the difference basically follows the predicted order. We find that the effect of emission infill (on line strength and ratio) can be larger than a factor of two, e.g., for \feii$\,\lambda2383$ and \mgiidoublet. After the emission-infill correction, we find that the line ratios are consistent in the spectra of all the sources. 

We stress that our method of estimating the emission infill is observation-driven, completely independent of any model. We only made the two assumptions about the unified emission and absorption profiles as elaborated at the beginning of this section. We also did not use any information about the strong quasar absorbers or the local SF regions. The agreement of the final line ratios is therefore not by construction, but rather the result of the method. As we expect that the gases inducing the absorption lines in different sources have similar origins, either supernova yields, stellar mass losses or other sources, and the line ratios in different source spectra should be in agreement, we consider our iterative fitting method successful in estimating the emission infill and determining the true absorption profiles.


\begin{figure}
\vspace{0.05in}
\epsscale{1.2}
\plotone{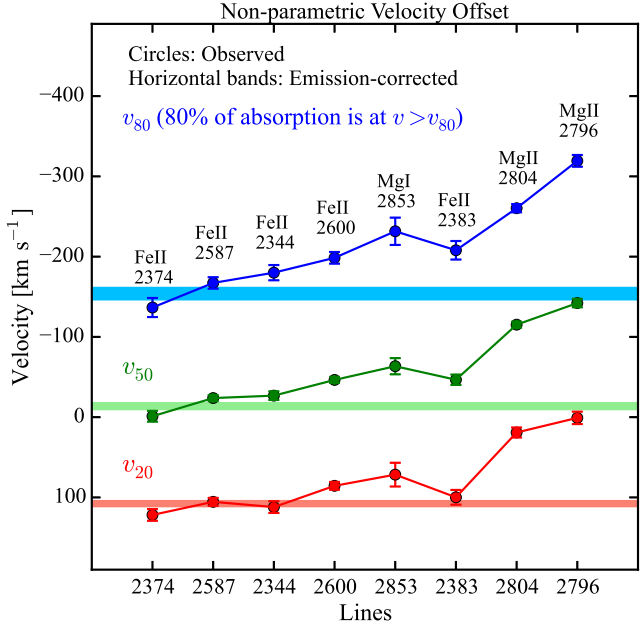}
\caption{The non-parametric characterization of the absorption velocity profiles. We show $v_{80}$ (\textit{blue}), $v_{50}$ (\textit{green}) and $v_{20}$ (\textit{red}), before (\textit{circles}) and after (\textit{horizontal bands}) the emission-infill correction.  The order of the lines is the same as in Figure~\ref{fig:absorptionlineratios}. The width of the horizontal bands indicate the uncertainties. All the uncertainties are determined by bootstrapping.
}
\label{fig:absorptionvelocity}
\vspace{0.05in}
\end{figure}

\vspace{0.05in}
\noindent $\bullet$ \textbf{Velocity offsets}
\vspace{0.05in}

To characterize the velocity profile, we define a velocity offset variable $v_{\rm xx}$ to be the velocity where a fraction of ${\rm xx}\,$per cent of the absorption is at velocity $v>v_{\rm xx}$\footnote{We note that this is a characterization of the \textit{total} absorption profile, including contributions from both outflows and ISM.}. In Figure~\ref{fig:absorptionvelocity}, we show $v_{80}$, $v_{50}$ and $v_{20}$ of the observed absorption profiles and of the unified profile. The order of lines is the same as in Figure~\ref{fig:absorptionlineratios}, given by Eqs.~\ref{eq:ordernonresonant} and \ref{eq:orderresonant}.

Figure~\ref{fig:absorptionvelocity} (see also Figure~\ref{fig:observedabsorption}) shows that, for the observed profiles, the velocity offsets of different lines are different, and the order follows the one predicted by the radiative transfer considerations in Eqs.~\ref{eq:ordernonresonant} and \ref{eq:orderresonant} and the difference in the emission and absorption profiles (point iii in Section~\ref{sec:model}).  The unified emission-corrected profile is still asymmetric, but with a smaller degree of blueshift: the $50\%$ velocity offset $v_{50}$ is about $-10\,\kms$, and $|v_{80}|$ is about $160\,\kms$, larger than $|v_{20}|\sim110\,\kms$. Our fitting result shows that without correcting for the emission infill, $v_{80}$, the $80\%$ velocity offset, can be overestimated by over a factor of two for the lines most severely affected, e.g., \feii$\,\lambda2383$ and \mgiidoublet.

\subsection{Discussion}\label{sec:inflow}

\begin{figure}
\vspace{0.02in}
\epsscale{1.20}
\plotone{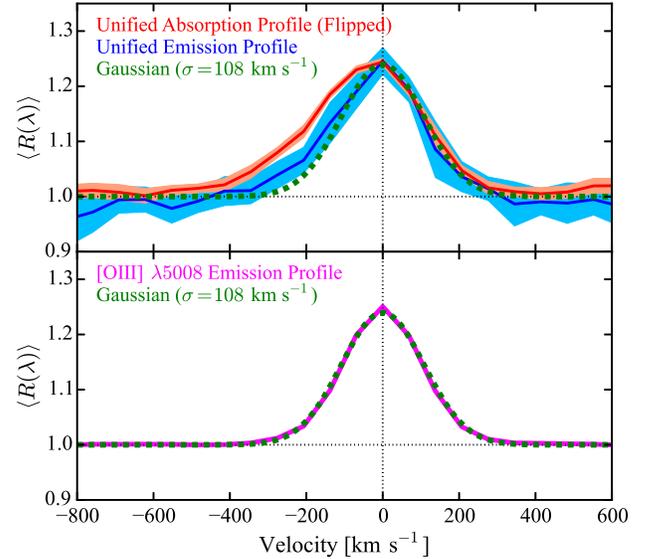}
\caption{\textit{Upper panel}: The comparison of the unified, emission-corrected absorption velocity profile (\textit{red}), the unified emission profile (\textit{blue}), and a Gaussian profile with the width of $108\,\kms$ (\textit{green}). For display purposes, we have flipped the absorption profile and also adjusted the profiles so that they have roughly the same peak value. The shaded regions show the $1\sigma$ bootstrapping uncertainties. \textit{Lower panel}: The comparison of the normalized \oiii$\,\lambda5008$ emission profile for ELGs (at $0.6<z\lesssim1.2$, \textit{magenta}) with the same Gaussian profile (\textit{green}) as in the upper panel. The uncertainties of the \oiii\ profile are about the same size as the line width.
}
\label{fig:oiiiprofile}
\vspace{0.05in}
\end{figure}

We now compare the observations, including the results of our emission-infill correction method, with the predictions of the spherical outflow model presented in Section~\ref{sec:model}. We go over the predictions point by point.


\vspace{0.05in}
\noindent [i.] Aperture dependence -- The physical aperture size of the eBOSS ELG spectra is about $15\,\kpc$, while that of the \textit{HST} FOS/GHRS spectra of the local SF regions is smaller than $40\,\pc$. In the spectra of the local SF regions, we do not detect the non-resonant emission that otherwise persist in the ELG ones. This agrees with the model and is because the \textit{HST} FOS/GHRS aperture size is too small to capture the extended fluorescent emission.

\vspace{0.05in}
\noindent [ii.] Net effect -- The net effect is always absorption due to occultation in the model, even if the aperture encloses all the emission scattered into the line of sight. In the data, when we sum up all the absorption and emission in a given set of channels, the net result is absorption.

\vspace{0.05in}
\noindent [iii.] Velocity profiles -- The outflow model predicts that both the emission and (emission-infill corrected) absorption profiles are blueshifted, with the absorption more so than the emission. Figures~\ref{fig:emissionexcess} and \ref{fig:absorptionvelocity} quantify the asymmetry and the blueshift of the profiles individually, though in different ways. To compare the two profiles directly, we show them together in Figure~\ref{fig:oiiiprofile}. For display purposes, we have flipped the absorption profile and also normalized them so that they roughly have the same peak value. The absorption profile is more blueshifted than the emission, as in the model. 

As discussed in the basics of the model, the composite spectra include effects from not only the outflows, but also the inflows, the motions of the ISM in the galaxy, the instrumental resolution as well as the redshift precision. If the aperture is large and collects all the re-emitted photons along the line of sight, and if the outflowing gas is extended to much larger scale than the galaxy, we expect the red side of the emission profiles to be broader than nebular lines, extended to further red. On the other hand, if a substantial fraction of the gas is falling in onto the galaxy at high velocities, we also expect the red side of both the emission and absorption profiles to be broader. In the lower panel of Figure~\ref{fig:oiiiprofile}, we show the \oiii$\,\lambda5008$ profile of the NUV sample. We note that ELGs at $z\gtrsim1$ do not have \oiii\ coverage, but they account for a small fraction ($10\%$) of the sample. We find the \oiii$\,\lambda5008$ profile is well-represented by a Gaussian profile and the best-fit width ($\sigma$) is about $108\,\kms$. We overplot this Gaussian profile in both panels for comparison. 

The mean spectral resolution of the BOSS spectrographs is about $60-70\,\kms$ and the average redshift precision of the eBOSS ELGs at redshift $0.6<z<1.2$ is about $20\,\kms$. The width of the nebular emission line profile ($108\,\kms$) must therefore be dominated by the intrinsic rotation and disordered motion of the ISM along the line of sight, which account for about $85\,\kms$ when we subtract the spectral resolution and redshift precision by quadrature. This is consistent with measurements of kinematic properties and the Tully-Fisher relation for bright galaxies at these redshifts \citep[\eg][]{vogt96a, weiner06a, miller11a}.

We find that, on the red side, both the emission and absorption profiles are consistent with \oiii$\,\lambda5008$ within the uncertainties.
This suggests that, within the uncertainties of our data, we have not observed evidence for outflows on scales larger than the galaxy nor the evidence for inflows. The larger-scale outflows would extend the emission profile to higher velocities (farther on the red side), and the inflows would extend both the emission and absorption profile. However, the former (that we did not see evidence for larger-scale outflows) could be because the aperture size ($15\,\kpc$) is not sufficiently large, while the latter could be because the inflow velocities (e.g., $<200\,\kms$, \citealt{rubin12a}) are comparable to the ISM motions and the emission/absorption effects of the inflows and the ISM are blended together.

\begin{figure*}[t]
\vspace{0.1in}
\epsscale{0.55}
\plotone{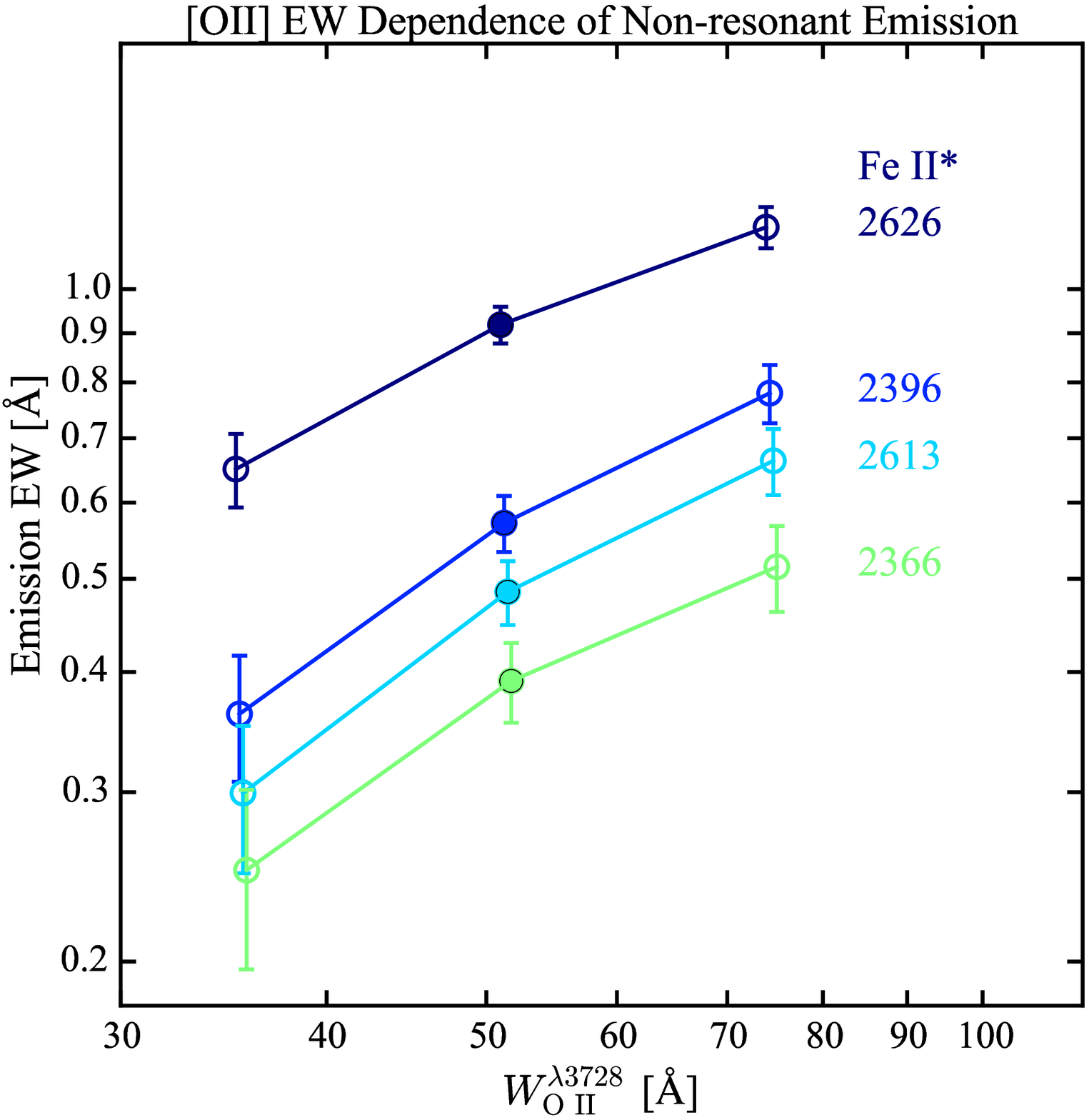}
\epsscale{0.55}
\plotone{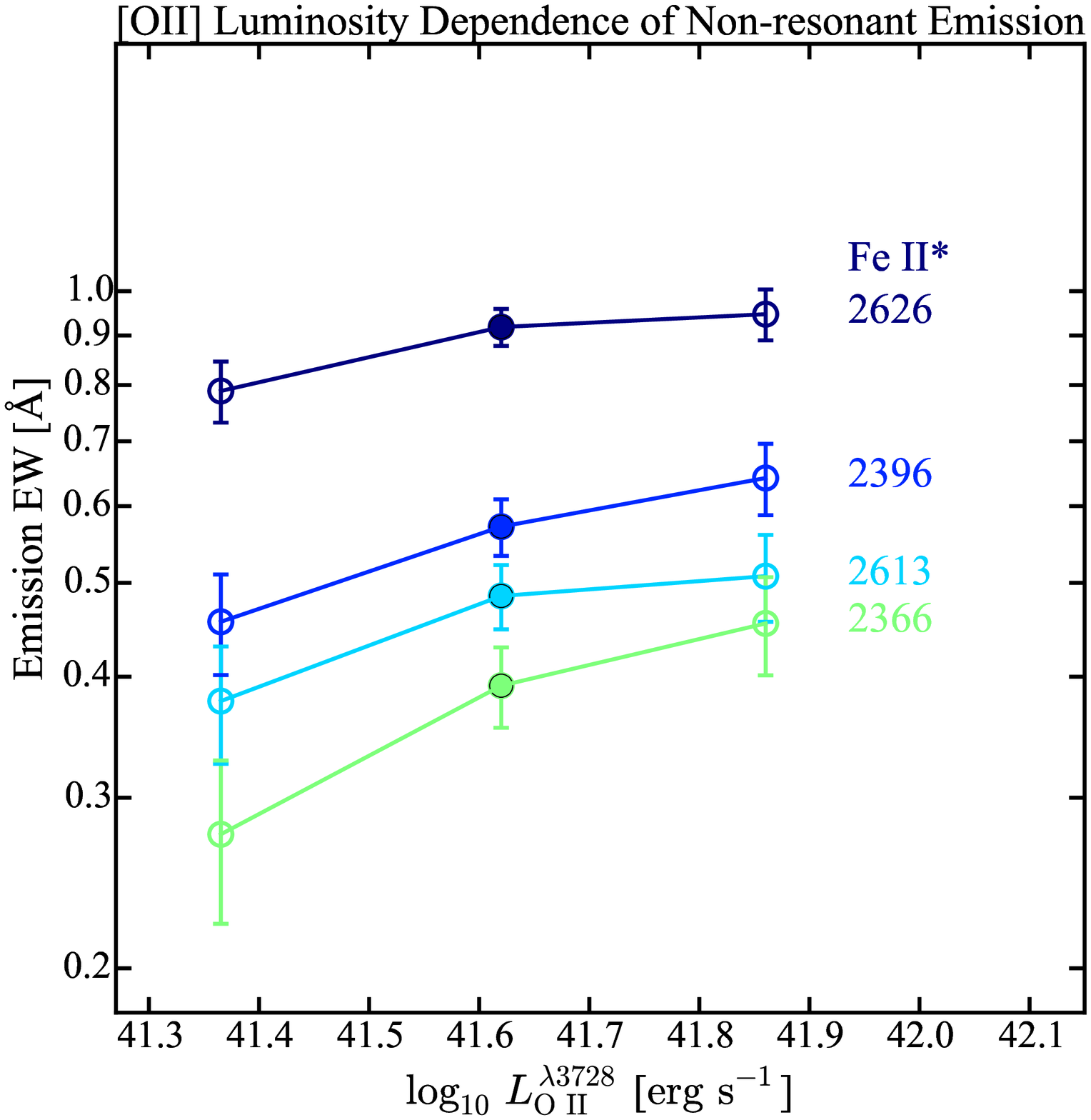}
\caption{The dependences of the rest equivalent width of the non-resonant emission lines on the \oiidoublet\ rest equivalent width (\textit{left}) and luminosity (\textit{right}). Note both the x axes are in logarithmic scale.
}
\label{fig:emissionrewoii}
\vspace{0.05in}
\end{figure*}

If we assume the emission-corrected absorption on the red side solely originates from the ISM, we can decompose the total absorption into a part due to the ISM with a symmetric profile and the other due to the outflows \citep[\eg][]{weiner09a}\footnote{We note \citet{weiner09a} applied this method to the \textit{observed} \mgii\ absorption as they did not consider the emission infill.}. Applying this decomposition method to the unified absorption-corrected profile, we obtain a blueshifted excess profile, more extended than the emission excess at $v<0\,\kms$ as shown in Figure~\ref{fig:emissionexcess}, with maximum velocity about $-600\,\kms$. The ratio of the excess to the subtracted symmetric profile, which represents the ratio of the amount of outflowing gas to that of the ISM assuming the decomposition is ideal, is about 1:3. However, because there could be a broad distribution of outflow velocities, i.e., a large $\sigma(r)$ in the outflow model even without the ISM contribution, it is likely that the symmetric component also includes a large contribution from the outflowing and/or inflowing gas, in which case the contribution from the ISM is much smaller than 3/4.

\vspace{0.05in}
\noindent [iv.] (Degree of) Emission infill -- The model predicts that, in large-aperture spectra, there is emission filling in on top of the resonant absorption, and because the emission profile is less blueshifted, the infill results in an observed absorption profile that is more blueshifted. The degree of the emission infill depends on the transition probabilities of the permitted channels and the degree of saturation for those without non-resonant channels. We have demonstrated that the data agree with these predictions in Figure~\ref{fig:observedabsorption}-\ref{fig:absorptionvelocity}. In particular, we show that the blueshifts, line ratios and velocity offsets of the observed profiles follow the same order as predicted (Eqs.~\ref{eq:ordernonresonant} and \ref{eq:orderresonant}). The observed $v_{80}$, the $80\%$ velocity offset, can be overestimated by a factor of two compared to that in the emission-corrected profile. After the emission correction, the line ratios in the spectra of ELGs are consistent with those of strong quasar absorbers and local SF regions.

\begin{figure}
\epsscale{1.20}
\plotone{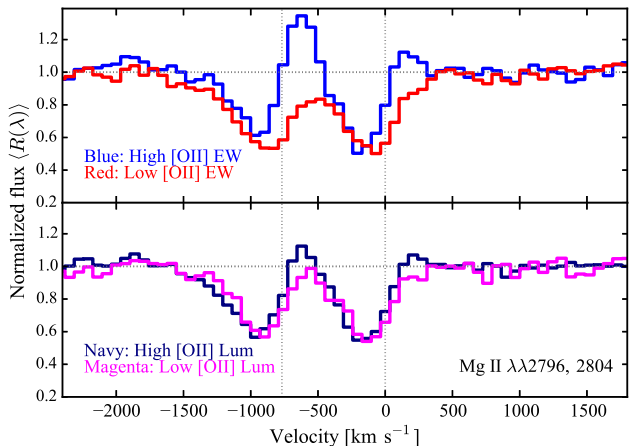}
\caption{The dependences of the observed absorption velocity profiles of the \mgii\ lines on the \oiidoublet\ rest equivalent width (\textit{upper panel}) and luminosity (\textit{lower panel}). 
}
\label{fig:absorptionoiimgii}
\vspace{0.05in}
\end{figure}

The different degrees of blueshift for different lines were also suggested by \citet{prochaska11a} and were observed in the Keck spectra of SFGs by \citet[][see their Figure $7$]{erb12a}. \citet[][see also \citealt{kornei13a}]{erb12a} also found a variety of \mgii\ profiles in their individual spectra, with some showing emission that might originate from \hyii\ regions. In our model, we have ignored such contribution. With our data, we cannot yet quantify the effect of the emission from \hyii\ regions on the line profiles in the composite analysis.

We do not observe P-cygni-like profiles in the composite spectrum of the full sample. This is likely because the difference of the emission and absorption profiles on the red side is not large (Figure~\ref{fig:oiiiprofile}) and the amount of emission infill is not sufficient, since P-cygni-like profiles require a large amount of emission infill that is more extended on the red side. In the next section, when studying the \oiidoublet\ dependence, we show that the \mgii\ absorption features P-cygni-like profiles for the subsample with the higher \oiidoublet\ rest equivalent width.

\vspace{0.05in}

In summary, we conclude that our statistical, spherical outflow model can simultaneously explain the multiple observed properties of emission and absorption features in the NUV. 

\section{Correlations with \oiidoublet}\label{sec:oii}

Observations have shown that outflow properties, such as the velocity,  depend on galaxy properties \citep[\eg][]{rupke05a, tremonti07a}. From the eBOSS pilot observations, we can measure the \oiidoublet\ properties of the ELGs. We here study the dependences of the emission and absorption lines in the NUV on the total rest equivalent width ($W^{\lambda3728}_{\rm [O\,II]}$) and luminosity ($L^{\lambda3728}_{\rm [O\,II]}$) of the \oii\ doublet. 

For each variable, we divide the NUV sample into two subsamples, split at the median values ($\left<W^{\lambda3728}_{\rm [O\,II]}\right>=51.4\,{\rm \AA}$ and $\left<\log_{10} L^{\lambda3728}_{\rm [O\,II]}\right>=41.6\,{\rm dex}$). We then perform the same analysis as for the full sample, including constructing the composite spectra, calculating the unified emission profiles, estimating the emission-infill and determining the true absorption profiles. We present some of the details in Appendix~\ref{app:oii}, including the distributions of $W^{\lambda3728}_{\rm [O\,II]}$ and $\log_{10} L^{\lambda3728}_{\rm [O\,II]}$ (Figure~\ref{fig:oiiewlumdist}), the observed emission/absorption profiles (Figures~\ref{fig:emissionprooii} and \ref{fig:absorptionprooii}), and the (unnormalized) emission-corrected absorption profiles (Figure~\ref{fig:unifiedabsorptionprooii}). We here discuss in detail the emission strength, the observed absorption profiles of \mgii, the emission-corrected absorption strength, and the unified velocity profiles.

\begin{figure*}[b]
\epsscale{0.53}
\plotone{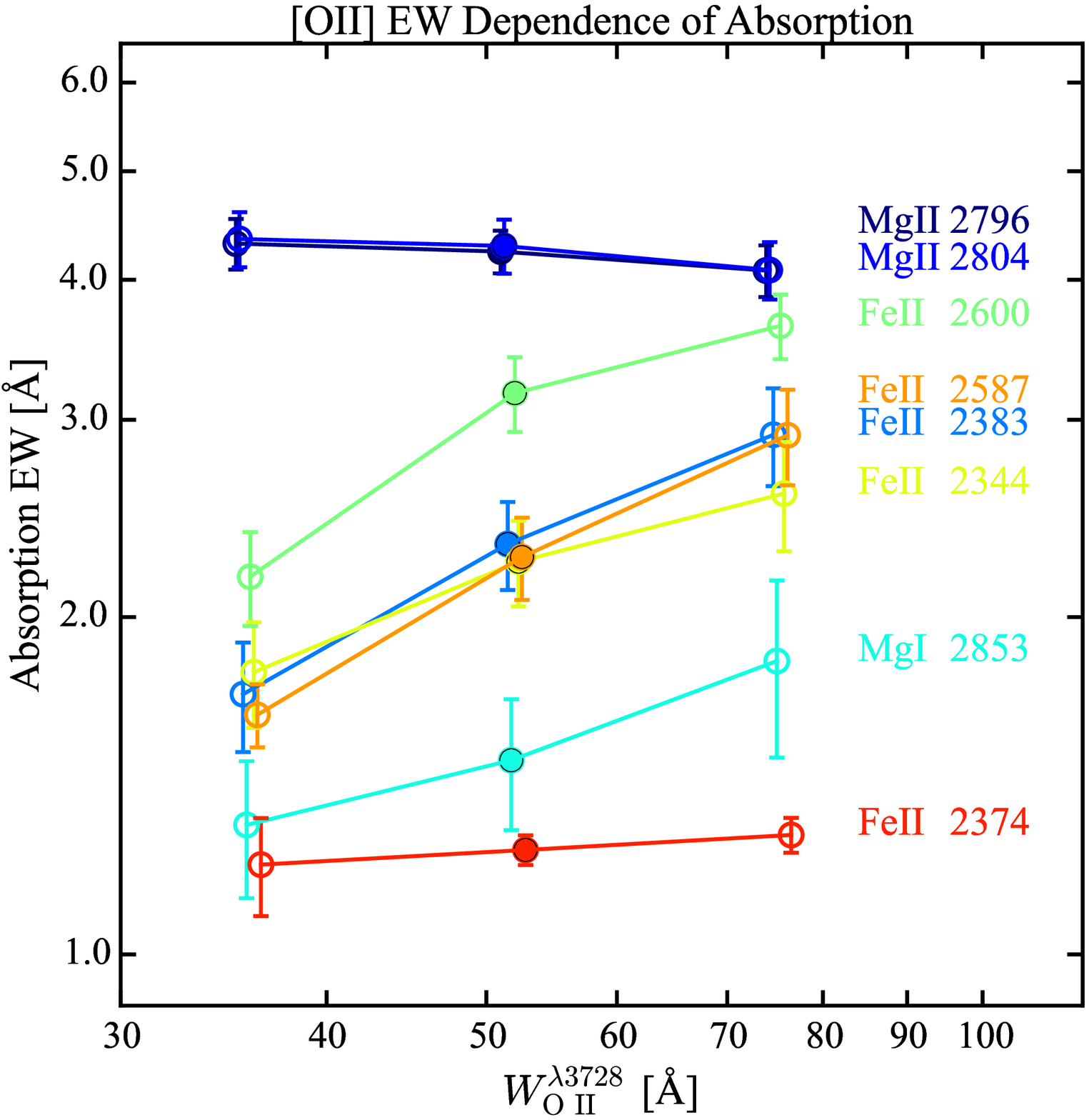}
\epsscale{0.53}
\plotone{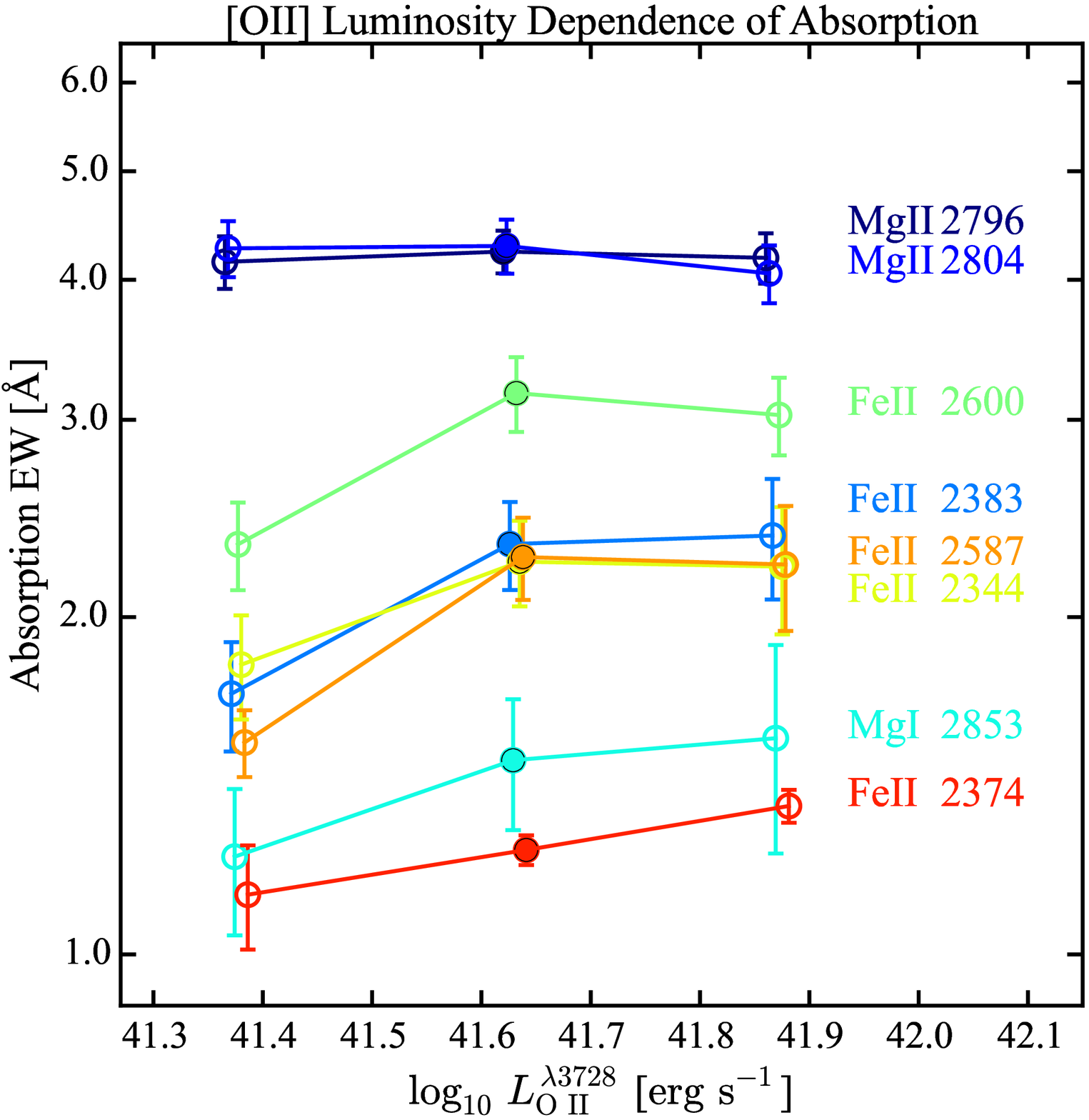}
\caption{The dependences of the rest equivalent width of the emission-corrected absorption lines on the \oiidoublet\ rest equivalent width (\textit{left}) and luminosity (\textit{right}). The color scales are the same as in Figure~\ref{fig:observedabsorption}, based on the orders given by Eqs.~\ref{eq:ordernonresonant} and \ref{eq:orderresonant}.
}
\label{fig:absorptionoii}
\vspace{0.05in}
\end{figure*}

\begin{figure*}[t]
\epsscale{0.53}
\plotone{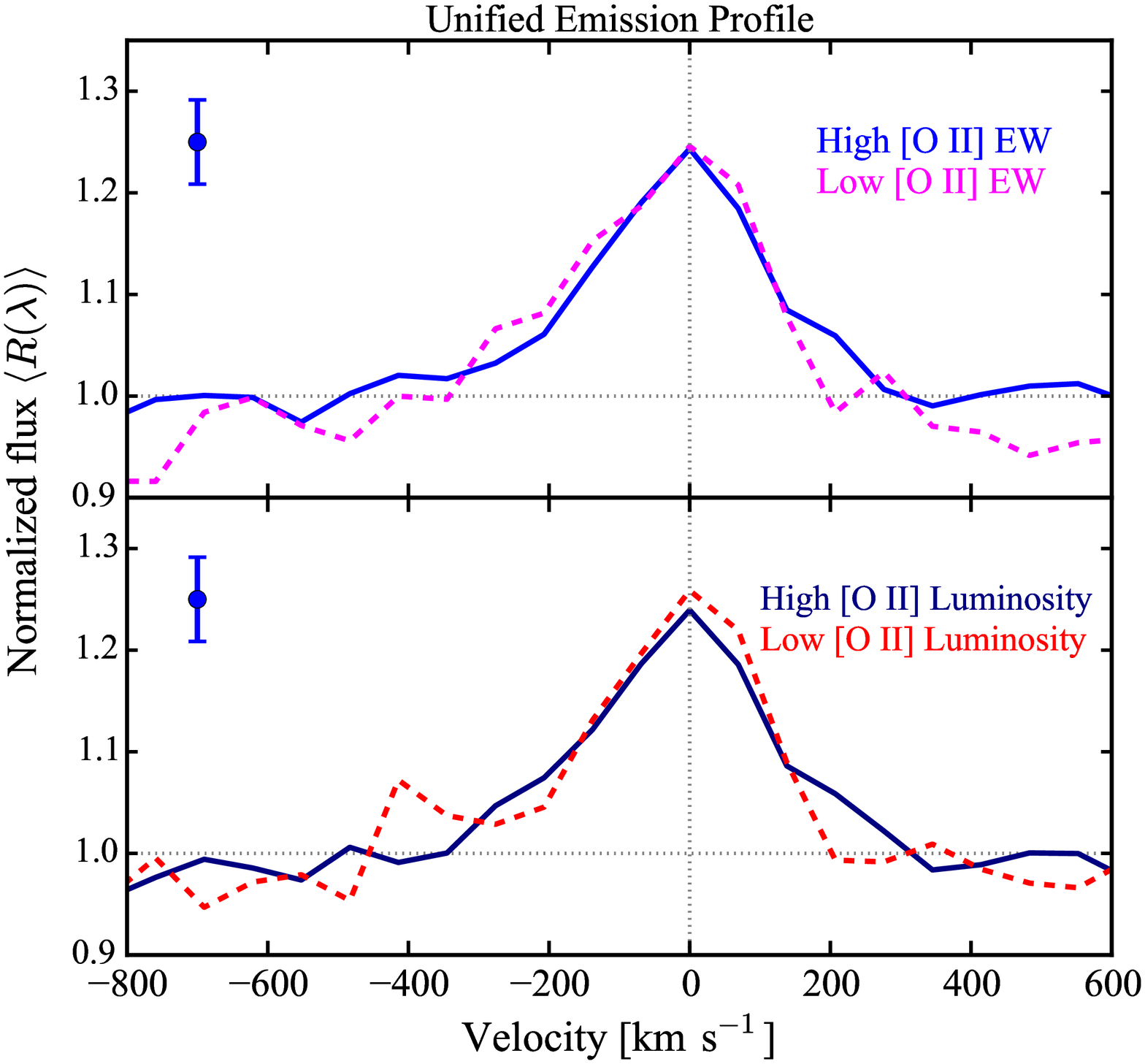}
\epsscale{0.53}
\plotone{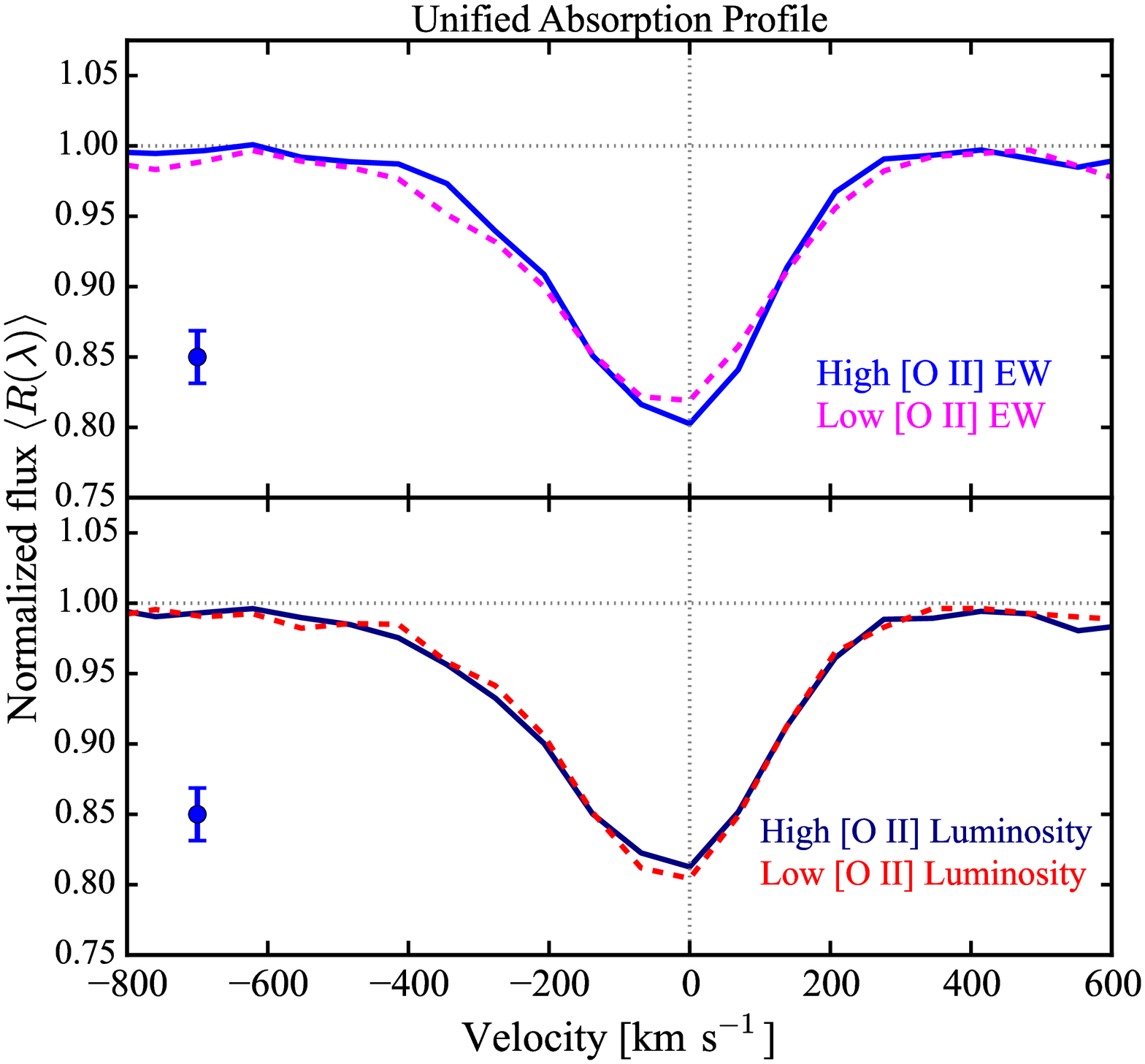}
\caption{The dependences of the unified emission profile (\textit{left}) and the unified emission-corrected absorption profile (\textit{right}) on the \oiidoublet\ rest equivalent width (\textit{upper panels}) and luminosity (\textit{lower panels}). 
}
\label{fig:profileoii}
\vspace{0.05in}
\end{figure*}

\vspace{0.05in}
\noindent $\bullet$ Emission strength -- Figure~\ref{fig:emissionrewoii} shows the dependences of the non-resonant emission strength on the \oiidoublet\ rest equivalent width and luminosity. Note for comparison, we have added the data points measured from the full sample in solid symbols, which are correlated with the measurements based on the two subsamples. Within the range probed, we find that the emission strength (in rest equivalent width) scales almost linearly with $W^{\lambda3728}_{\rm [O\,II]}$.
In the right panel, we show that the emission equivalent width also positively depends on the \oii\ luminosity, but to a lesser degree. 

\vspace{0.05in}
\noindent $\bullet$ Observed absorption profiles -- 
In Figure~\ref{fig:absorptionoiimgii}, we present the observed \mgii\ profiles, which show the strongest dependence on \oii\ among the absorption lines (Figure~\ref{fig:absorptionprooii}). The correlation appears to be stronger with the rest equivalent width than with the luminosity. This is due to the stronger dependence of the emission infill on the rest equivalent width, as suggested by Figure~\ref{fig:emissionrewoii}. The observed \mgii\ profile for the subsample with higher $W^{\lambda3728}_{\rm [O\,II]}$ has a P-cygni-like shape, indicating a large amount of emission infill.

\vspace{0.05in}
\noindent $\bullet$ Emission-corrected absorption strength -- Figure~\ref{fig:absorptionoii} shows the dependences of the emission-corrected absorption rest equivalent width on the \oii\ properties. Except for the saturated \mgii\ lines, other lines are positively correlated with both $W^{\lambda3728}_{\rm [O\,II]}$ and $L^{\lambda3728}_{\rm [O\,II]}$, with the dependence tentatively stronger for the former. 

\vspace{0.05in}
\noindent $\bullet$ Unified velocity profiles -- Figure~\ref{fig:profileoii} presents the unified emission and absorption profiles as a function of \oii\ rest equivalent width and luminosity. Within the uncertainties, for both profiles, we do not find a dependence on either $W^{\lambda3728}_{\rm [O\,II]}$ or $L^{\lambda3728}_{\rm [O\,II]}$, although the P-cygni-like shape of the observed \mgii\ absorption for the subsample with higher $W^{\lambda3728}_{\rm [O\,II]}$ requires the emission profile to be more extended on the red side than the absorption, unlike for the main sample (Figure~\ref{fig:oiiiprofile}). Larger samples in the future will help pin down these dependences with high S/N. 

\vspace{0.05in}
Among all the correlations, we find that the strongest one is between the rest equivalent widths of non-resonant emission and \oii, which also results in the strong dependence of the observed absorption profiles, especially of the \mgii\ lines, on the \oii\ rest equivalent width.
To the first order, the \oii\ luminosity is an indicator of SFR, while the \oii\ rest equivalent width is an indicator of specific SFR. Our results suggest that the properties of the emission and, to a lesser degree, the absorption are stronger functions of specific SFR than of SFR. However, considering the uncertainties due to our sample size, the exact correlations between the properties of the spectral features in the NUV and those of galaxies and their implications for galaxy evolution remain to be determined.

\section{Summary}\label{sec:summary}

The pilot observations of the emission-line galaxy (ELG) program in the Extended Baryon Oscillation Spectroscopic Survey (eBOSS) in SDSS-IV have obtained a sample of $8620$ ELGs at $0.6<z<1.2$, providing a good opportunity for investigations of the near-ultraviolet (NUV) part of the spectral energy distributions (SEDs) of star-forming galaxies (SFGs). We constructed median composite continuum-normalized spectra to study the emission and absorption features in the NUV. Our main results are:

\begin{itemize}
\item The median composite spectra of the ELGs feature non-resonant \feii$^*$ emission and resonant absorption due to \mgi, \mgii\ and \feii. Both the emission and absorption profiles are asymmetric, preferentially blueshifted, indicating ubiquitous outflows driven by star formation at $0.6<z<1.2$.
\item We found a variety of velocity profiles for the observed absorption lines with different degrees of blueshift. 
\item Comparing the ELG spectra with those of intervening quasar absorption-line systems in the same redshift range, we found they feature the same absorption lines but with different line ratios.
\item We compared the eBOSS ELG spectra with the NUV spectra of the local star-forming regions taken with the Faint Object Spectrograph (FOS) and Goddard High-resolution Spectrograph (GHRS) on \textit{HST}. The physical aperture size of the eBOSS ELG spectra at $0.6<z<1.2$ is about $15\,\kpc$, while the aperture size of the FOS/GHRS spectra of the local SF regions is less than $40\,\pc$. We found the FOS/GHRS spectra also display the same (though weaker) absorption lines, but do not exhibit the non-resonant \feii$^*$ emission. We also found different ratios for the resonant absorption lines.
\item We introduced a statistical, spherical outflow model, in which the observed non-resonant emission is the fluorescent (re-emitted) photons after the occurrence of absorption that are scattered into the line of sight. The model predicts that there is scattered resonant emission filling in on top of absorption, and the amount of emission infill depends on the transition probabilities of the allowed channels, resulting in the variety of the observed absorption profiles.
\item We developed an observation-driven, model-independent method to estimate the emission infill and reveal the true absorption profile. We showed that after the emission correction, the absorption line ratios in the ELG spectra are consistent with those in the spectra of strong quasar absorbers and local star-forming regions. 
\item We demonstrated that the outflow model can explain simultaneously the multiple observed properties of the emission and absorption features in the NUV, including i) the aperture dependence, ii) the net effect, iii) the emission velocity profiles and the emission-infill corrected absorption profiles, and iv) the variety of the observed absorption profiles and the degree of emission infill.
\item Finally, we investigated the dependence of NUV features on the \oiidoublet\ rest equivalent width and luminosity and found that the strongest correlation is between the non-resonant emission strength (in rest equivalent width) and the \oii\ rest equivalent width.
\end{itemize}

Our observations provided strong evidence for ubiquitous galactic-scale outflows driven by star formation. 
Our analysis also demonstrated that the NUV window is an informative region in the spectrum. The series of emission and absorption lines provides a new means to probe the gas physics. 
The model we introduced \citep[see also][]{rubin11a, prochaska11a, scarlata15a} has many important implications, such as the dependences of non-resonant emission on the aperture, outflow, and galaxy (occultation) sizes, and points the future investigations of outflow physics into new directions. For instance, it is of great interest to explore the surface brightness profile of the non-resonant emission, e.g., through narrow-band imaging or spatially-resolved spectroscopy \citep[\eg][]{rubin11a, martin13a}, to further study the scale dependence of outflows. 

Besides using the ``down-the-barrel'' spectra to probe the gas physics directly associated with the galaxies, we can also employ the cross-correlation techniques developed recently \citep[\eg][]{steidel10a, zhu14a}, using absorption information induced in background source spectra, to probe the circumgalactic medium of foreground sources. Combining the two different types of observation will produce a more complete picture of the baryon processes in galaxy formation and evolution.

The sample size of the NUV spectroscopic datasets will grow by orders-of-magnitude in the next decade. At the conclusion of the ELG program, eBOSS will obtain spectra for about $200,000$ ELGs at $z\gtrsim0.6$, a sample over $20$ times larger than the one used in this paper. DESI \citep[][]{schlegel11a, levi13a} and PFS \citep[][]{takada14a} will obtain higher-resolution spectra with larger telescopes for about $20$ million ELGs at higher redshift ($z>1$), where more lines are redshifted into the optical. Based upon the details revealed in the composite spectra of less than $10,000$ galaxies with the $2.5$-meter SDSS telescope, we expect that NUV spectroscopy will play an important role in future investigations of the properties and evolution of galaxies.

\acknowledgments

This work started when G.B.Z. was visiting Princeton University in December 2014 and he would like to thank Michael Strauss and Jim Gunn for their hospitality.
He also thanks Bruce Draine, Tim Heckman, Claus Leitherer, and Jason X. Prochaska for useful discussions. 
G.B.Z. acknowledges support provided by NASA through Hubble Fellowship grant \#HST-HF2-51351 awarded by the Space Telescope Science Institute, which is operated by the Association of Universities for Research in Astronomy, Inc., under contract NAS 5-26555.

J.C. acknowledges financial support from MINECO (Spain) under project number AYA2012-31101.
J-P.K. and T.D. acknowledge support from the LIDA ERC advanced grant.
A.R. acknowledges funding from the P2IO LabEx (ANR-10-LABX-0038) in the framework ``Investissements d'Avenir'' (ANR-11-IDEX-0003-01) managed by the French National Research Agency (ANR).

This paper represents an effort by both the SDSS-III and SDSS-IV collaborations. 
Funding for the Sloan Digital Sky Survey IV has been provided by the Alfred P. Sloan Foundation, the U.S. Department of Energy Office of Science, and the Participating Institutions. SDSS-IV acknowledges support and resources from the Center for High-Performance Computing at the University of Utah. The SDSS web site is www.sdss.org.

SDSS-IV is managed by the Astrophysical Research Consortium for the Participating Institutions of the SDSS Collaboration including the Brazilian Participation Group, the Carnegie Institution for Science, Carnegie Mellon University, the Chilean Participation Group, the French Participation Group, Harvard-Smithsonian Center for Astrophysics, Instituto de Astrof\'isica de Canarias, The Johns Hopkins University, Kavli Institute for the Physics and Mathematics of the Universe (IPMU)/University of Tokyo, Lawrence Berkeley National Laboratory, Leibniz Institut f\"ur Astrophysik Potsdam (AIP), Max-Planck-Institut f\"ur Astronomie (MPIA Heidelberg), Max-Planck-Institut f\"ur Astrophysik (MPA Garching), Max-Planck-Institut f\"ur Extraterrestrische Physik (MPE), National Astronomical Observatory of China, New Mexico State University, New York University, University of Notre Dame, Observat\'ario Nacional/MCTI, The Ohio State University, Pennsylvania State University, Shanghai Astronomical Observatory, United Kingdom Participation Group, Universidad Nacional Aut\'onoma de M\'exico, University of Arizona, University of Colorado Boulder, University of Oxford, University of Portsmouth, University of Utah, University of Virginia, University of Washington, University of Wisconsin, Vanderbilt University, and Yale University.

Some of the data \citep{leitherer11a} are based on observations made with the NASA/ESA \textit{Hubble Space Telescope}, obtained from the Data Archive at the Space Telescope Science Institute, which is operated by the Association of Universities for Research in Astronomy, Inc., under NASA contract NAS 5-26555.

\bibliographystyle{apj}

\appendix
\section{Atomic data}\label{app:atomic}
\begin{deluxetable*}{lrrlllllll}[h]
\tabletypesize{\scriptsize}
\tablecolumns{10}
\tablecaption{The Ionization Potential\tablenotemark{$a$}}
\tablehead{
 \colhead{Atom} &  \colhead{$N_{\rm p}$} &  \colhead{$N_{\rm n}$} & \colhead{I$\rightarrow$II} & \colhead{II$\rightarrow$III} & \colhead{III$\rightarrow$IV} & \colhead{IV$\rightarrow$V} & \colhead{V$\rightarrow$VI} & \colhead{VI$\rightarrow$VII} & \colhead{VII$\rightarrow$VIII}}
\startdata
H   &  1   &  0   &  13.59843 &  \nodata  & \nodata  & \nodata  & \nodata   & \nodata  & \nodata     \\
He  &  2   &  2   &  24.58739 &  54.41776 & \nodata  & \nodata  & \nodata   & \nodata  & \nodata     \\
Li  &  3   &  4   &  5.391715 &  75.64009 & 122.4544 & \nodata  & \nodata   & \nodata  & \nodata     \\
Be  &  4   &  5   &  9.322699 &  18.21115 & 153.8962 & 217.7186 & \nodata   & \nodata  & \nodata     \\
B   &  5   &  6   &  8.298019 &  25.15483 & 37.93058 & 259.3715 & 340.2260  & \nodata  & \nodata     \\
C   &  6   &  6   &  11.26030 &  24.3845  & 47.88778 & 64.49358 & 392.0905  & 489.9932 & \nodata     \\
N   &  7   &  7   &  14.53413 &  29.60125 & 47.4453  & 77.4735  & 97.89013  & 552.0673 & 667.0461    \\
O   &  8   &  8   &  13.61805 &  35.12111 & 54.93554 & 77.4135  & 113.8989  & 138.1189 & 739.3268    \\
F   &  9   &  10  &  17.42282 &  34.97081 & 62.7080  & 87.175   & 114.249   & 157.1631 & 185.1868    \\
Ne  &  10  &  10  &  21.56454 &  40.96296 & 63.4233  & 97.1900  & 126.247   & 157.934  & 207.271     \\
Na  &  11  &  12  &  5.139077 &  47.28636 & 71.6200  & 98.936   & 138.404   & 172.23   & 208.504     \\
Mg  &  12  &  12  &  7.646235 &  15.03527 & 80.1436  & 109.2654 & 141.33    & 186.76   & 225.02      \\
Al  &  13  &  14  &  5.985768 &  18.82855 & 28.44764 & 119.9924 & 153.825   & 190.49   & 241.76      \\
Si  &  14  &  14  &  8.151683 &  16.34585 & 33.493   & 45.14179 & 166.767   & 205.267  & 246.32      \\
P   &  15  &  16  &  10.48669 &  19.76949 & 30.20264 & 51.44387 & 65.02511  & 220.430  & 263.57      \\
S   &  16  &  16  &  10.36001 &  23.33788 & 34.86    & 47.222   & 72.5945   & 88.0529  & 280.954     \\
Cl  &  17  &  18  &  12.96763 &  23.81364 & 39.80    & 53.24    & 67.68     & 96.94    & 114.2013    \\
Ar  &  18  &  22  &  15.75961 &  27.62967 & 40.735   & 59.58    & 74.84     & 91.29    & 124.41      \\
K   &  19  &  20  &  4.340664 &  31.625   & 45.8031  & 60.917   & 82.66     & 99.44    & 117.56      \\
Ca  &  20  &  20  &  6.113155 &  11.87172 & 50.91315 & 67.2732  & 84.34     & 108.78   & 127.21      \\
Sc  &  21  &  24  &  6.56149  &  12.79977 & 24.75684 & 73.4894  & 91.95     & 110.68   & 137.99      \\
Ti  &  22  &  26  &  6.82812  &  13.5755  & 27.49171 & 43.26717 & 99.299    & 119.533  & 140.68      \\
V   &  23  &  28  &  6.746187 &  14.618   & 29.3111  & 46.709   & 65.28165  & 128.125  & 150.72      \\
Cr  &  24  &  28  &  6.76651  &  16.48630 & 30.959   & 49.16    & 69.46     & 90.6349  & 160.29      \\
Mn  &  25  &  30  &  7.434018 &  15.63999 & 33.668   & 51.2     & 72.41     & 95.604   & 119.203     \\
Fe  &  26  &  30  &  7.902468 &  16.1992  & 30.651   & 54.91    & 75.0      & 98.985   & 124.98      \\
Co  &  27  &  32  &  7.88101  &  17.0844  & 33.50    & 51.27    & 79.50     & 102.0    & 128.9       \\
Ni  &  28  &  31  &  7.639877 &  18.16884 & 35.187   & 54.92    & 76.06     & 108.0    & 132.0       \\
Cu  &  29  &  35  &  7.72638  &  20.29239 & 36.841   & 57.38    & 79.8      & 103.0    & 139.0       \\
Zn  &  30  &  35  &  9.394199 &  17.96439 & 39.7233  & 59.573   & 82.6      & 108.0    & 133.9 

\enddata
\tablenotetext{$a$}{In unit of ${\rm eV}$. Data are taken from the NIST-ASD database \citep{NIST_ASD}. Only most abundant isotopes are included.}
\vspace{0.1in}
\label{tbl:ionization}
\end{deluxetable*}

\begin{deluxetable*}{lcccllll}[h]
\tabletypesize{\scriptsize}
\tablecolumns{8}
\tablecaption{The list of lines between $2200\,{\rm \AA}$ and $7500\,{\rm \AA}$ associated with star-forming galaxies}
\tablehead{
 \colhead{Name} &  \colhead{Wavelength} &  \colhead{$A_{ul}$} & \colhead{$f_{lu}$} & \colhead{$E_l$} & \colhead{$E_u$} & \colhead{Multiplet} & \colhead{Ref\tablenotemark{$a$}} \\
 \colhead{ } & \colhead{\AA\ (Vac.)} & \colhead{$\mathrm{s}^{-1}$} & \colhead{ } & \colhead{$\mathrm{cm}^{-1}$} & \colhead{$\mathrm{cm}^{-1}$} & \colhead{No.} & \colhead{}
 }
\startdata
\feii$\,\lambda2250$     & 2249.88  &  3.00E+06  &  1.82E-03  &  0           &   44446.905  &   UV5   &  1,2  \\
\feii$\,\lambda2261$     & 2260.78  &  3.18E+06  &  2.44E-03  &  0           &   44232.540  &   UV4   &  1,2  \\
\feii$^*\,\lambda2270$   & 2269.52  &  4.00E+05  &  3.10E-04  &  384.7872    &   44446.905  &   UV5   &  1,2  \\
\feii$^*\,\lambda2281$   & 2280.62  &  4.49E+06  &  4.38E-03  &  384.7872    &   44232.540  &   UV4   &  1,2  \\
\ciii$\,\lambda2298$     & 2297.58  &  1.38E+08  &  1.82E-01  &  102352.04   &   145876.13  &   UV8   &  3  \\
\ciisemi$\,\lambda2324$  & 2324.21  &  \nodata   &  \nodata   &  0           &   43025.3    &   UV0.01   &  3,4  \\
\ciisemi$\,\lambda2325$  & 2325.40  &  \nodata   &  \nodata   &  0           &   43003.3    &   UV0.01   &  3,4  \\
\ciisemi$\,\lambda2326$  & 2326.11  &  \nodata   &  \nodata   &  63.42       &   43053.6    &   UV0.01   &  3,4  \\
\ciisemi$\,\lambda2328$  & 2327.64  &  \nodata   &  \nodata   &  63.42       &   43025.3    &   UV0.01   &  3,4  \\
\ciisemi$\,\lambda2329$  & 2328.83  &  \nodata   &  \nodata   &  63.42       &   43003.3    &   UV0.01   &  3,4  \\
\feii$\,\lambda2344$     & 2344.21  &  1.73E+08  &  1.14E-01  &  0           &   42658.244  &   UV3   &  1,2  \\
\feii$^*\,\lambda2366$   & 2365.55  &  5.90E+07  &  4.95E-02  &  384.7872    &   42658.244  &   UV3   &  1,2  \\
\feii$\,\lambda2374$     & 2374.46  &  4.25E+07  &  3.59E-02  &  0           &   42114.838  &   UV2   &  1,2  \\
\feii$^*\,\lambda2381$   & 2381.49  &  3.10E+07  &  3.51E-02  &  667.6829    &   42658.244  &   UV3   &  1,2  \\
\feii$\,\lambda2383$     & 2382.76  &  3.13E+08  &  3.20E-01  &  0           &   41968.070  &   UV2   &  1,2  \\
\feii$^*\,\lambda2396$   & 2396.35  &  2.59E+08  &  2.79E-01  &  384.7872    &   42114.838  &   UV2   &  1,2  \\
\neiv$\,\lambda2423$     & 2422.56  &  \nodata   &   \nodata  &  0           &   41278.89   &   UV1   &  5  \\
\neiv$\,\lambda2425$     & 2425.14  &  \nodata   &   \nodata  &  0           &   41234.43   &   UV1   &  5  \\
\oii$\,\lambda2471.0$    & 2470.97  &  \nodata   &   \nodata  &  0           &   40470.00   &   1.01F   &  3,6  \\
\oii$\,\lambda2471.1$    & 2471.09  &  \nodata   &   \nodata  &  0           &   40468.01   &   1.01F   &  3,6  \\
\mnii$\,\lambda2577$     & 2576.88  &  2.80E+08  &  3.58E-01  &  0           &   38806.691  &   UV1   &  7,8,9  \\
\feii$\,\lambda2587$     & 2586.65  &  8.94E+07  &  7.17E-02  &  0           &   38660.054  &   UV1   &  1,2  \\
\mnii$\,\lambda2594$     & 2594.50  &  2.76E+08  &  2.79E-01  &  0           &   38543.122  &   UV1   &  7,8,9  \\
\feii$\,\lambda2600$     & 2600.17  &  2.35E+08  &  2.39E-01  &  0           &   38458.993  &   UV1   &  1,2  \\
\mnii$\,\lambda2606$     & 2606.46  &  2.69E+08  &  1.96E-01  &  0           &   38366.232  &   UV1   &  7,8,9  \\
\feii$^*\,\lambda2613$   & 2612.65  &  1.20E+08  &  1.22E-01  &  384.7872    &   38660.054  &   UV1   &  1,2  \\
\feii$^*\,\lambda2626$   & 2626.45  &  3.52E+07  &  4.55E-02  &  384.7872    &   38458.993  &   UV1   &  1,2  \\
\feii$^*\,\lambda2632$   & 2632.11  &  6.29E+07  &  8.70E-02  &  667.6829    &   38660.054  &   UV1   &  1,2  \\
\mgii$\,\lambda2796$     & 2796.35  &  2.60E+08  &  6.08E-01  &  0           &   35760.88   &   UV1   &  10,11  \\
\mgii$\,\lambda2804$     & 2803.53  &  2.57E+08  &  3.03E-01  &  0           &   35669.31   &   UV1   &  10,11  \\
\mgi$\,\lambda2853$      & 2852.96  &  4.91E+08  &  1.80E-00  &  0           &   35051.264  &   UV1   &  10,11  \\
\tiii$\,\lambda3067$     & 3067.24  &  3.47E+07  &  4.89E-02  &  0           &   32602.626  &   5   &  10  \\
\tiii$\,\lambda3074$     & 3073.86  &  1.71E+08  &  1.21E-01  &  0           &   32532.354  &   5   &  10  \\
\hei$\,\lambda3189$      & 3188.67  &  5.64E+06  &  2.58E-02  &  159855.97   &   191217.05  &   3   &  12  \\
\tiii$\,\lambda3230$     & 3230.12  &  2.93E+07  &  6.87E-02  &  0           &   30958.585  &   2   &  10  \\
\tiii$\,\lambda3243$     & 3242.92  &  1.47E+08  &  2.32E-01  &  0           &   30836.425  &   2   &  10  \\
\tiii$\,\lambda3385$     & 3384.73  &  1.39E+08  &  3.58E-01  &  0           &   29544.454  &   1   &  10  \\
\oii$\,\lambda3727$      & 3727.10  &  \nodata   &   \nodata  &  0           &   26830.57   &   1F   & 3,6  \\
\oii$\,\lambda3730$      & 3729.86  &  \nodata   &   \nodata  &  0           &   26810.55   &   1F   & 3,6  \\
\neiii$\,\lambda3870$    & 3869.86  &  \nodata   &   \nodata  &  0           &   25840.72   &   1F   & 13  \\
\hei$\,\lambda3890$      & 3889.74  &  9.47E+06  &  6.45E-02  &  159855.97   &   185564.60  &   2    & 12  \\
\neiii$\,\lambda3969$    & 3968.59  &  \nodata   &   \nodata  &  642.876     &   25840.72   &   1F   & 13  \\
\caii$\,\lambda3935$     & 3934.77  &  1.40E+08  &  6.49E-01  &  0           &   25414.40   &   1   &  14  \\
\caii$\,\lambda3970$     & 3969.59  &  1.36E+08  &  3.21E-01  &  0           &   25191.51   &   1   &  14  \\
\suii$\,\lambda4070$     & 4069.75  &  \nodata   &   \nodata  &  0           &   24571.54   &   1F   &  15  \\
\oiii$\,\lambda4364$     & 4364.44  &  \nodata   &   \nodata  &  20273.27    &   43185.74   &   2F   & 3,6  \\
\hei$\,\lambda4473$      & 4472.76  &  2.46E+07  &  1.23E-01  &  169086.91   &   191444.48  &   14   & 12  \\
\oiii$\,\lambda4960$     & 4960.30  &  \nodata   &   \nodata  &  113.178     &   20273.27   &   1F  &  3,6  \\
\oiii$\,\lambda5008$     & 5008.24  &  \nodata   &   \nodata  &  306.174     &   20273.27   &   1F  &  3,6  \\
\ni$\,\lambda5199$       & 5199.35  &  \nodata   &   \nodata  &  0           &   19233.177  &   1F   & 3,4  \\
\ni$\,\lambda5202$       & 5201.70  &  \nodata   &   \nodata  &  0           &   19224.464  &   1F   & 3,4  \\
\mgi$\,\lambda5169$      & 5168.76  &  1.13E+07  &  1.35E-01  &  21850.405   &   41197.403  &   2   &  10,11  \\
\mgi$\,\lambda5174$      & 5174.12  &  3.37E+07  &  1.35E-01  &  21870.464   &   41197.403  &   2   &  10,11  \\
\mgi$\,\lambda5185$      & 5185.05  &  5.61E+07  &  1.36E-01  &  21911.178   &   41197.403  &   2   &  10,11  \\
\hei$\,\lambda5877$      & 5877.29  &  7.07E+07  &  6.10E-01  &  169086.91   &   186101.55  &   11   & 12  \\
\nai$\,\lambda5892$      & 5891.58  &  6.16E+07  &  6.41E-01  &  0           &   16973.366  &   1   &  11  \\
\nai$\,\lambda5898$      & 5897.56  &  6.14E+07  &  3.20E-01  &  0           &   16956.170  &   1   &  11  \\
\oi$\,\lambda6302$       & 6302.05  &  \nodata   &   \nodata  &  0           &   15867.862  &   1F   & 3,6  \\
\oi$\,\lambda6366$       & 6365.54  &  \nodata   &   \nodata  &  158.265     &   15867.862  &   1F   & 3,6  \\
\oi$\,\lambda6394$       & 6393.50  &  \nodata   &   \nodata  &  226.977     &   15867.862  &   1F   & 3,6  \\
\nii$\,\lambda6550$      & 6549.86  &  \nodata   &   \nodata  &  48.7        &   15316.2    &   1F   & 3,4  \\
\nii$\,\lambda6585$      & 6585.27  &  \nodata   &   \nodata  &  130.8       &   15316.2    &   1F   & 3,4  \\
\hei$\,\lambda6680$      & 6680.00  &  6.37E+07  &  7.10E-01  &  171134.90   &   186104.96  &   46   & 12  \\
\suii$\,\lambda6718$     & 6718.29  &  \nodata   &   \nodata  &  0           &   14884.73   &   2F   &  15  \\
\suii$\,\lambda6733$     & 6732.67  &  \nodata   &   \nodata  &  0           &   14852.94   &   2F   &  15  \\
\hei$\,\lambda7067$      & 7067.20  &  2.78E+07  &  6.95E-02  &  169086.91   &   183236.79  &   10   & 12  \\
\ariii$\,\lambda7138$    & 7137.77  &  \nodata   &   \nodata  &  0           &   14010.00   &   1F  &  16 \\
\oii$\,\lambda7322$      & 7322.01  &  \nodata   &   \nodata  &  26810.55    &   40468.01   &   2F   & 3,4  \\
\oii$\,\lambda7332$      & 7331.69  &  \nodata   &   \nodata  &  26830.57    &   40470.00   &   2F   & 3,4  \\
\oii$\,\lambda7333$      & 7332.75  &  \nodata   &   \nodata  &  26830.57    &   40468.01   &   2F   & 3,4  \\

\enddata
\tablenotetext{$a$}{References: 1. \citet{fuhr06a}, 2. \citet{nave13a}, 3. \citet{gallagher93a}, 4. \citet{wiese07a}, 5. \citet{kramida99a}, 6. \citet{wiese96a}, 7. \citet{kramida13a}, 8. \citet{aldenius09a}, 9. \citet{denhartog11a}, 10. \citet{ruffoni10a}, 11. \citet{kelleher08a}, 12. \citet{wiese09a}, 13. \citet{kramida06a}, 14. \citet{safronova11a}, 15. \citet{podobedova09a}, 16. \citet{saloman10a}.}
\label{tbl:linelist}
\end{deluxetable*}

The spectroscopic analysis presented in this paper requires accurate atomic data.
In this appendix, we present the data we gathered for completeness and future references.

We collect most of the data from the National Institute of Standards and Technology Atomic Spectra Database \citep[NIST-ASD,][]{NIST_ASD} and have cross-checked the information with either the original or new references we found in the literature. 

Most of the conventions used in spectroscopy are well established and we follow them whenever possible.  
However, we also found some ambiguous terms that may cause confusion, mostly due to historical reasons. 
We try to be as specific as possible on those occasions. 
We use \citet[][]{moore50a, moore72a}, \citet[][]{cohen87a}, and \citet[][]{drainebook} as the main references. 
We briefly summarize the nomenclature we adopt below.  We only include those definitions that are most relevant to our discussion or may cause confusion. 
For a complete description, we refer the reader to the references above.
We note our intention is not to give a comprehensive review or to resolve all the disagreements between the nomenclature in observational astronomy and that in quantum theory, but rather to make our presentation clear while maintaining consistency with the literature in astronomy. 

\vspace{0.05in}
\noindent $\bullet$ \textbf{Nomenclature}
\vspace{-0.05in}

\begin{itemize}
\item[$\circ$] Roman numerals: To specify the spectrum of a $z$-fold ionized atom, we use a small capital roman numeral corresponding to $z+1$ written with a thin space following the chemical symbol. For example, \hyi\ denotes the spectrum of neutral hydrogen (${\rm H}^0$), \mgii\ that of singly-ionized magnesium (${\rm Mg}^+$), \ciii\ that of doubly-ionized carbon (${\rm C}^{2+}$). Recently, this convention has also been used by many authors to specify the ionized atom itself, not only the spectrum.
\item[$\circ$] Electron configuration: An atomic electron configuration is indicated symbolically by
\beq
(nl)^k(n'l')^{k'} \,\mathrm{,}
\eeq
in which $k,~k',~...$ are the numbers of electrons with principal quantum numbers $n,~n',~...$ and orbital angular momentum quantum numbers $l,~l',~...$, respectively. The orbital angular momentum numbers $l = 0,~1,~2,~3$, etc. are indicated by s, p, d, f, g, h, i and k, and the parentheses are often omitted. For example, the electron configuration of the ground state of ${\rm Fe}^+$ is given by $1\mathrm{s}^22\mathrm{s}^22\mathrm{p}^63\mathrm{s}^23\mathrm{p}^63\mathrm{d}^6\mathrm{4}s$. In practice, we only show the last two or three subshells, e.g., $3\mathrm{d}^6\mathrm{4}s$ for ${\rm Fe}^+$.
\item[$\circ$] Spin-orbit ($L$-$S$) coupling: In $L$-$S$ coupling, a given combination of ($L$, $S$) defines an atomic \textit{term}. A given set of ($L$, $S$, $J$) defines an atomic \textit{level}, and is indicated by
\beq
 ^{2S+1}L_J^p \,\mathrm{,}
\eeq
where $p$ represents parity and is blank for even parity and ``o'' for odd.
\citet[][]{cohen87a} defines an atomic \textit{state} as a level with a given set of $L$, $S$, $J$ and $M_J$. We find this definition is not used frequently in astronomy. In this paper, we use $term$ and $state$ interchangeably. We suggest to use \textit{microstate} or \textit{quantum state} for a given set of $L$, $S$, $J$ and $M_J$ (or $L$, $S$, $M_S$ and $M_L$). With our choices, the ground $term$ ($state$) of ${\rm Fe}^+$ has $L=2~(\mathrm{D})$ and $S=5/2$ with configuration $3\mathrm{d}^6\mathrm{4}s$ and even parity, and is split into five $levels$ with $J=1/2,~3/2,~5/2,~7/2,~9/2$, among which the lowest (ground) level is $ ^6\mathrm{D}_{9/2}$. The first excited term (state) of ${\rm Fe}^+$ (that is relevant in the NUV spectroscopy) has $L=2~(\mathrm{D})$ and $S=5/2$ with configuration $3\mathrm{d}^64\mathrm{p}$ and odd parity, and is split into five levels with $J=1/2,~3/2,~5/2,~7/2,~9/2$, among which the lowest level is $ ^6\mathrm{D}_{9/2}^\mathrm{o}$. Note each level has a degeneracy of $g=2J+1$ and the total multiplicity of a given term is $(2S+1)(2J+1)$.

For a multiple-(sub)shell system, there are multiple terms with the same $L$ and $S$ for the outmost electron(s) due to different terms of electrons in the inner (sub)shells. \citet[][]{moore45a} labeled different terms with a front lowercase letter ``a'' - ``z'' to distinguish them and this labeling method has been used since. We also use this convention, with the labeling letters extracted from the NIST-ASD database.
\item[$\circ$] Resonance: We refer to a transition between two different terms as a resonant transition if and only if the lower level of the transition is the lowest level of the ground term. For a permitted non-resonant transition, we label it with a right superscript asterisk ``$*$''. 
This definition is different from that by \citet[][]{morton03a}, who considered transitions whose lower term is the ground term as resonant transitions, regardless of which level. 
Note we only apply this to permitted transitions, so forbidden (semi-forbidden) lines like \oiii$\,\lambda5008$ (\ciisemi$\,\lambda2326$), although the lower energy level is not the lowest level, are labeled with the brackets ``[\,]'' (``]'') only\footnote{According to the selection rules, the transitions in \feii\ UV4 and UV5 shown here, $\lambda2250$, $\lambda2261$, $(^*)\lambda2270$ and $(^*)\lambda2281$ are semi-forbidden. Because the selection rules apply to emission lines and historically \feii\ lines were more often observed in absorption, these lines are usually not marked with ``]''. Nonetheless, their transition probabilities are several orders-of-magnitude higher than other semi-forbidden lines. We treat them as permitted transitions here.}.
We refer to a transition between two different levels, i.e., with different $J$-values, within the same term as a fine-structure transition. Fine-structure transitions are forbidden since the configuration does not change and thus $\Delta l=0$.
\item[$\circ$] Multiplet groups: \citet[][]{moore45a} labeled the multiplets first in the order of increasing excitation energy of the lower term then the higher term. At wavelength $\lambda\gtrsim3000\,{\rm \AA}$, the multiplets were included in the Revised Multiplet Table \citep[][]{moore72a}, and at shorter wavelength in the Ultraviolet Multiplet Table \citep[][]{moore50a, moore52a}. Though the data have improved over the years, this convention has been used by many authors. Newly-identified multiplet groups were usually added between the original groups by a decimal system \citep[\eg][]{gallagher93a} and forbidden lines are denoted with ``F'' following the Multiplet Number. 
We notice that, in practice, the Multiplet Numbers are rarely used for transitions in the optical.
Nonetheless, for consistency with the literature, we follow the same convention, and present the Multiplet Numbers taken from \citet{moore50a, moore52a, moore72a} or new references we found in the literature. 
\end{itemize}

\vspace{0.00in}
\noindent $\bullet$ \textbf{Ionization potentials}
\vspace{0.05in}

Table~\ref{tbl:ionization} presents the ionization potential for elements with atomic number up to $30$, in unit of ${\rm eV}$. We have included the most abundant isotopes ionized up to seven times. The columns $N_p$ and $N_n$ are the numbers of protons and neutrons. When available, we show seven significant figures.

\vspace{0.05in}
\noindent $\bullet$ \textbf{Line list}
\vspace{0.05in}

In Table~\ref{tbl:linelist}, we list the lines associated with star-forming galaxies between $2200\,{\rm \AA}$ and $7500\,{\rm \AA}$.
We detect most of them in the composite spectra of the emission-line galaxies from the eBOSS pilot observations. We do not list the categories of origin and refer the reader to Section~\ref{sec:nuvcomposite} for a summary of typical origins for the lines in the NUV. We present vacuum wavelength and some of the lines are therefore labeled as $1\,$\AA\ longer than if labeled with the air wavelength, such as \oiii$\,\lambda5008$. For useful formulae for the conversion between vacuum and air wavelengths, we refer the reader to \citet{ciddor96a}.
We have omitted the hydrogen lines, a recent compilation of which can be found in \citet{wiese09a}, and the G4300-band, which is a blend of many lines predominantly by ${\rm CH}$ and \fei\ \citep[\eg][]{worthey94a}.

For permitted transitions, we list the Einstein $A$ coefficient ($A_{ul}$) and oscillator strength ($f_{lu}$), which are essential quantities for the radiative transfer analysis presented in this paper. For all the transitions, we list the energies of the lower ($E_l$) and higher levels ($E_u$), relative to the ground level (of the ground term). The energies are in unit of ${\rm cm}^{-1}$, and can be converted to ${\rm eV}$ when multiplied by $1.2398$E-4. The transition wavelength (in \AA) can be derived with $1\mathrm{E}8/(E_u-E_l)$.

\vspace{0.05in}
\noindent $\bullet$ \textbf{Energy-level diagrams}
\vspace{0.05in}

\begin{figure}
\vspace{0.3in}
\epsscale{0.575}
\plotone{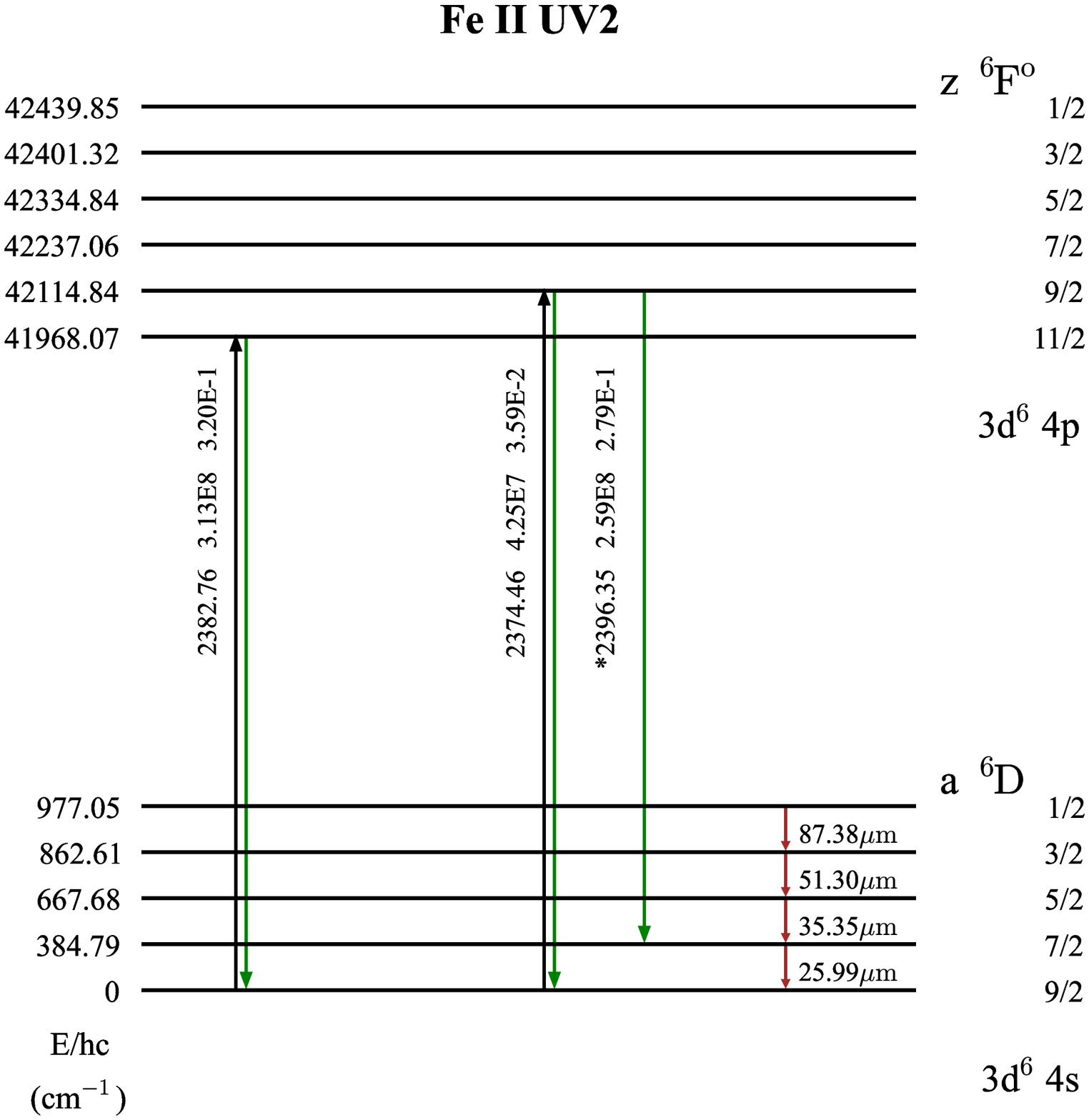}
\epsscale{0.575}
\plotone{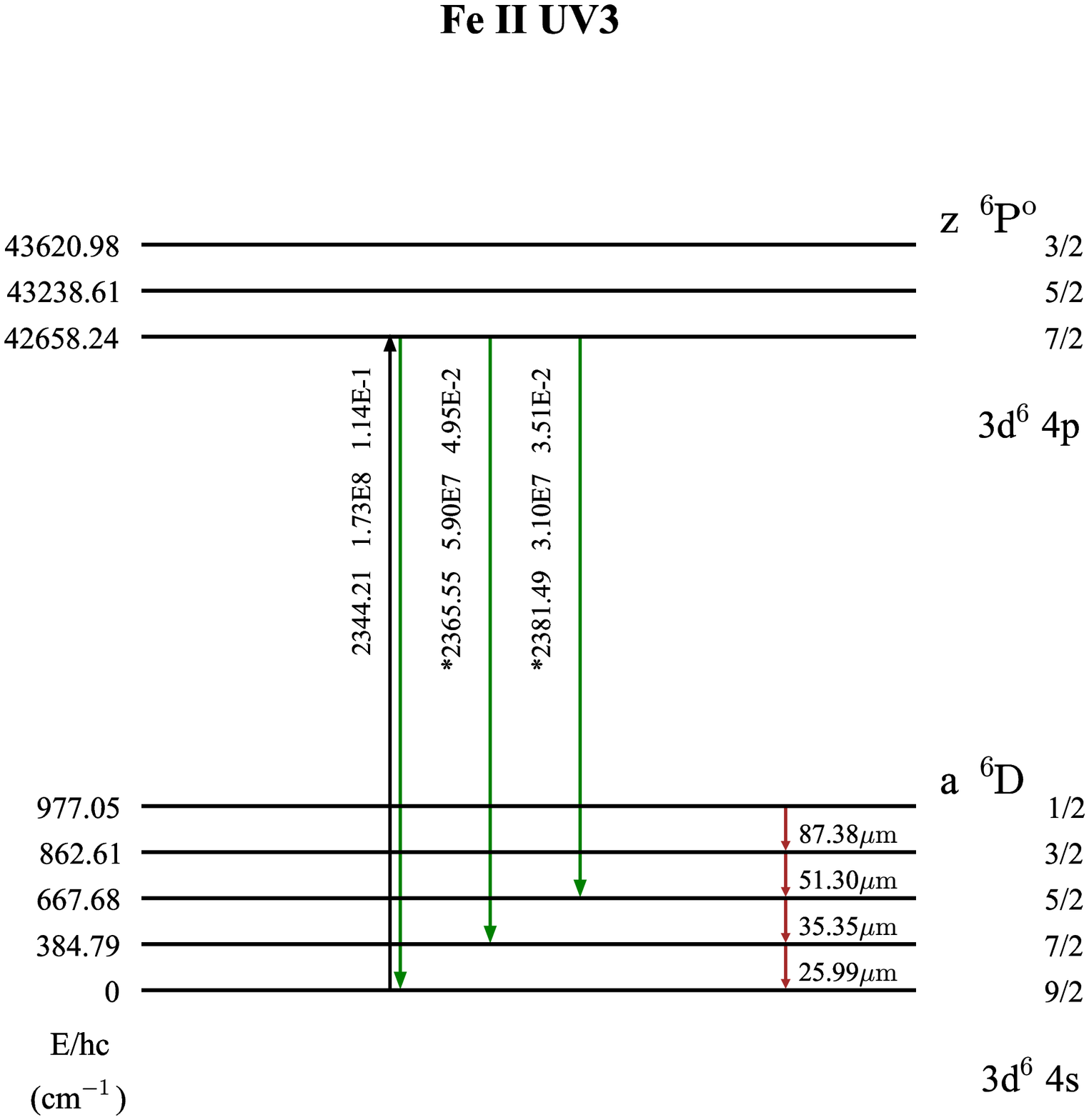}
\caption{The energy-level diagrams for the \feii\ UV2 and UV3 groups. The numbers next to the arrows represent the vacuum wavelength, the Einstein $A$ coefficient, and the oscillator strength.
}
\label{fig:feiiuv23}
\end{figure}

\begin{figure}
\vspace{0.3in}
\epsscale{0.575}
\plotone{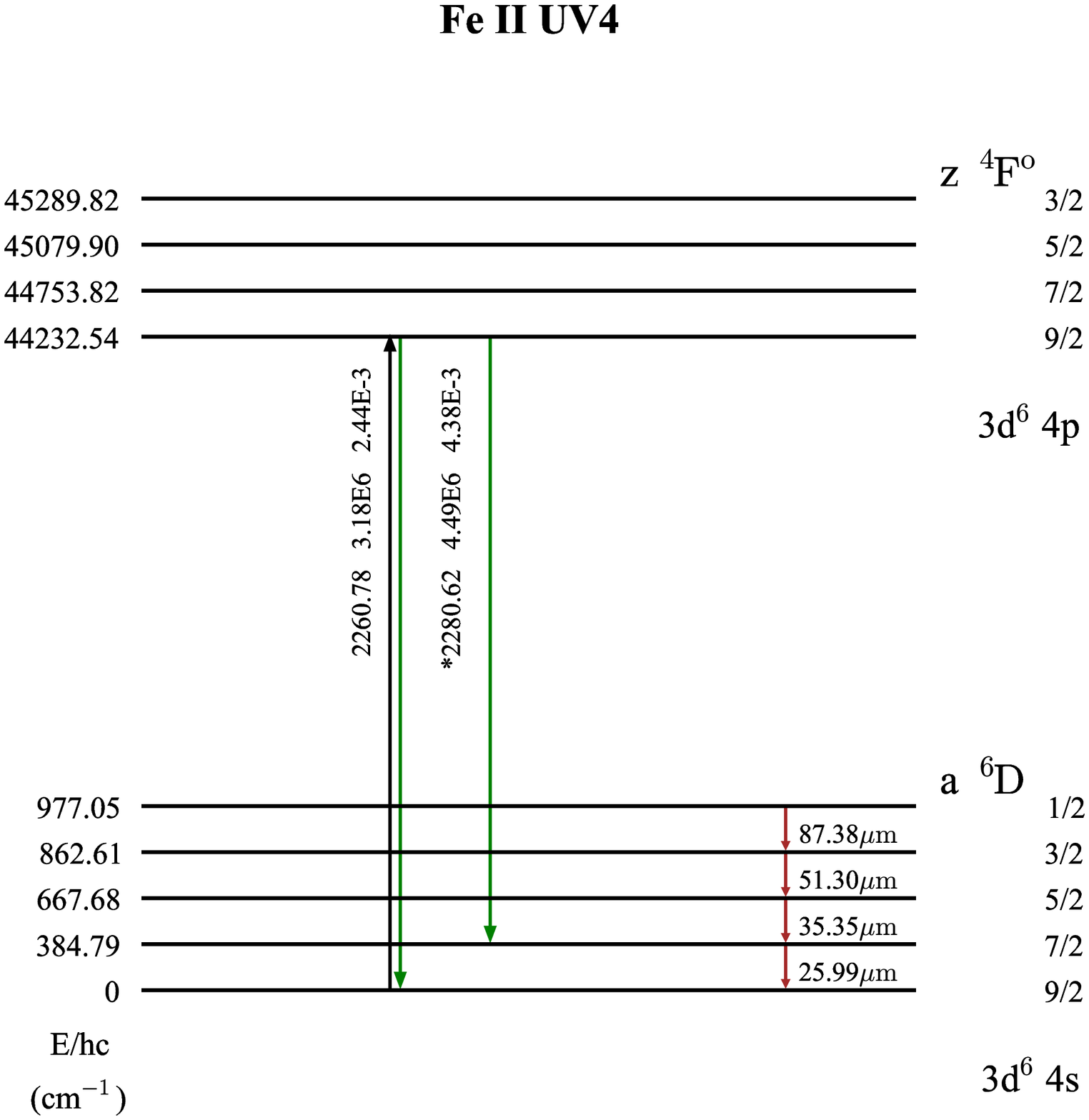}
\epsscale{0.575}
\plotone{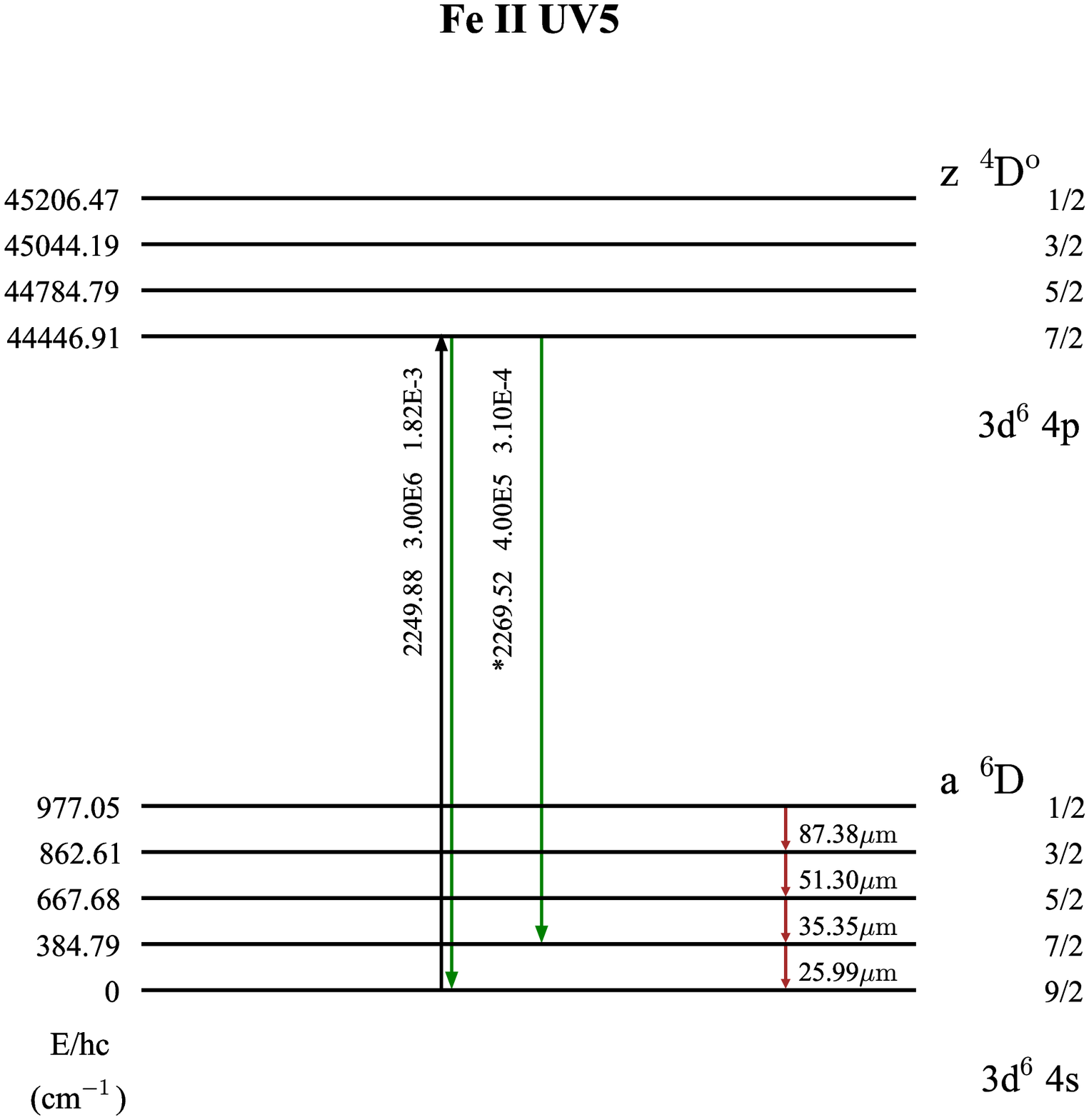}
\caption{The energy-level diagrams for the \feii\ UV4 and UV5 groups.
}
\label{fig:feiiuv45}
\end{figure}

\begin{figure}
\vspace{0.2in}
\epsscale{0.5}
\plotone{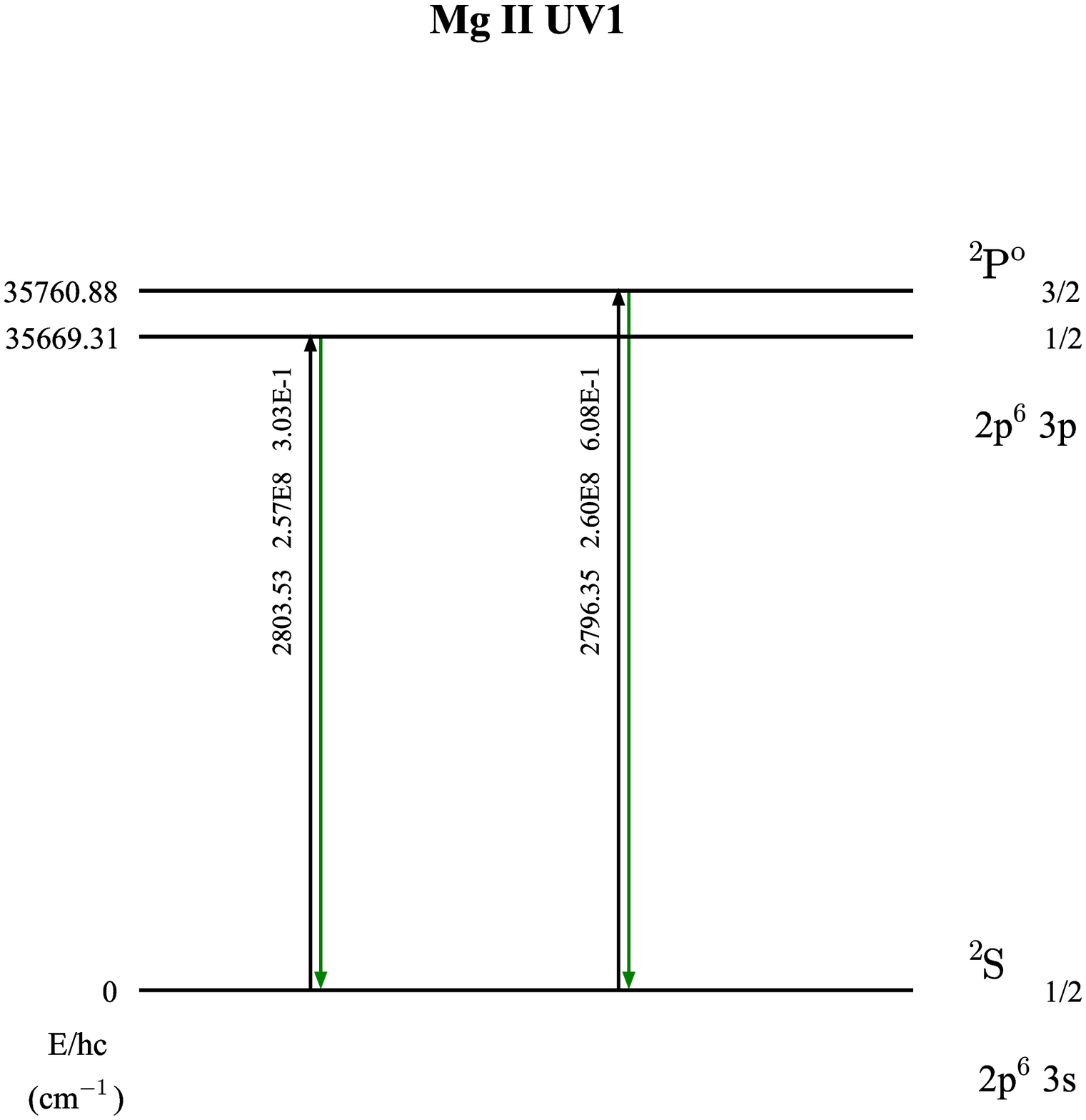}
\plotone{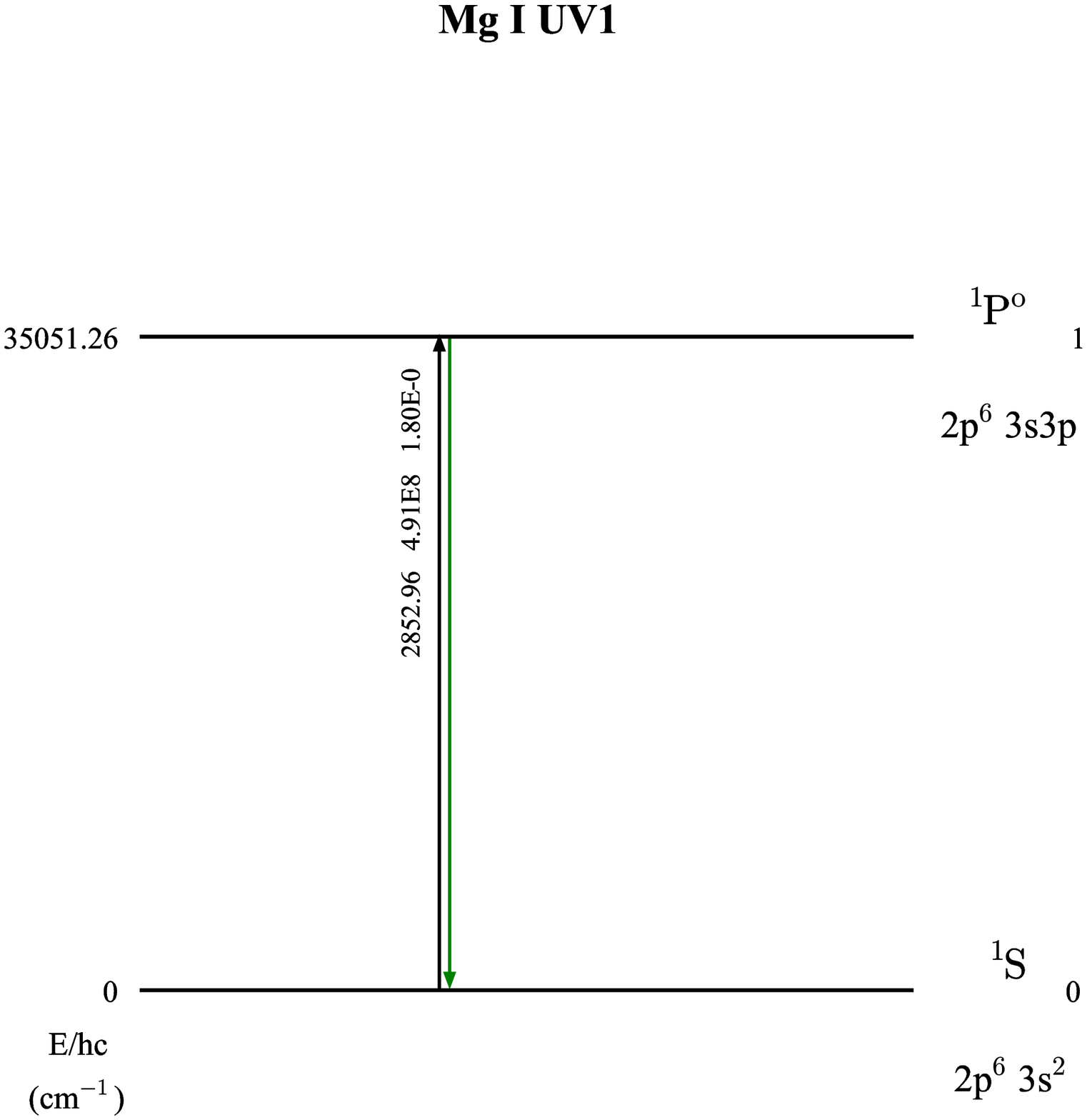}
\caption{The energy-level diagrams for the \mgii\ UV1 and \mgi\ UV1 groups.
}
\label{fig:mgiiuv1}
\end{figure}

\begin{figure}
\vspace{0.2in}
\epsscale{0.5}
\plotone{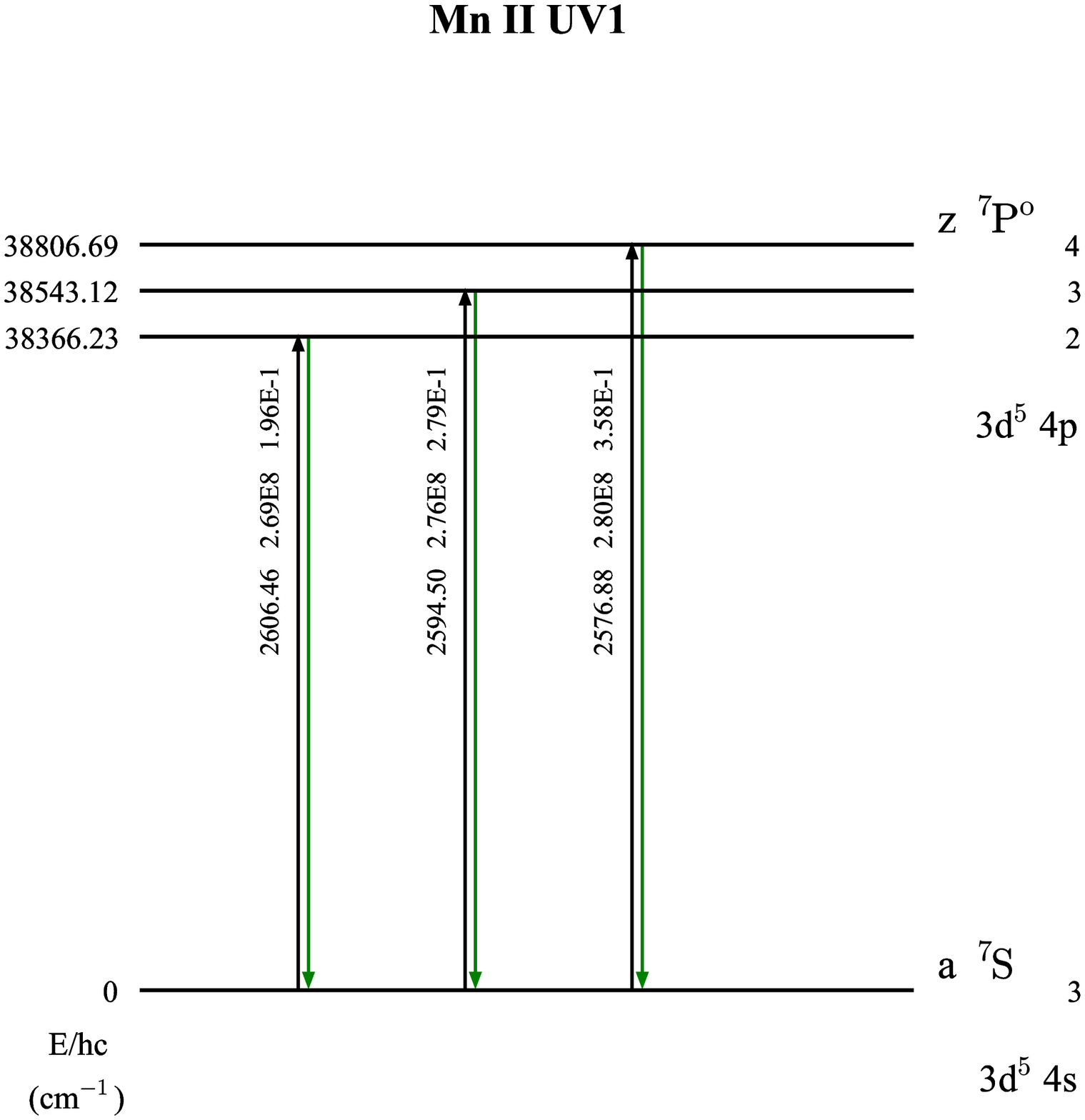}
\plotone{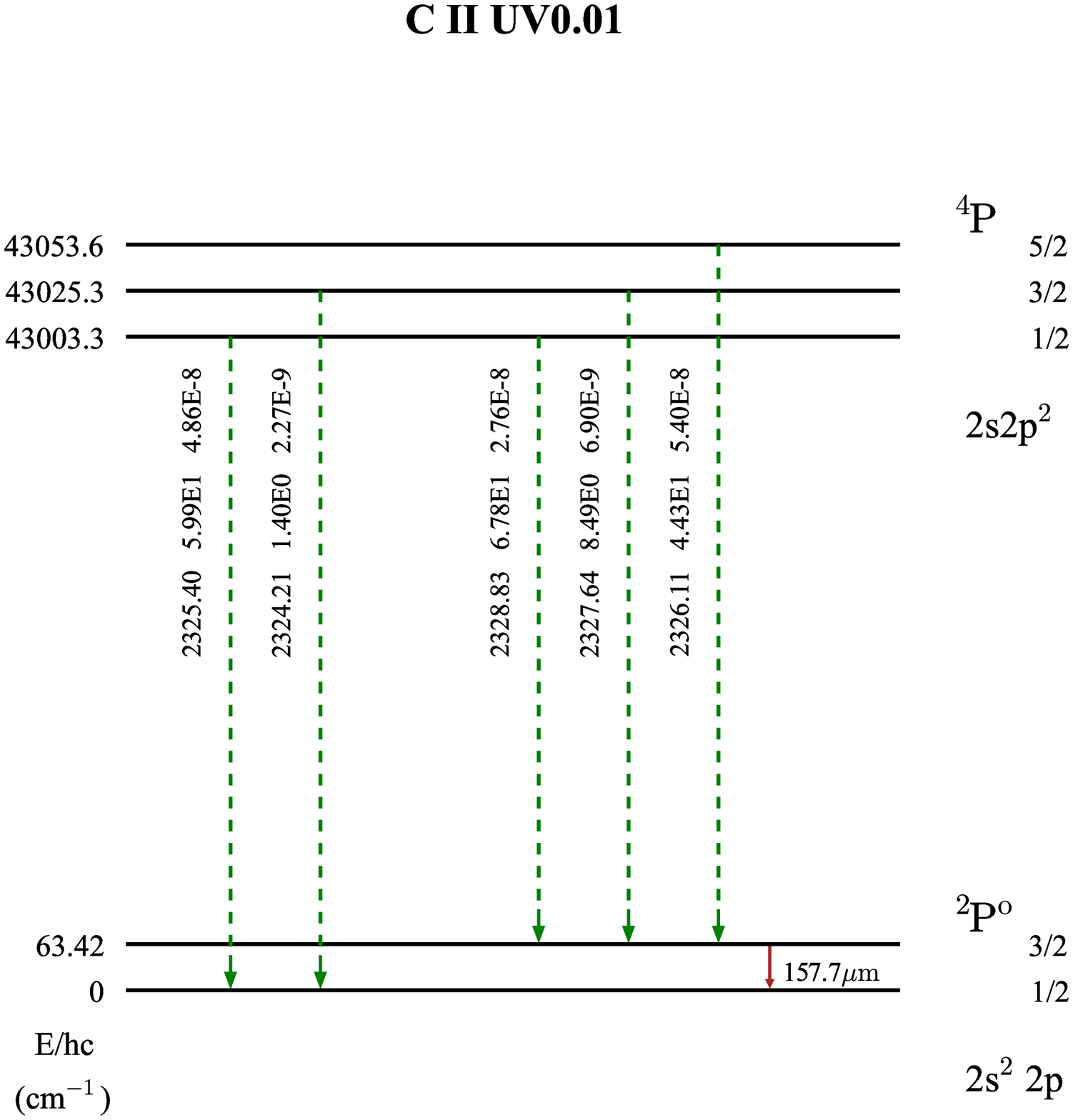}
\caption{The energy-level diagrams for the \mnii\ UV1 and \cii\ UV0.01 groups. Note although we show the Einstein $A$ coefficient and the Oscillator strength of the \cii\ transitions, they are very low for these semi-forbidden lines.
}
\label{fig:mniiuv1}
\end{figure}

\begin{figure}
\epsscale{0.35}
\plotone{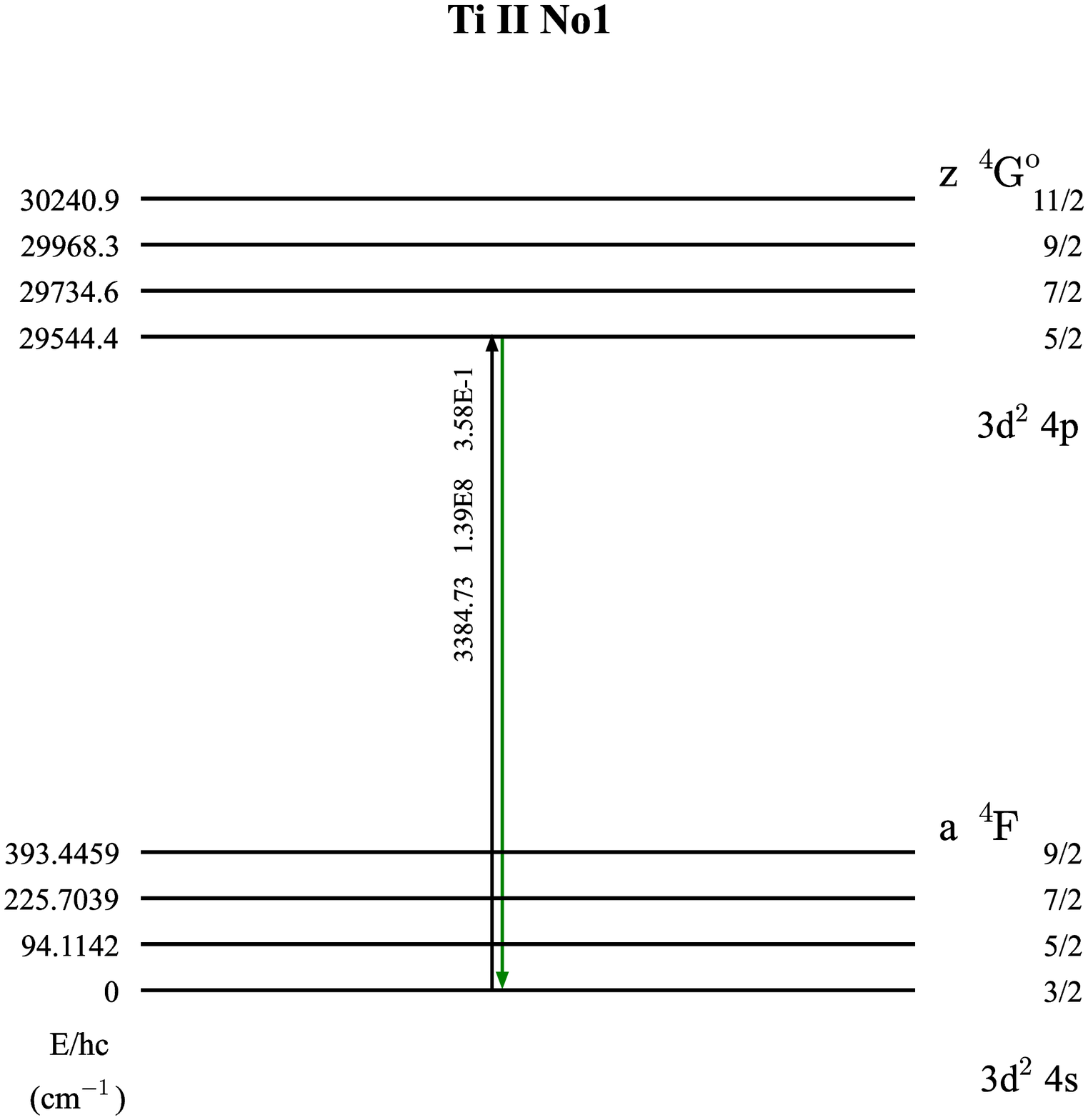}
\plotone{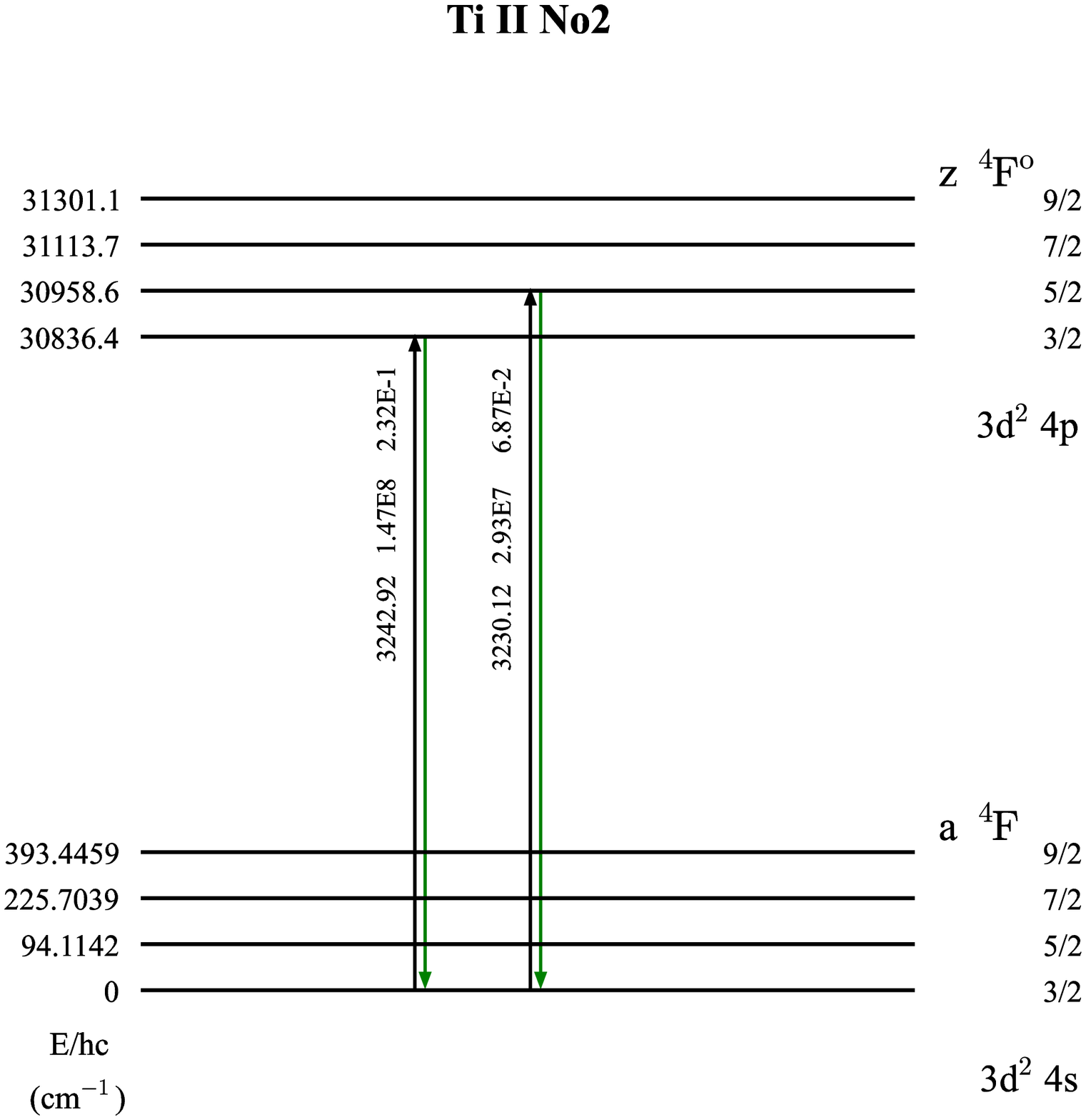}
\plotone{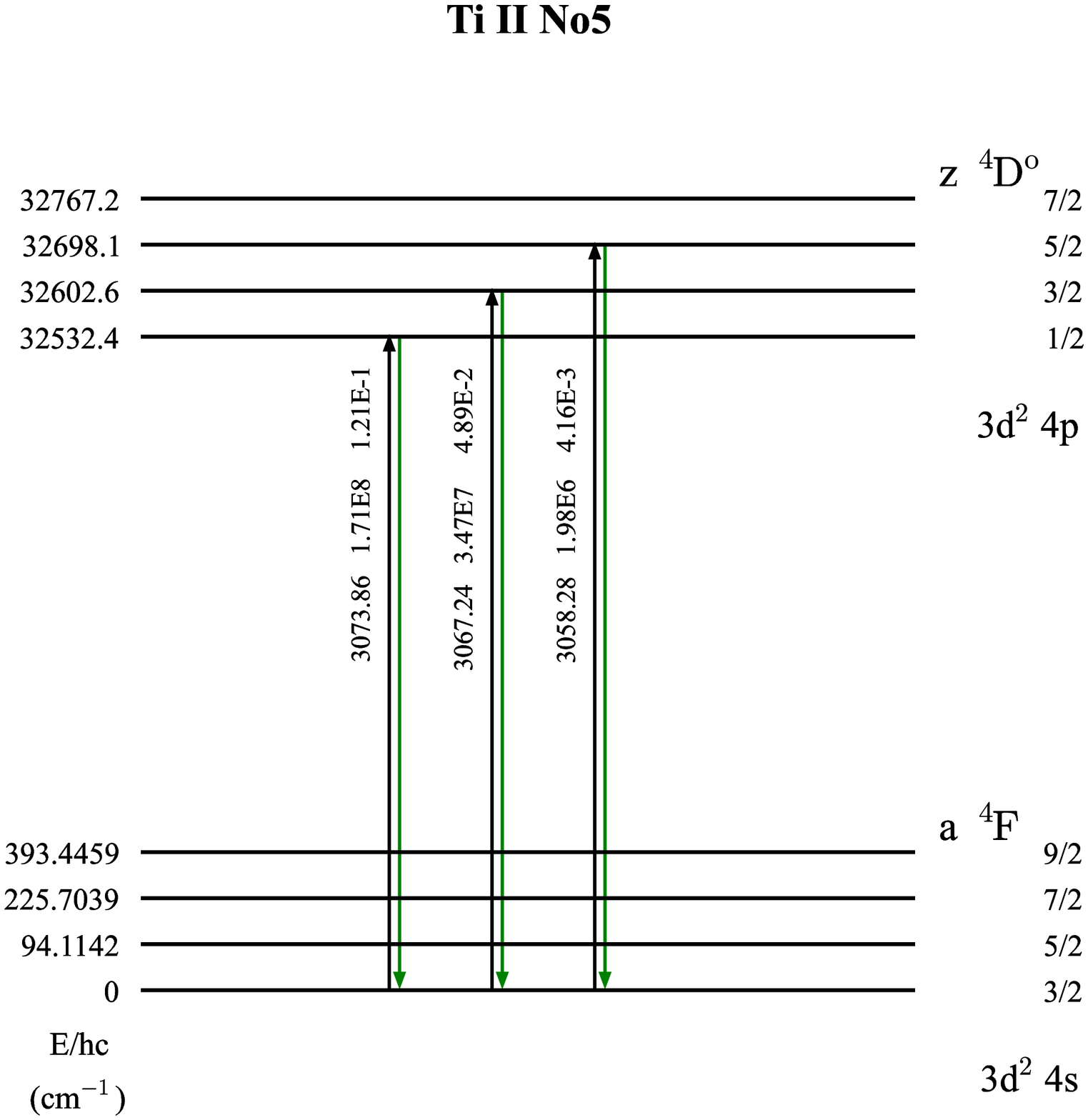}
\caption{The energy-level diagrams for the \tiii\ No. 1, 2 \& 5 groups.
}
\label{fig:tiii125}
\end{figure}

The relationships between the different resonant and non-resonant channels are essential to our analysis of radiative transfer processes. We illustrate these relationships with the energy-level diagrams of the corresponding multiplet groups. The diagram for the \feii\ UV1 group is presented in Section~\ref{sec:nuvcomposite} (Figure~\ref{fig:feiiuv1}). Here we present those for \feii\ UV2-UV5 (Figures~\ref{fig:feiiuv23} \& \ref{fig:feiiuv45}), \mgii\ UV1 and \mgi\ UV1 (Figure~\ref{fig:mgiiuv1}). We also include \cii\ UV0.01 and \mnii\ UV1 (Figure~\ref{fig:mniiuv1}), and \tiii\ No. 1, 2 \& 5 (Figures~\ref{fig:tiii125}) as they merit further attention once we obtain more data in the near future. For other lines identified in the composite spectra of ELGs, such as \oii$\,\lambda\lambda2471.0, 2471.1$, their energy-level diagrams can be found in standard textbooks (\eg Appendix E in \citealt{drainebook}).

The terms and symbols were introduced in the nomenclature above. For each transition, we show the vacuum wavelength, the Einstein $A$ coefficient, and the oscillator strength. The black lines with upward arrows represent the observed absorption lines. The green lines with downward arrows show the observed or expected emission lines, with solid indicating permitted lines, dashed semi-forbidden (\ciisemi). The brown lines with downward arrows represent fine-structure transitions.

\section{Comparison of ELGs with local star-forming regions}\label{app:localsf}

\begin{figure*}
\epsscale{1.0}
\plotone{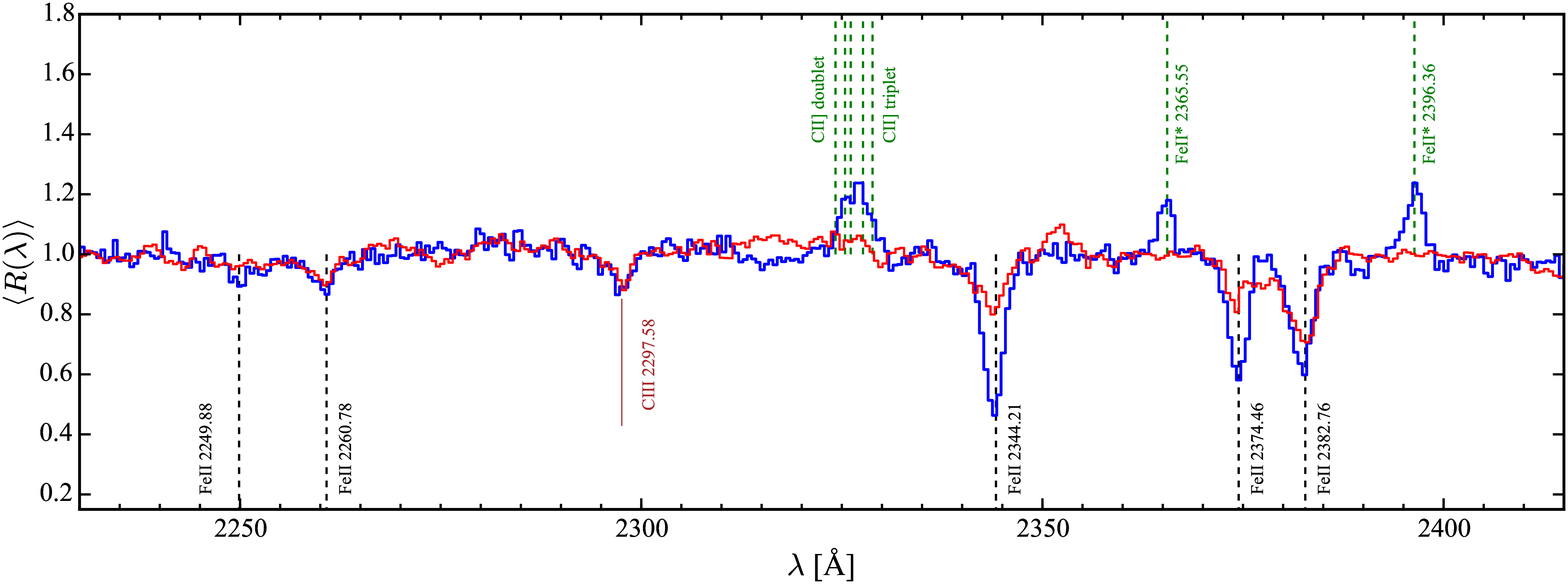}
\vspace{-0.05in}
\plotone{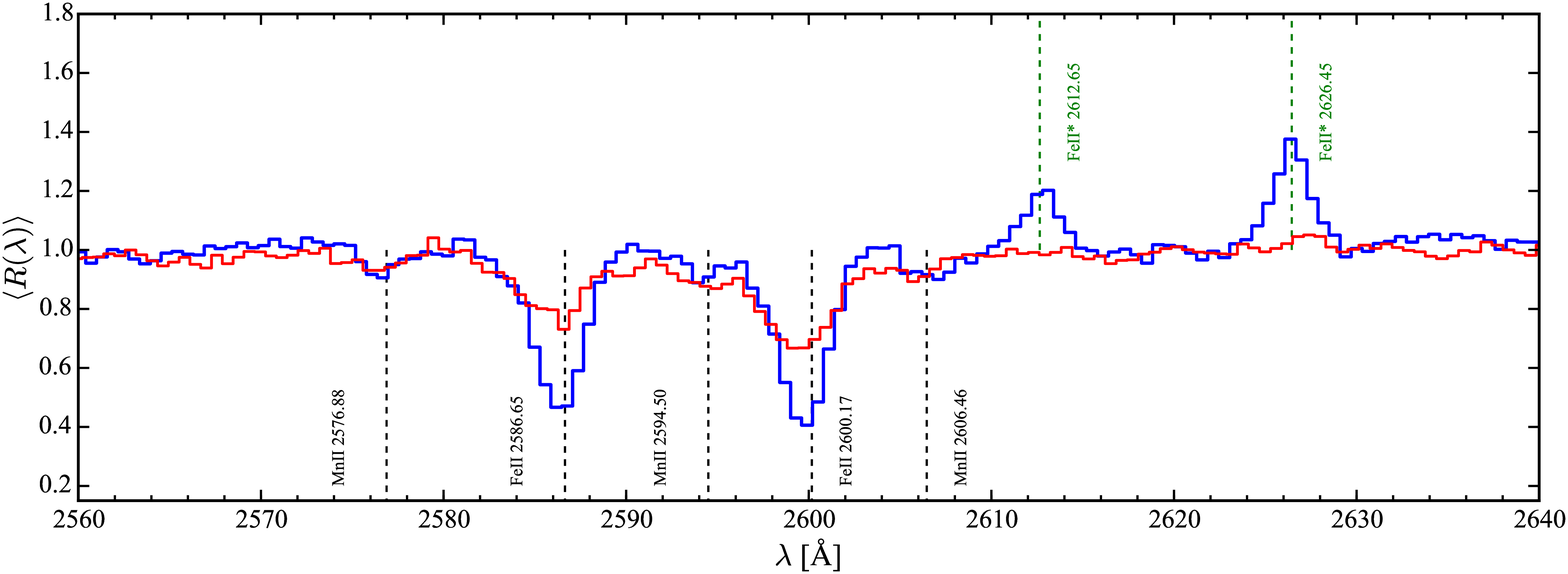}
\vspace{-0.05in}
\plotone{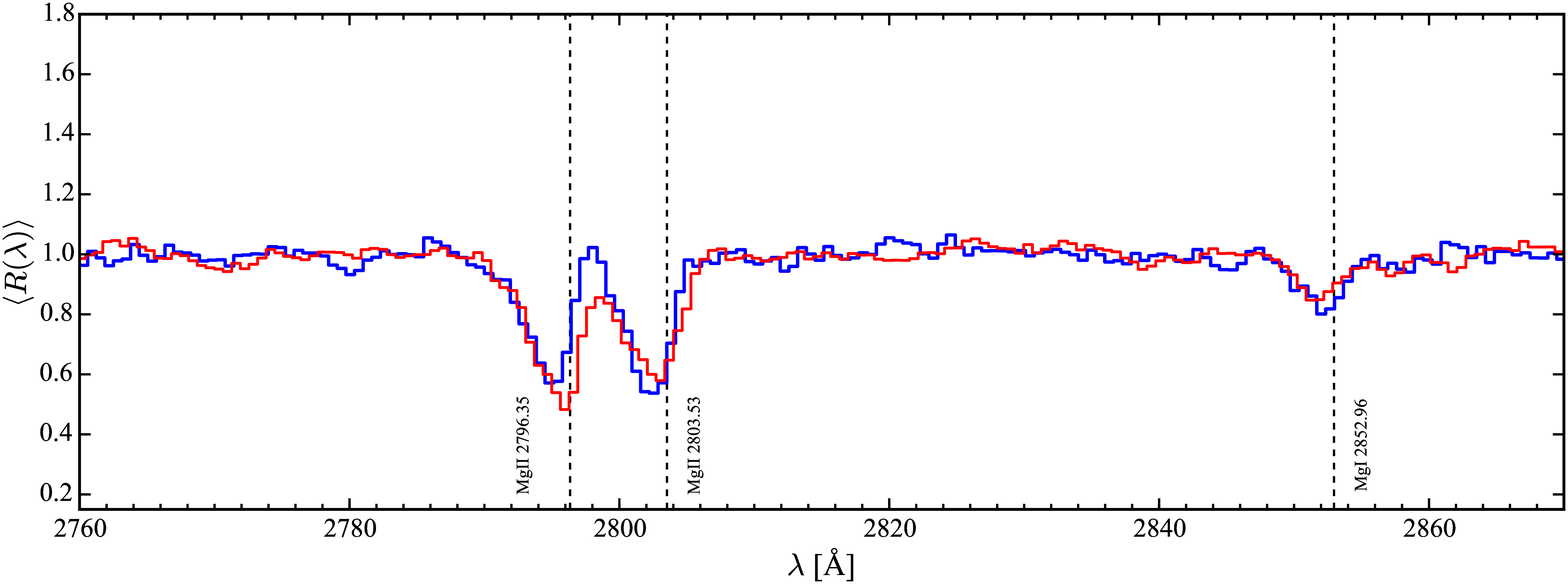}
\caption{The composite continuum-normalized spectrum of $8620$ ELGs at $0.6<z<1.2$ (\textit{blue}) in eBOSS, compared with the composite spectrum of local star-forming regions (\textit{red}) taken with the FOS/GHRS on \textit{HST}.}
\label{fig:localsffine}
\end{figure*}

In Section~\ref{sec:localsf}, we compare the ELG composite spectrum (in blue) with the composite of the nine local SF regions (in red) at $2200\,{\rm \AA}<\lambda<2900\,{\rm \AA}$ in Figure~\ref{fig:localsf}. In Figure~\ref{fig:localsffine}, we zoom in on the most prominent features to show the details, as we did in Figure~\ref{fig:nuvfine} for the comparison with quasar absorption-line systems. Because the UV atlas by \citet{leitherer11a} does not have wavelength coverage longer than $3200\,{\rm \AA}$, we do not compare the spectra at $3000\,{\rm \AA}<\lambda<4000\,{\rm \AA}$ as in the bottom panel of Figure~\ref{fig:nuvfine}. As mentioned in Section~\ref{sec:localsf}, we have corrected the original wavelength in \citet{leitherer11a} by $-0.7\,{\rm \AA}$ based on the position of \ciii$\,\lambda2298$. The main differences are in the non-resonant emission lines and the absorption line ratios. See the main text for details.

\section{The dependences of the NUV features on \oii\ equivalent width and luminosity}\label{app:oii}

The wavelength coverage of eBOSS allows us to study the dependences of the NUV features in the ELG spectra on the \oiidoublet\ properties. We divide the NUV sample of $8620$ ELGs into half subsamples based on the \textit{total} rest equivalent width and luminosity of the \oiidoublet\ doublet. In Section~\ref{sec:oii}, we discuss in depth the emission strength, the observed absorption profiles of \mgii, the emission-corrected absorption strength, and the unified velocity profiles. Here we present some additional details of the analysis. 

\begin{figure*}[t]
\vspace{0.2in}
\epsscale{0.57}
\plotone{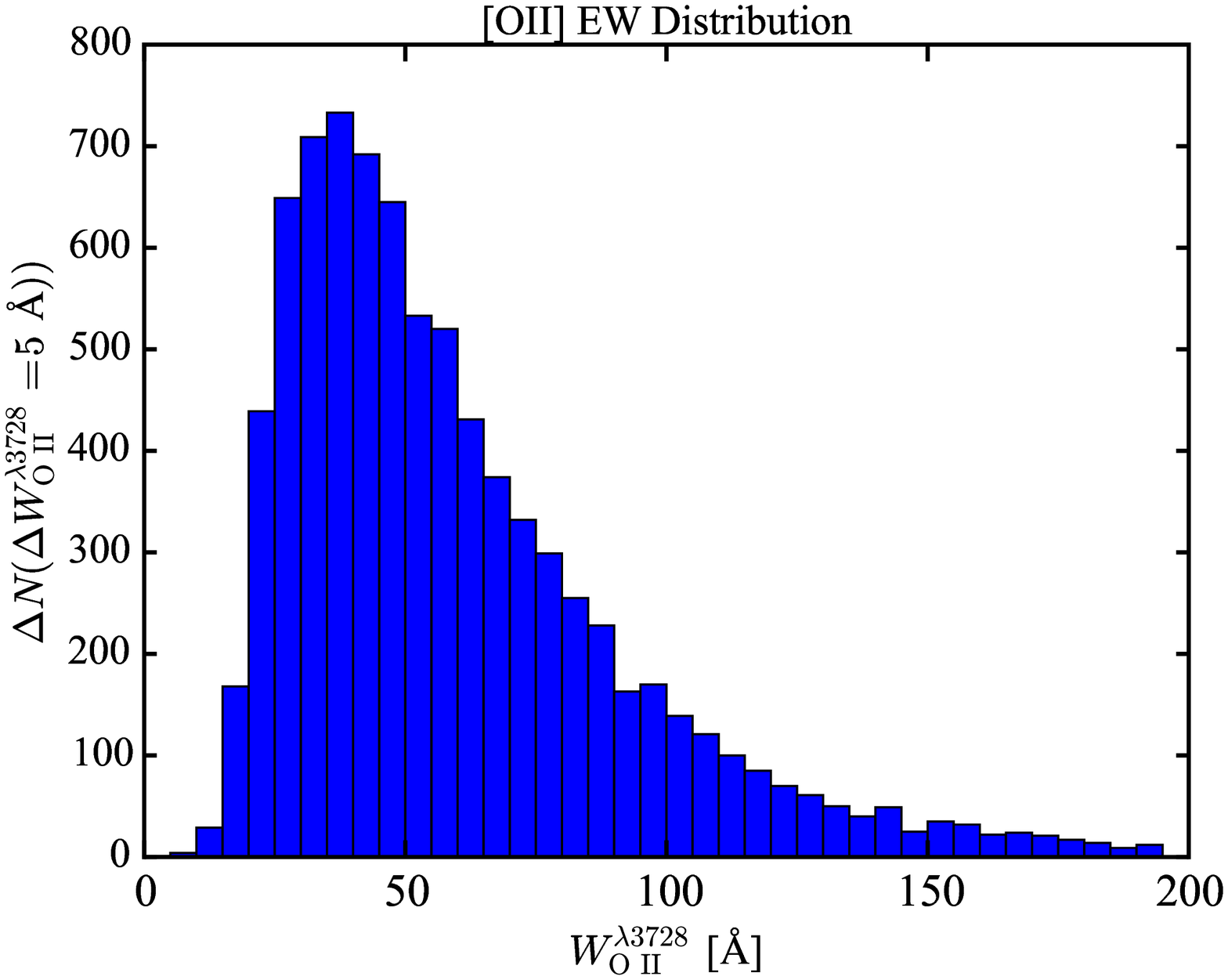}
\plotone{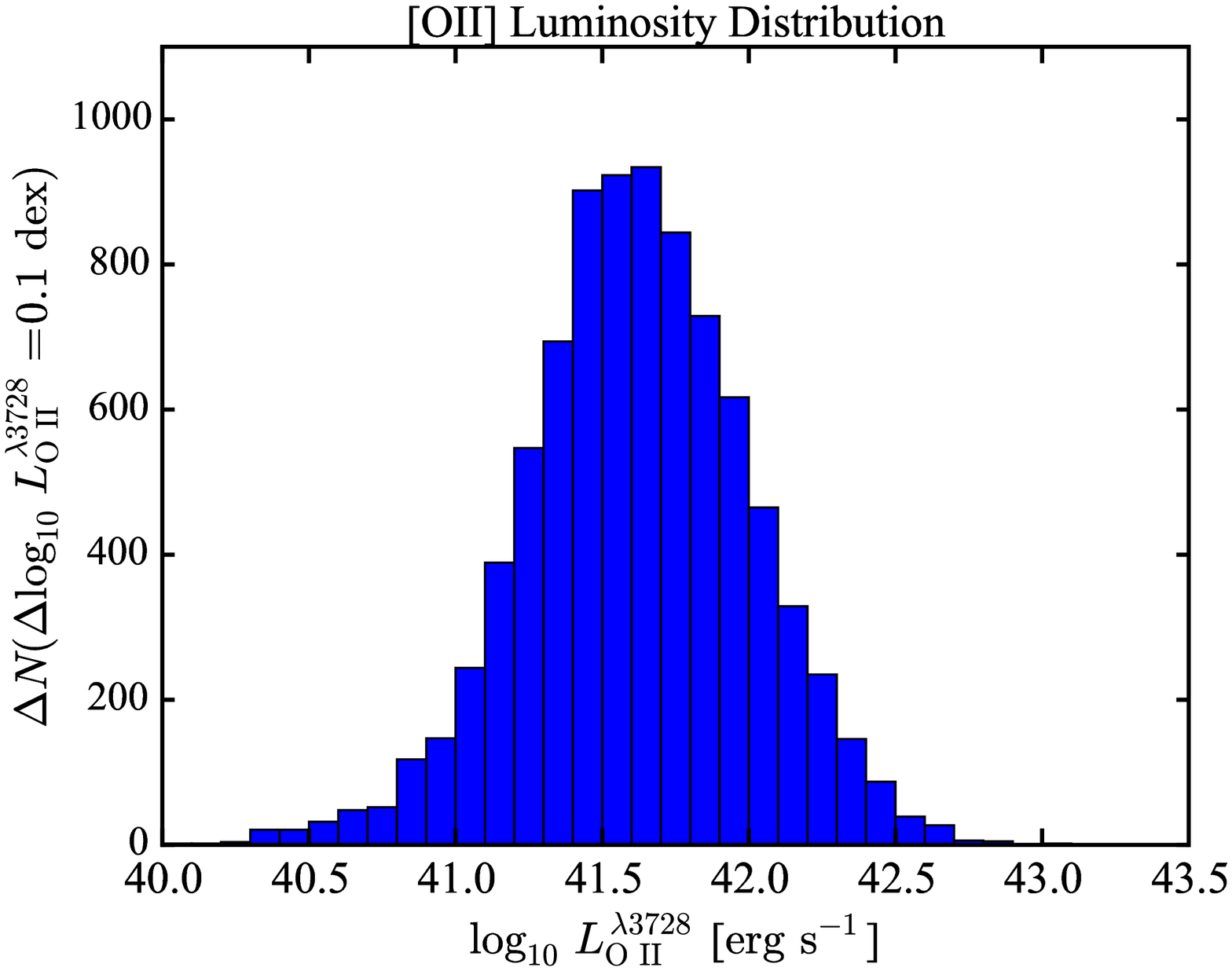}
\caption{The distributions of the total \oiidoublet\ rest equivalent width (\textit{left}) and luminosity (\textit{right}, in logarithmic scale). To calculate the luminosity, we assume the $\Lambda$CDM cosmological model.
}
\label{fig:oiiewlumdist}
\end{figure*}

In Figure~\ref{fig:oiiewlumdist}, we present the distributions of total rest equivalent width ($W^{\lambda3728}_{\rm [O\,II]}$) and luminosity (in logarithmic scale, $\log_{10} L^{\lambda3728}_{\rm [O\,II]}$). The median value of $W^{\lambda3728}_{\rm [O\,II]}$ is about $51.4\,{\rm \AA}$ and the median of $\log_{10} L^{\lambda3728}_{\rm [O\,II]}$ is about $41.6\,{\rm dex}$.

\begin{figure*}[t]
\epsscale{0.57}
\plotone{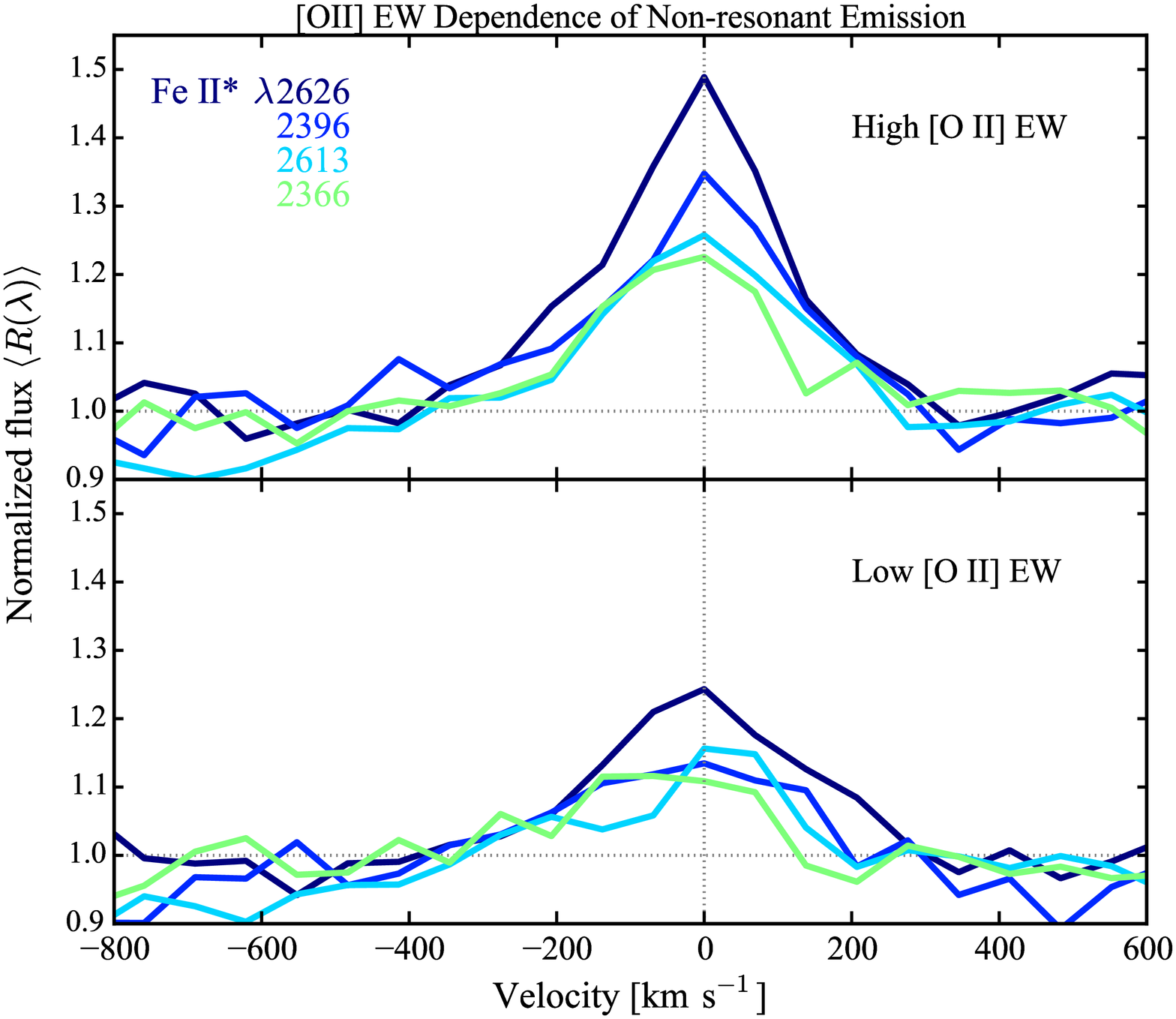}
\epsscale{0.57}
\plotone{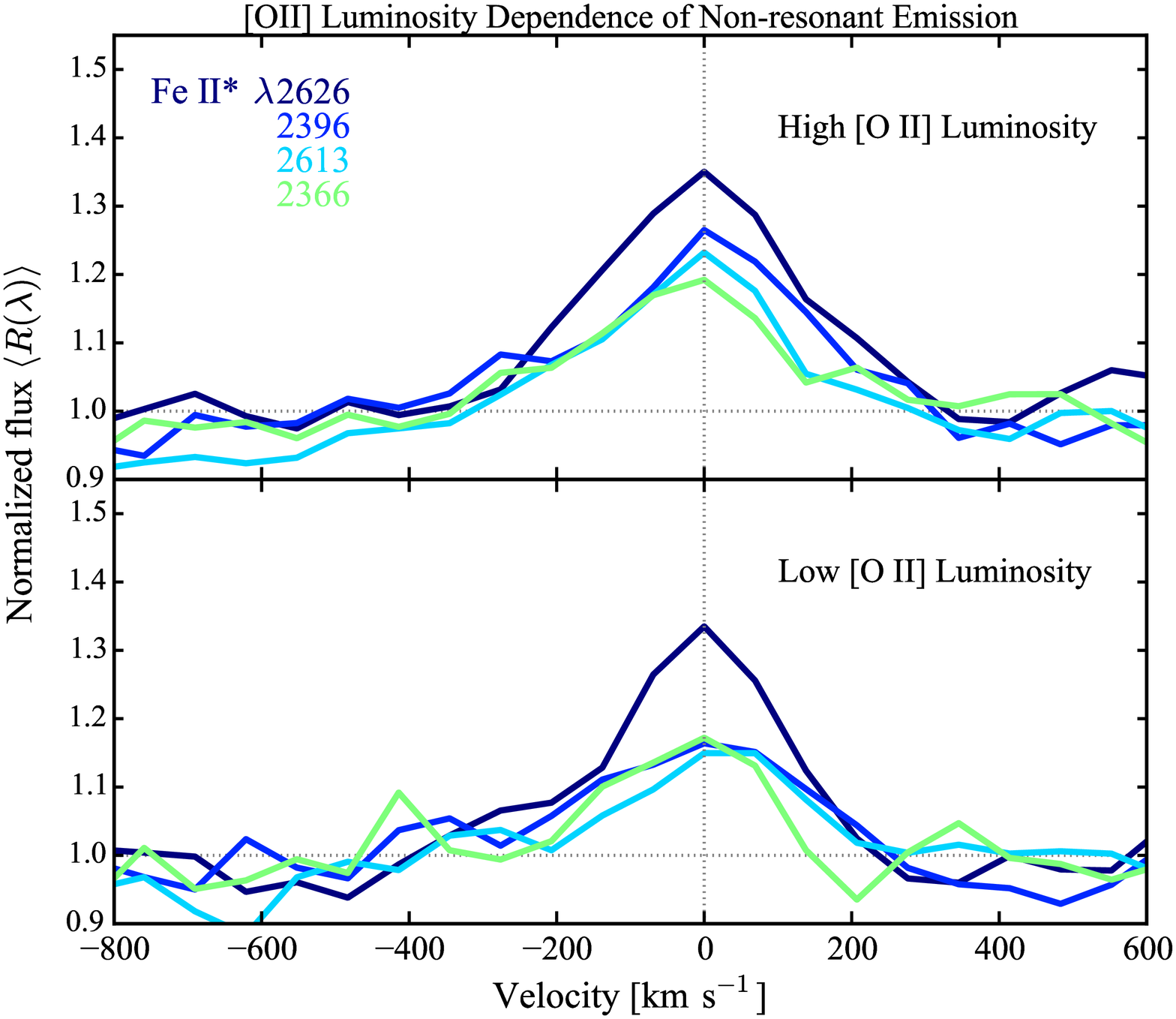}
\caption{The dependences of the emission velocity profiles on the \oiidoublet\ rest equivalent width (\textit{left}) and luminosity (\textit{right}). 
}
\label{fig:emissionprooii}
\end{figure*}

Figure~\ref{fig:emissionprooii} presents the observed velocity profiles of the non-resonant \feii$^*$ emission. The dependence on the \oii\ rest equivalent width is larger than on the luminosity.

\begin{figure*}[t]
\vspace{0.3in}
\epsscale{0.5}
\plotone{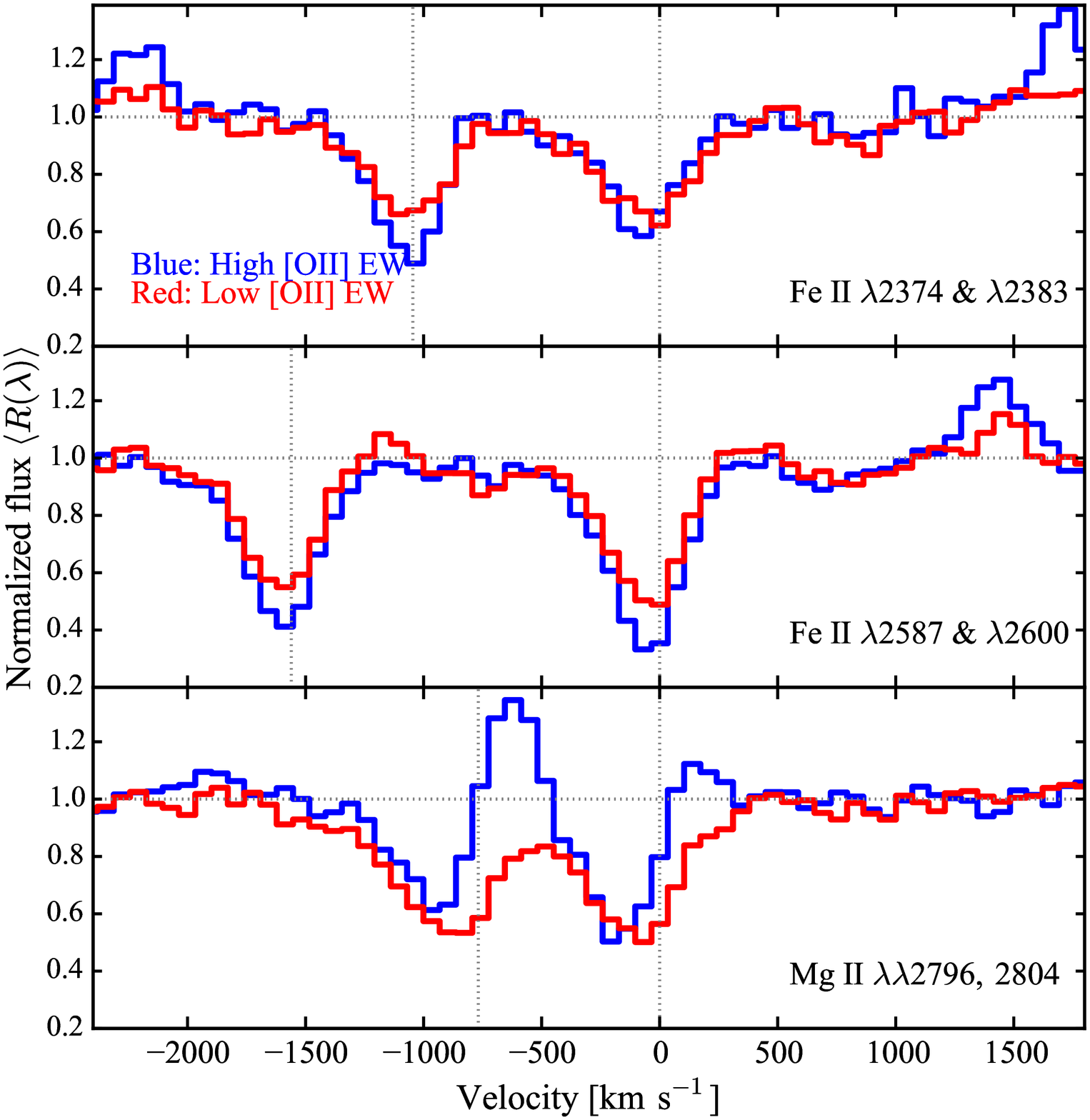}
\epsscale{0.5}
\plotone{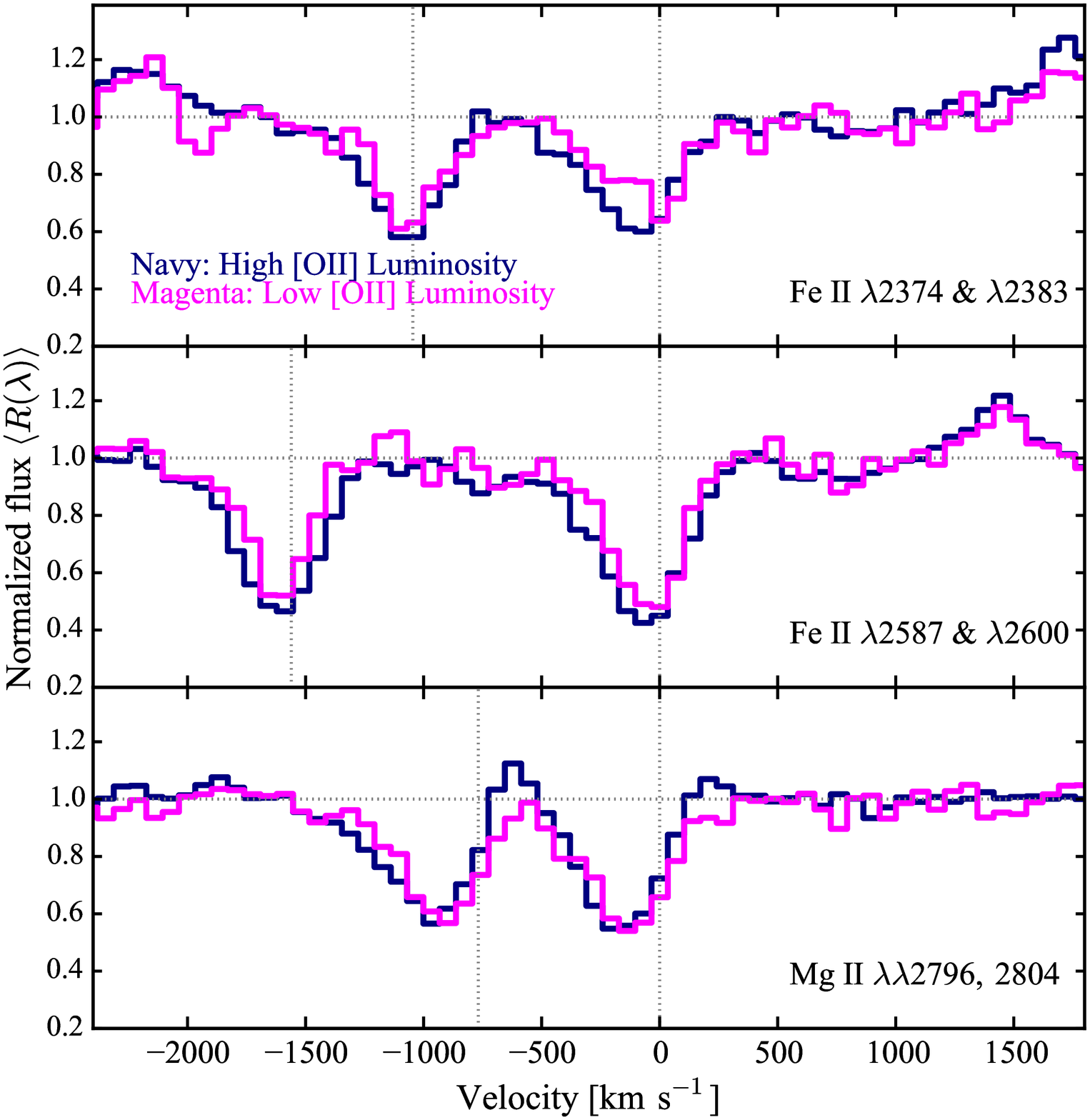}
\caption{The dependences of the observed absorption velocity profiles on the \oiidoublet\ rest equivalent width (\textit{left}) and luminosity (\textit{right}). 
}
\label{fig:absorptionprooii}
\end{figure*}

We show examples of the \textit{observed} absorption profiles for the different subsamples in Figure~\ref{fig:absorptionprooii}. The observed profiles are affected by the scattered emission filling in on top of the true absorption. The largest correlation is between the \mgii\ profiles and the \oii\ rest equivalent width, which we discuss in Section~\ref{sec:oii}.

\begin{figure*}[t]
\epsscale{0.57}
\plotone{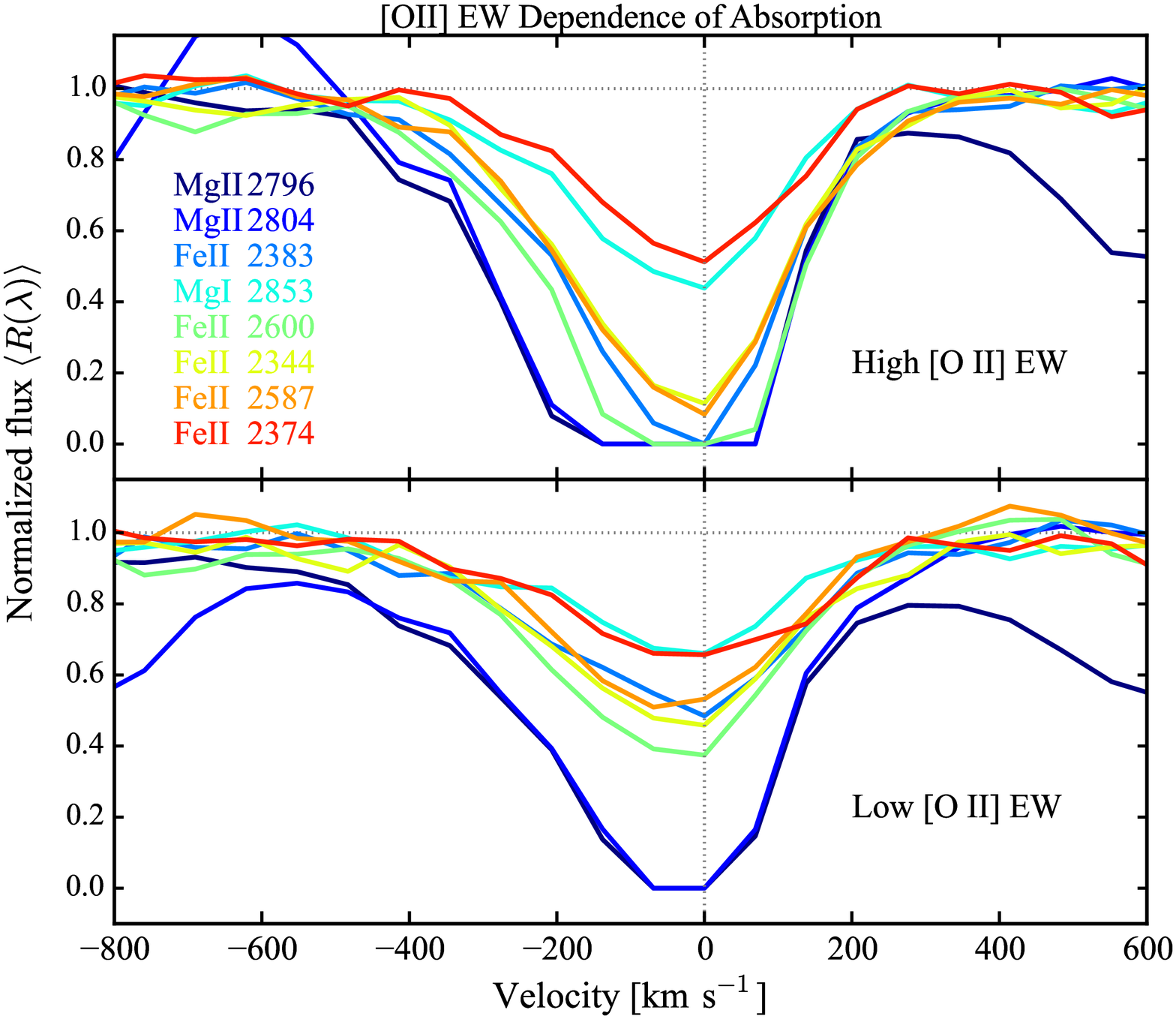}
\epsscale{0.57}
\plotone{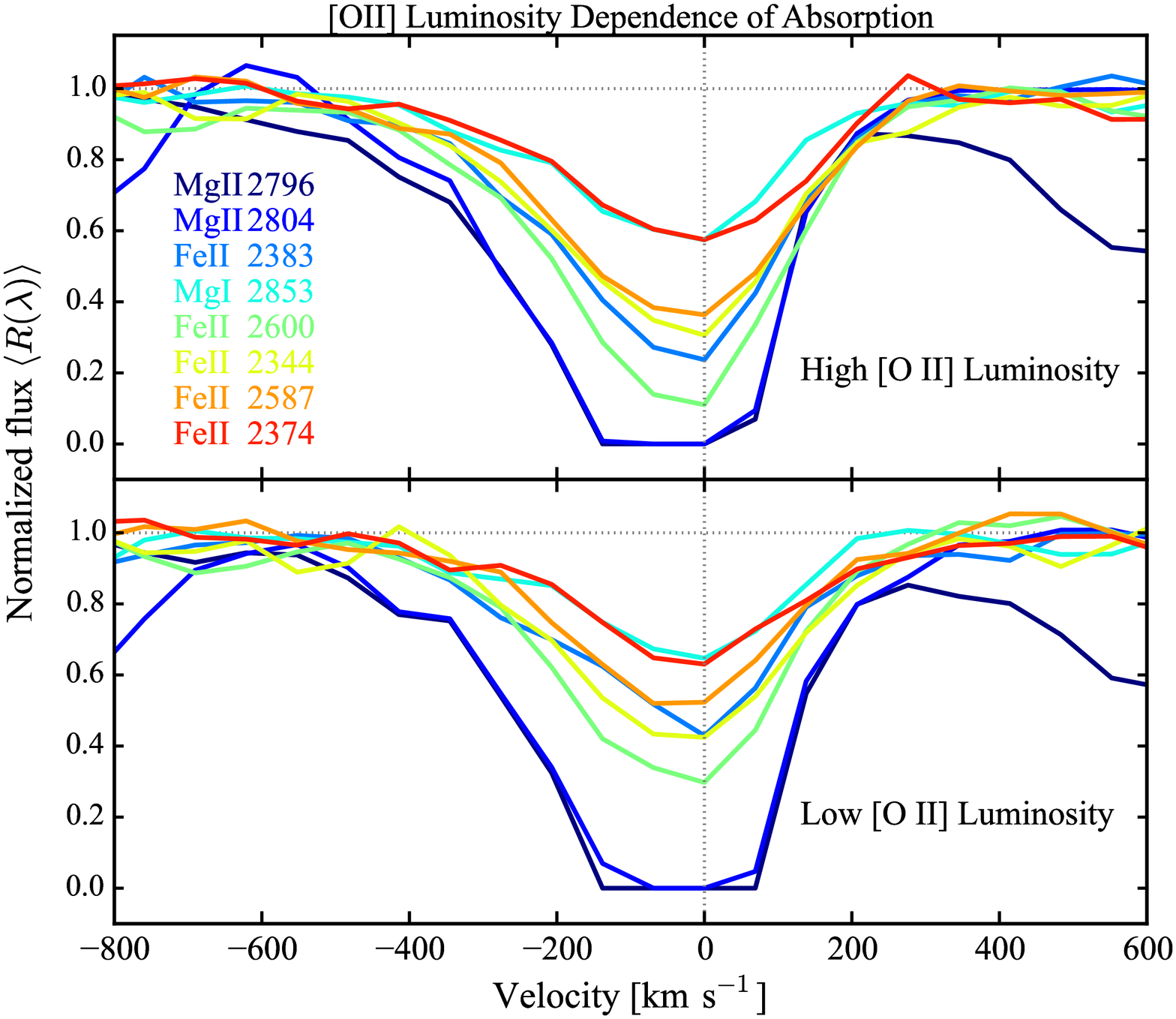}
\caption{The dependences of the emission-corrected absorption velocity profiles on the \oiidoublet\ rest equivalent width (\textit{left}) and luminosity (\textit{right}). The color scales are the same as in Figure~\ref{fig:observedabsorption}, based on the orders given by Eqs.~\ref{eq:ordernonresonant} and \ref{eq:orderresonant}.
}
\label{fig:unifiedabsorptionprooii}
\end{figure*}

Figure~\ref{fig:unifiedabsorptionprooii} presents the emission-corrected absorption profiles, based on the observation-driven iterative method described in Section~\ref{sec:trueabsorption}. We show the correlations between the rest equivalent width of the emission-corrected absorption with the \oiidoublet\ properties in Figure~\ref{fig:absorptionoii}.

\end{document}